\documentclass{notices} 
\usepackage{amsmath}
\usepackage{amsthm}
\usepackage{amsfonts}
\usepackage{amssymb}
\usepackage{latexsym} 
\usepackage{epsfig}
\usepackage{graphicx}

\usepackage{calc}  
\usepackage[matrix,tips,graph,curve,web]{xy}

\makeatletter

\def\@maketitle{%
  \newpage
  \null
  \let \footnote \thanks
    {\normalfont\sffamily\bfseries\Large\noindent\@title \par}%
    \vskip 1em%
    {\normalfont\sffamily 
        \noindent
        \@author
        \par}
  \par
  \vskip 4em}
\def\@seccntformat#1{\csname the#1\endcsname{.\ }}
\renewcommand\section{\@startsection {section}{1}{\z@}%
                                   {-3.0ex \@plus -1ex \@minus -.2ex}%
                                   {1.5ex \@plus.2ex}%
                                   {\normalfont\large\bfseries}}
\renewcommand\subsection{\@startsection{subsection}{2}{\z@}%
                                     {-2.75ex\@plus -1ex \@minus -.2ex}%
                                     {1.5ex \@plus .2ex}%
                                   {\normalfont\large}}
\def\fnum@figure{\normalfont\footnotesize\figurename~\thefigure}

\renewcommand\tableofcontents{%
    \section*{\contentsname
        \@mkboth{%
           \MakeUppercase\contentsname}{\MakeUppercase\contentsname}}%
    \@starttoc{toc}%
    }
\renewcommand*\l@part[2]{%
  \ifnum \c@tocdepth >-2\relax
    \addpenalty\@secpenalty
    \addvspace{2.25em \@plus\p@}%
    \begingroup
      \setlength\@tempdima{3em}%
      \parindent \z@ \rightskip \@pnumwidth
      \parfillskip -\@pnumwidth
      {\leavevmode
       \large \bfseries #1\hfil \hb@xt@\@pnumwidth{\hss #2}}\par
       \nobreak
       \if@compatibility
         \global\@nobreaktrue
         \everypar{\global\@nobreakfalse\everypar{}}%
      \fi
    \endgroup
  \fi}
\renewcommand*\l@section[2]{%
  \ifnum \c@tocdepth >\z@
    \addpenalty\@secpenalty
    \addvspace{1.0em \@plus\p@}%
    \setlength\@tempdima{1.5em}%
    \begingroup
      \parindent \z@ \rightskip \@pnumwidth
      \parfillskip -\@pnumwidth
      \leavevmode \sffamily\bfseries
      \advance\leftskip\@tempdima
      \hskip -\leftskip
      #1\nobreak\hfil \nobreak\hb@xt@\@pnumwidth{\hss #2}\par
    \endgroup
  \fi}
\renewcommand*\l@subsection{\sffamily\@dottedtocline{2}{1.5em}{2.3em}}
\renewcommand*\l@subsubsection{\@dottedtocline{3}{3.8em}{3.2em}}
\renewcommand*\l@paragraph{\@dottedtocline{4}{7.0em}{4.1em}}
\renewcommand*\l@subparagraph{\@dottedtocline{5}{10em}{5em}}
\makeatother


\theoremstyle{plain}
\newtheorem{theorem}{Theorem}[section]
\newtheorem{corollary}[theorem]{Corollary}
\newtheorem{lemma}[theorem]{Lemma}

\theoremstyle{definition}
\newtheorem{definition}[theorem]{Definition}

\newtheorem{question}{Question}

  {\begin{list}{}%
    {%
    \settowidth{\labelwidth}{#1}%
    \setlength{\itemindent}{0pt}%
    \setlength{\labelsep}{1em}%
    \setlength{\leftmargin}{\labelwidth+\parindent+\labelsep}%
    \setlength{\itemsep}{0pt}%
    \setlength{\parsep}{.6ex}}}%
  {\end{list}}

  {\begin{list}{}%
    {%
    \settowidth{\labelwidth}{#1}%
    \setlength{\itemindent}{0pt}%
    \setlength{\labelsep}{1em}%
    \setlength{\leftmargin}{\labelwidth+\labelsep}%
    \setlength{\itemsep}{0pt}%
    \setlength{\parsep}{.6ex}}}%
  {\end{list}}

\makeatletter
\newenvironment{subequations*}{
  \begingroup 
  \let\protect\@nx
  \edef\@tempa{\def\@nx\theparentequation{\theequation}}%
  \@xp\endgroup\@tempa
  \setcounter{parentequation}{\value{equation}}%
  \setcounter{equation}{0}%
  \def\theequation{\theparentequation\alph{equation}}%
  \ignorespaces
}{%
  \setcounter{equation}{\value{parentequation}}%
  \global\@ignoretrue
}

\newcommand{\sgn}{\rm sgn\,}

\def\d/{/\mspace{-6.0mu}/}

\usepackage{amssymb,amsfonts,amsmath,amsthm,url}
\usepackage{dblfloatfix}
\usepackage{paralist,xspace, setspace}
\usepackage[text={6.5in,9in},centering]{geometry}

\usepackage{epsfig,graphicx,wrapfig}

\theoremstyle{definition}

\def\F{\mathcal{F}}

\def\od{\stackrel{\mathrm{def}}{=}}
\def\U{\mathcal{U}}
\def\RR{\mathbb{R}}
\def\supp{\operatorname{supp}}

\def\sgn{\operatorname{sgn}}
\def\FP{\operatorname{FP}}
\def\FPcore{\operatorname{FP_{core}}}
\def\tilFP{\operatorname{\widehat{FP}}}
\def\N{\operatorname{\mathcal{N}}}

\usepackage{multicol}
\usepackage{color}
\definecolor{gold}{rgb}{0.85,.66,0}
\definecolor{cherry}{rgb}{0.9,.1,.2}
\definecolor{burgundy}{rgb}{0.8,.2,.2}
\definecolor{orangered}{rgb}{0.85,.3,0}
\definecolor{orange}{rgb}{0.85,.4,0}
\definecolor{olive}{rgb}{.45,.4,0}
\definecolor{lime}{rgb}{.6,.9,0}
\definecolor{green}{rgb}{.2,.7,0}
\definecolor{darkgreen}{rgb}{.1,.5,0}
\definecolor{grey}{rgb}{.4,.4,.2}
\definecolor{brown}{rgb}{.4,.2,.1}
\definecolor{blue}{rgb}{0,.0, .81}
\definecolor{bluepurple}{rgb}{.3, .0, .7}
\definecolor{black}{rgb}{0,0,0}

\def\black#1{\textcolor{black}{#1}}
\def\carina#1{\black{#1}} 

\begin{document}
\title{Graph rules for recurrent neural network dynamics: extended version}
\author{Carina Curto$^{1}$ and Katherine Morrison$^{2}$\\
\begin{footnotesize}
$^{1}$The Pennsylvania State University, 
$^{2}$The University of Northern Colorado\\
\end{footnotesize}
January 29, 2023}

\maketitle

\tableofcontents
\bigskip

\section{Introduction}

Neurons in the brain are constantly flickering with activity, which can be spontaneous or 
in response to stimuli \cite{Luczak-Neuron}.  Because of positive feedback loops and the potential 
for runaway excitation, real neural networks often possess an abundance of inhibition that serves to 
shape and stabilize the dynamics \cite{Yuste-inhibition, Karnani-inhibition, Yuste-CPG}. 
The excitatory neurons in such networks exhibit intricate patterns of connectivity, whose structure controls the allowed patterns of activity.
A central question in neuroscience is thus: how does network connectivity shape dynamics? 

For a given model, this question becomes a mathematical challenge. The goal is to develop a theory that 
directly relates properties of a nonlinear dynamical system to its underlying graph. Such a theory can provide insights
and hypotheses about how network connectivity constrains activity in real brains. It also opens up new possibilities
for modeling neural phenomena in a mathematically tractable way. 

Here we describe a class of inhibition-dominated neural networks corresponding to directed graphs,
and introduce some of the theory that has been developed to study them. The heart of the theory is a set of 
parameter-independent {\em graph rules} that enables us to directly predict features of the dynamics from 
combinatorial properties of the graph. Specifically, graph rules allow us to constrain, and in some cases fully determine, the collection of stable and unstable fixed points of a network based solely on graph structure. 

Stable fixed points are themselves static attractors of the network, and have long been used as a model of stored memory patterns \cite{Hopfield1982, Hopfield1984}. In contrast, unstable fixed points have been shown to play an important role in shaping {\it dynamic} (non-static) attractors, such as limit cycles \cite{core-motifs}. By understanding the fixed points of simple networks, and how they relate to the underlying architecture, we can gain valuable insight into the high-dimensional nonlinear dynamics of neurons in the brain.

For more complex architectures, built from smaller component subgraphs, we present a series of {\em gluing rules} 
that allow us to determine all fixed points of the network by gluing together those of the components.
These gluing rules are reminiscent of sheaf-theoretic constructions, with fixed points playing the role of sections over subnetworks. 

First, we review some basics of recurrent neural networks and a bit of historical context.

\begin{figure}[!h]
\begin{center}
\includegraphics[width=3.2in]{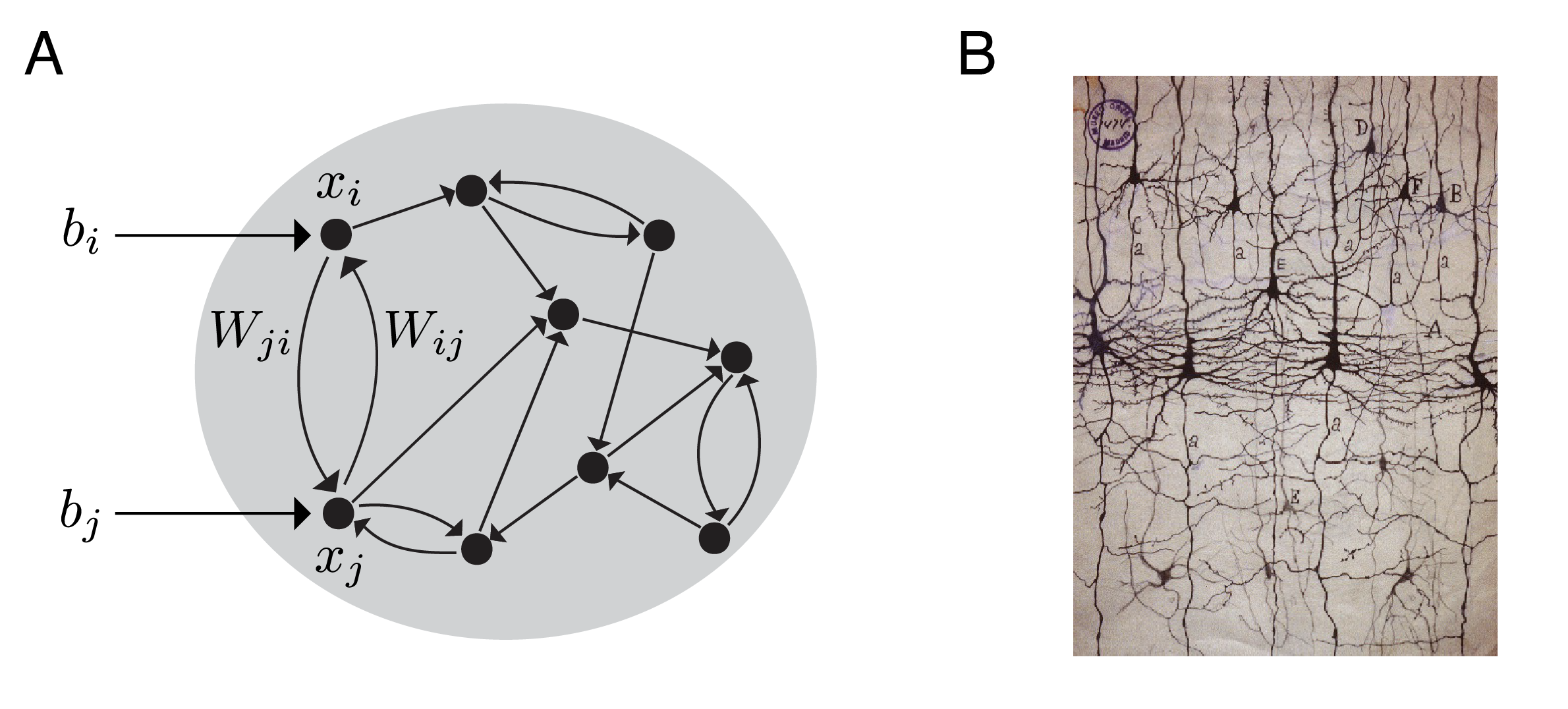}
\end{center}
\caption{(A) Recurrent network setup. (B) A Ramón y Cajal drawing of real cortical neurons.}
\label{fig:network-cartoon}
\vspace{-.1in}
\end{figure}

\paragraph{Basic network setup.}
A {\it recurrent neural network} is a directed graph $G$ together with a prescription for the dynamics on the vertices, which represent neurons (see Figure~\ref{fig:network-cartoon}A). To each vertex $i$ we associate a function $x_i(t)$ that tracks the activity level of neuron $i$ as it evolves in time. To each ordered pair of vertices $(i,j)$ we assign a weight, $W_{ij}$, governing the strength of the influence of neuron $j$ on neuron $i$. In principle, there can be a nonzero weight between any two nodes, with the graph $G$ providing constraints on the allowed values $W_{ij}$, depending on the specifics of the model. 

The dynamics often take the form of a system of ODEs, called a {\em firing rate model} 
\cite{Dayan-Abbott,ErmentroutTerman,AppendixE}:
\begin{eqnarray}
\tau_i\dfrac{dx_i}{dt} &=& - x_i + \varphi\left(\sum_{j=1}^n W_{ij} x_j + b_i\right), \label{network-setup}\\
&=& - x_i + \varphi(y_i), \nonumber
\end{eqnarray}
for  $i = 1,\ldots,n.$ The various terms in the equation are illustrated in Figure~\ref{fig:network-cartoon}, and can be thought of as follows:
\begin{itemize}
\item $x_i = x_i(t)$ is the firing rate of a single neuron $i$ (or the average activity of a subpopulation of neurons);
\item $\tau_i$ is the ``leak'' timescale, governing how quickly a neuron's activity exponentially decays to zero in the absence of external or recurrent input;
\item $W$ is a real-valued matrix of synaptic interaction strengths, with $W_{ij}$ representing the strength of the connection from neuron $j$ to neuron $i$;
\item $b_i = b_i(t)$ is a real-valued external input to neuron $i$ that may or may not vary with time; 
\item $y_i = y_i(t) = \sum_{j=1}^n W_{ij} x_j(t) + b_i(t)$ is the total input to neuron $i$ as a function of time; and
\item $\varphi: \RR \to \RR$ is a nonlinear, but typically monotone increasing function. 
\end{itemize}

Of particular importance for this article is the family of {\it threshold-linear networks} (TLNs). In this case, the nonlinearity is chosen to be the popular threshold-linear (or ReLU) function,
\vspace{-.05in}
$$\varphi(y) = [y]_+ = \max\{0,y\}.\vspace{-.05in}$$
TLNs are common firing rate models that have been used in computational neuroscience for decades \cite{AppendixE,Tsodyks-JN-1997,Seung-Nature,Fitzgerald2022}. The use of threshold-linear units in neural modeling dates back at least to 1958 \cite{Hartline-Ratliff-1958}. In the last 20 years, TLNs have also been shown to be surprisingly tractable mathematically \cite{XieHahnSeung, HahnSeungSlotine, net-encoding, pattern-completion, CTLN-preprint, book-chapter, fp-paper, stable-fp-paper, seq-attractors}, though much of the theory remains under-developed. We are especially interested in {\em competitive} or {\em inhibition-dominated} TLNs, where the $W$ matrix is non-positive so the effective interaction between any pair of neurons is inhibitory. In this case, the activity remains bounded despite the lack of saturation in the nonlinearity \cite{CTLN-preprint}. These networks produce complex nonlinear dynamics and can possess a remarkable variety of attractors \cite{CTLN-preprint,book-chapter,seq-attractors, core-motifs}.

Firing rate models of the form~\eqref{network-setup} are examples of {\it recurrent} networks because the $W$ matrix allows for all pairwise interactions, and there is no constraint that the architecture (i.e., the underlying graph $G$) be feedforward. Unlike deep neural networks, which can be thought of as classifiers implementing a clustering function, recurrent networks are primarily thought of as dynamical systems. And the main purpose of these networks is to model the dynamics of neural activity in the brain. The central question is thus:

\begin{question} \label{ques:Q1}
Given a firing rate model defined by~\eqref{network-setup} with network parameters $(W,b)$ and underlying graph $G$, what are the emergent network dynamics? What can we say about the dynamics from knowledge of $G$ alone?
\end{question}

\begin{figure}[!h]
\begin{center}
\includegraphics[width=3.3in]{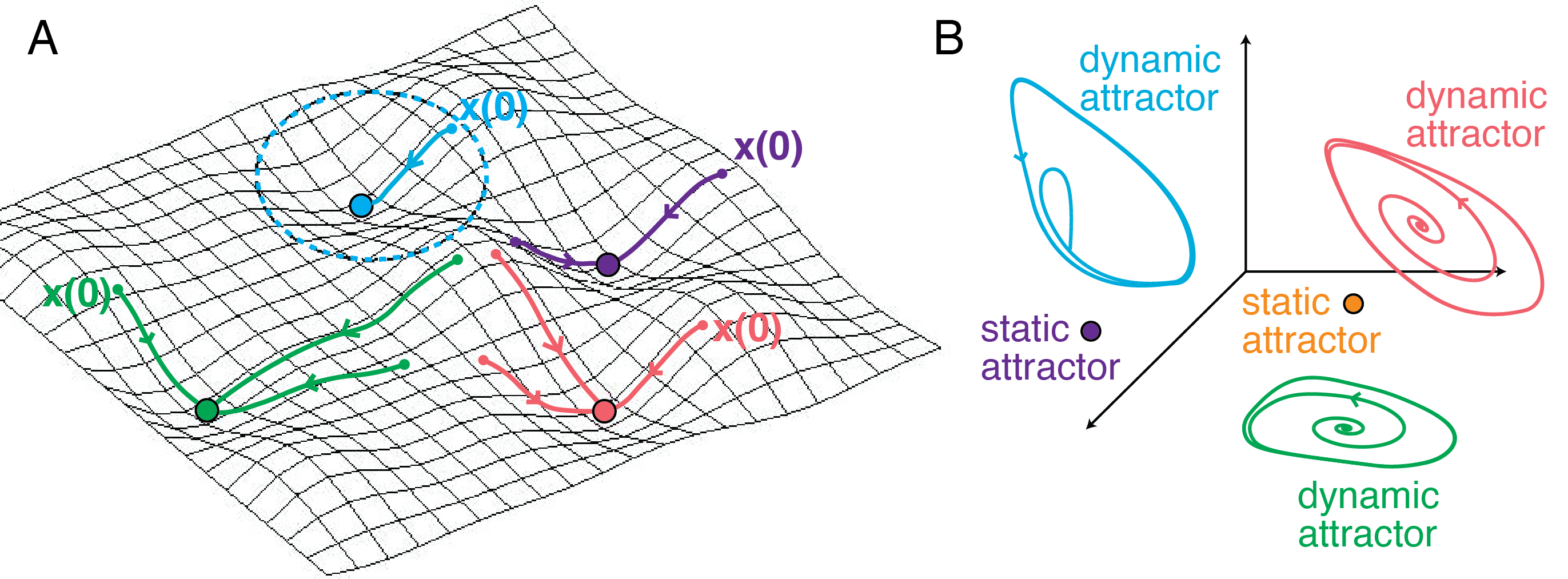}
\end{center}
\caption{{\bf Attractor neural networks.} (A) For symmetric Hopfield networks and symmetric inhibitory TLNs, trajectories are guaranteed to converge to stable fixed point attractors. Sample trajectories are shown, with the basin of attraction for the blue stable fixed point outlined in blue.   (B) For asymmetric TLNs, dynamic attractors can coexist with (static) stable fixed point attractors. }
\label{fig:energy-landscape}
\end{figure}

\vspace{-.05in}
We are particularly interested in understanding the {\it attractors} of such a network, including both stable fixed points and dynamic attractors such as limit cycles. The attractors are important because they comprise the set of possible asymptotic behaviors of the network in response to different inputs or initial conditions (see Figure~\ref{fig:energy-landscape}).

Note that Question~\ref{ques:Q1} is posed for a fixed connectivity matrix $W$, but of course $W$ can change over time (e.g., as a result of learning or training of the network). Here we restrict ourselves to considering constant $W$ matrices; this allows us to focus on understanding network dynamics on a fast timescale, assuming slowly varying synaptic weights. Understanding the dynamics associated to changing $W$ is an important topic, currently beyond the scope of this work.

\vspace{-.1in}
\paragraph{Historical interlude: memories as attractors.} Attractor neural networks became popular in the 1980s as models of associative memory encoding and retrieval. The best-known example from that era is the Hopfield model \cite{Hopfield1982, Hopfield1984}, originally conceived as a variant on the Ising model from statistical mechanics. In the Hopfield model, the neurons can be in one of two states, $s_i \in \{\pm 1\}$, and the activity evolves according to the discrete time update rule:
$$s_i(t+1) = \sgn\left(\sum_{j=1}^n W_{ij} s_j(t) - \theta_i\right).$$
Hopfield's famous 1982 result is that the dynamics are guaranteed to converge to a stable fixed point, provided the interaction matrix $W$ is {\em symmetric}: that is, $W_{ij} = W_{ji}$ for every $i,j \in \{1,\ldots,n\}$. Specifically, he showed that the ``energy'' function,
$$E = -\dfrac{1}{2} \sum_{i,j} W_{ij} s_i s_j + \sum_i \theta_i s_i,$$
decreases along trajectories of the dynamics, and thus acts as a Lyapunov function \cite{Hopfield1982}. The stable fixed points are local minima of the energy landscape (Figure~\ref{fig:energy-landscape}A). A stronger, more general convergence result for competitive neural networks was shown in \cite{CohenGrossberg1983}.

These fixed points are the only attractors of the network, and they represent the set of memories encoded in the network. Hopfield networks perform a kind of {\em pattern completion}: given an initial condition $s(0)$, the activity evolves until it converges to one of multiple stored patterns in the network. If, for example, the individual neurons store black and white pixel values, this process could input a corrupted image and recover the original image, provided it had previously been stored as a stable fixed point in the network by appropriately selecting the weights of the $W$ matrix.  The novelty at the time was the nonlinear phenomenon of multistability: namely, that the network could encode many such stable equilibria and thus maintain an entire catalogue of stored memory patterns. The key to Hopfield's convergence result was the requirement that $W$ be a symmetric interaction matrix. Although this was known to be an unrealistic assumption for real (biological) neural networks, it was considered a tolerable price to pay for guaranteed convergence. One did not want an associative memory network that wandered the state space indefinitely without ever recalling a definite pattern.

Twenty years later, Hahnloser, Seung, and others followed up and proved a similar convergence result in the case of  symmetric inhibitory threshold-linear networks \cite{HahnSeungSlotine}. 
\carina{ Specifically, they found a Lyapunov-like function
 $$L = \dfrac{1}{2}x^T(I-W)x - b^Tx,$$
following the notation in~\eqref{network-setup} with $\varphi(y) = [y]_+$. For fixed $b$, it can easily be shown that $L$ is strictly decreasing along trajectories of the TLN dynamics, and minima of $L$ correspond to steady states -- provided $W$ is symmetric and $I-W$ is copositive \cite[Theorem 1]{HahnSeungSlotine}.}
More results on the collections of stable fixed points that can be simultaneously encoded in a symmetric TLN can be found in \cite{flex-memory, net-encoding, pattern-completion}, including some unexpected connections to Cayley-Menger determinants and classical distance geometry. 

In all of this work, stable fixed points have served as the model for encoded memories. Indeed, these are the only types of attractors that arise for symmetric Hopfield networks or symmetric TLNs. Whether or not guaranteed convergence to stable fixed points is desirable, however, is a matter of perspective. For a network whose job it is to perform pattern completion or classification for static images (or codewords), as in the classical Hopfield model, this is exactly what one wants. But it is also important to consider memories that are temporal in nature, such as sequences and other dynamic patterns of activity. Sequential activity, as observed in central pattern generator circuits (CPGs) and spontaneous activity in hippocampus and cortex, is more naturally modeled by dynamic attractors such as limit cycles. This requires shifting attention to the {\it asymmetric} case, in order to be able to encode attractors that are not stable fixed points (Figure~\ref{fig:energy-landscape}B).

\begin{figure}[!h]
\begin{center}
\includegraphics[width=3.25in]{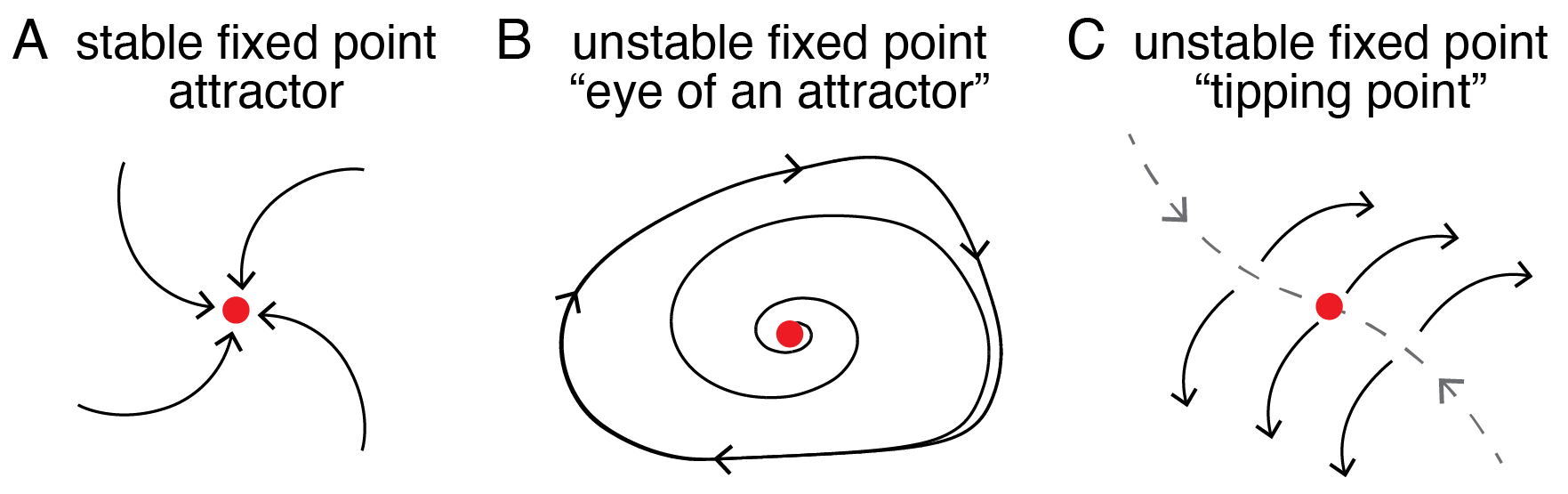}
\end{center}
\vspace{-.1in}
\caption{{\bf Stable and unstable fixed points.} (A) Stable fixed points are attractors of the network.  (B-C) Unstable fixed points are not themselves attractors, but certain unstable fixed points seem to correspond to dynamic attractors (B), while others function solely as tipping points between multiple attractors (C).}
\label{fig:fp-types}
\end{figure}

\vspace{-.15in}
\paragraph{Beyond stable fixed points.}
When the symmetry assumption is removed, TLNs can support a rich variety of dynamic attractors such as limit cycles, quasiperiodic attractors, and even strange (chaotic) attractors. Indeed, this richness can already be observed in a special class of TLNs called combinatorial threshold-linear networks (CTLNs), introduced in Section~\ref{sec:ctlns}. These networks are defined from directed graphs, and the dynamics are almost entirely determined by the graph structure. A striking feature of CTLNs is that the dynamics are shaped not only by the stable fixed points, but also the {\em unstable} fixed points. In particular, we have observed a direct correspondence between certain types of unstable fixed points and dynamic attractors (see Figure~\ref{fig:fp-types}) \cite{core-motifs}. This is reviewed in Section~\ref{sec:core-motifs}.

Despite exhibiting complex, high-dimensional, nonlinear dynamics, recent work has shown that TLNs -- and especially CTLNs -- are surprisingly tractable mathematically. Motivated by the relationship between fixed points and attractors, a great deal of progress has been made on the problem of relating fixed point structure to network architecture. In the case of CTLNs, this has resulted in a series of {\em graph rules}: theorems that allow us to rule in and rule out potential fixed points based purely on the structure of the underlying graph \cite{fp-paper, book-chapter, seq-attractors}. In Section~\ref{sec:graph-rules}, we give a novel exposition of graph rules, and introduce several {\em elementary graph rules} from which the others can be derived. 

Inhibition-dominated TLNs and CTLNs also display a remarkable degree of modularity. Namely, attractors associated to smaller networks can be embedded in larger ones with minimal distortion \cite{core-motifs}. This is likely a consequence of the high levels of background inhibition: it serves to stabilize and preserve local properties of the dynamics. These networks also exhibit a kind of compositionality, wherein fixed points and attractors of subnetworks can be effectively ``glued'' together into fixed points and attractors of a larger network. These local-to-global relationships are given by a series of theorems we call {\em gluing rules}, given in Section~\ref{sec:gluing-rules}.

\section{TLNs and hyperplane arrangements} \label{sec:hyperplanes}
For firing rate models with threshold-nonlinearity $\varphi(y) = [y]_+ = \max\{0,y\},$ the network equations~\eqref{network-setup} become
\begin{eqnarray}
\dfrac{dx_i}{dt} &=& -x_i + \left[\sum_{j=1}^n W_{ij}x_j+b_i \right]_+ \label{eq:TLN-dynamics}\\
&=&  -x_i + [y_i]_+, \nonumber
\end{eqnarray}
for  $i = 1,\ldots,n.$ We also assume $W_{ii} = 0$ for each $i$. Note that the leak timescales have been set to $\tau_i = 1$ for all $i$. We thus measure time in units of this timescale.

For constant $W$ matrix and input vector $b$, the equations
$$y_i = \sum_{j=1}^n W_{ij}x_j+b_i = 0,$$
define a hyperplane arrangement $\mathcal{H} = \mathcal{H}(W,b) = \{H_1,\ldots,H_n\}$ in $\RR^n$. The $i$-th hyperplane $H_i$ is defined by $y_i = \vec{n}_i \cdot x + b_i = 0$, with normal vector $\vec{n}_i = (W_{i1},\ldots,W_{in}),$ population activity vector $x = (x_1,\ldots,x_n)$, and affine shift $b_i$. If $W_{ij} \neq 0$, then $H_i$ intersects the $j$-th coordinate axis at the point $x_j = - b_i/W_{ij}$. $H_i$ is parallel to the $i$-th axis.

The hyperplanes $\mathcal{H}$ partition the positive orthant $\RR^n_{\geq 0}$ into chambers. Within the interior of any chamber, each point $x$ is on the plus or minus side of each hyperplane $H_i$. The equations thus reduce to a linear system of ODEs, with the equation for each $i = 1,\ldots,n$ being either 
\carina{
$$\dfrac{dx_i}{dt} = -x_i + y_i = -x_i + \sum_{j=1}^n W_{ij}x_j+b_i, \text{ if } y_i>0,$$ 
or
$$\dfrac{dx_i}{dt} = -x_i, \text{ if } y_i \leq 0.$$
In particular, TLNs are piecewise-linear dynamical systems with a different linear system, $L_\sigma$, governing the dynamics in each chamber \cite{CTLN-preprint}.}

A {\it fixed point} of a TLN~\eqref{eq:TLN-dynamics} is a point $x^* \in \RR^n$ that satisfies $dx_i/dt|_{x=x^*} = 0$ for each $i \in \{1, \ldots, n\}$. In particular, we must have
\begin{equation}\label{eq:at-fp}
x_i^* = [y_i^*]_+ \text{ for all } i = 1,\ldots,n,
\end{equation}
where $y_i^*$ is $y_i$ evaluated at the fixed point.
We typically assume a nondegeneracy condition on $(W,b)$ \cite{fp-paper,CTLN-preprint}, which guarantees that each linear system is nondegenerate and has a single fixed point. This fixed point may or may not lie within the chamber where its corresponding linear system applies. The fixed points of the TLN are precisely the fixed points of the linear systems that lie within their respective chambers. 

\begin{figure}[!h]
\begin{center}
\includegraphics[width=3.25in]{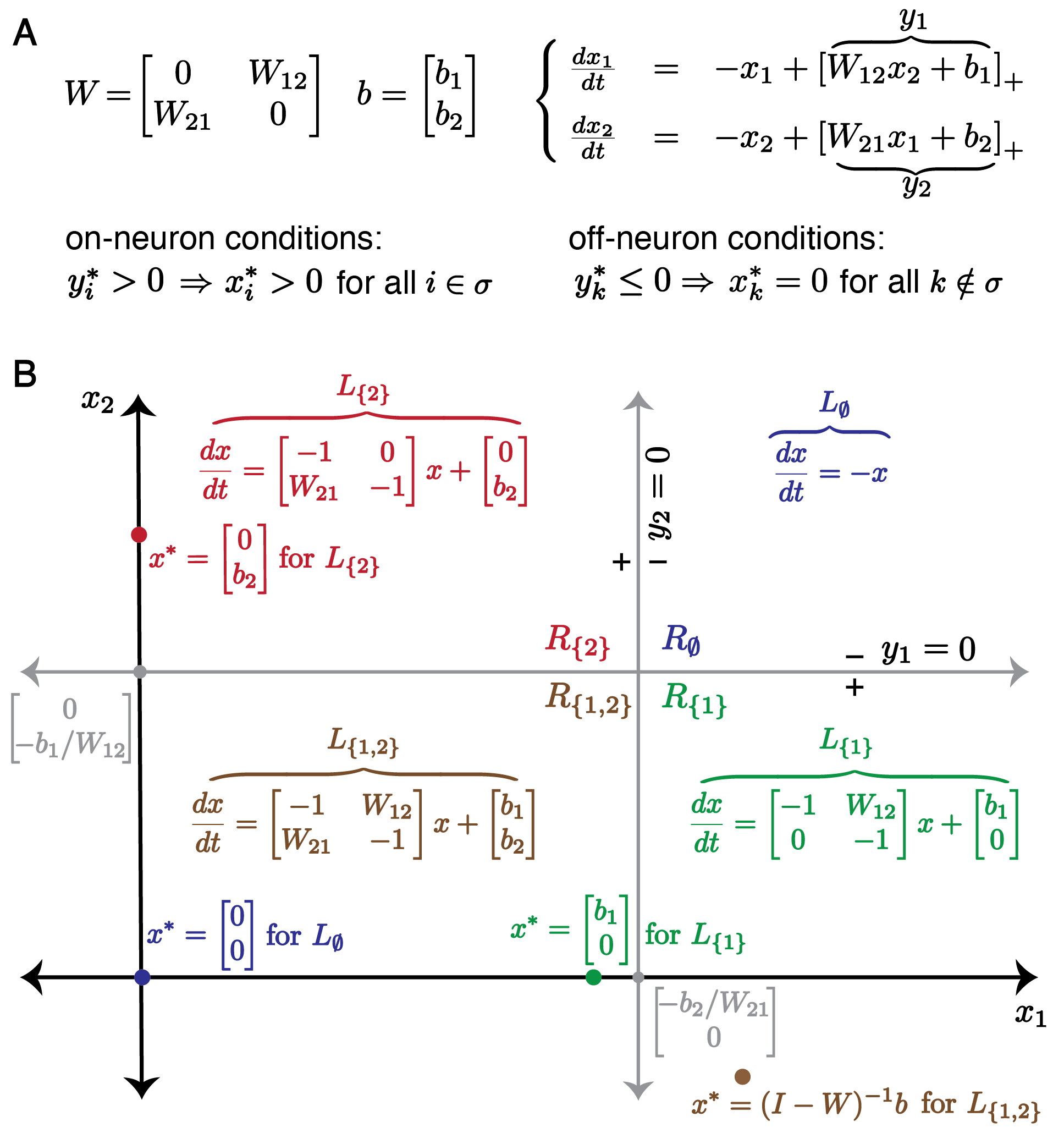}
\end{center}
\vspace{-.1in}
\caption{{\bf TLNs as a patchwork of linear systems.} (A) The connectivity matrix $W$, input $b$, and differential equations for a TLN with $n=2$ neurons. (B) The state space is divided into chambers (regions) $R_\sigma$, each having dynamics governed by a different linear system $L_\sigma$. The chambers are defined by the hyperplanes $\{H_i\}_{i=1,2}$, with $H_i$ defined by $y_i = 0$ (gray lines).}
\label{fig:patchwork}
\end{figure}

Figure~\ref{fig:patchwork} illustrates the hyperplanes and chambers for a TLN with $n=2$. Each chamber, denoted as a region $R_\sigma$, has its own linear system of ODEs, $L_\sigma,$ for $\sigma = \emptyset, \{1\}, \{2\},$ or $\{1,2\}$. The fixed point corresponding to each linear system is denoted by $x^*$, in matching color. Note that only chamber $R_{\{2\}}$ contains its own fixed point (in red). This fixed point, $x^* = [0, b_2]^T$, is thus the only fixed point of the TLN.

\begin{figure*}[!h]
\vspace{-.1in}
\begin{center}
\includegraphics[width=6.25in]{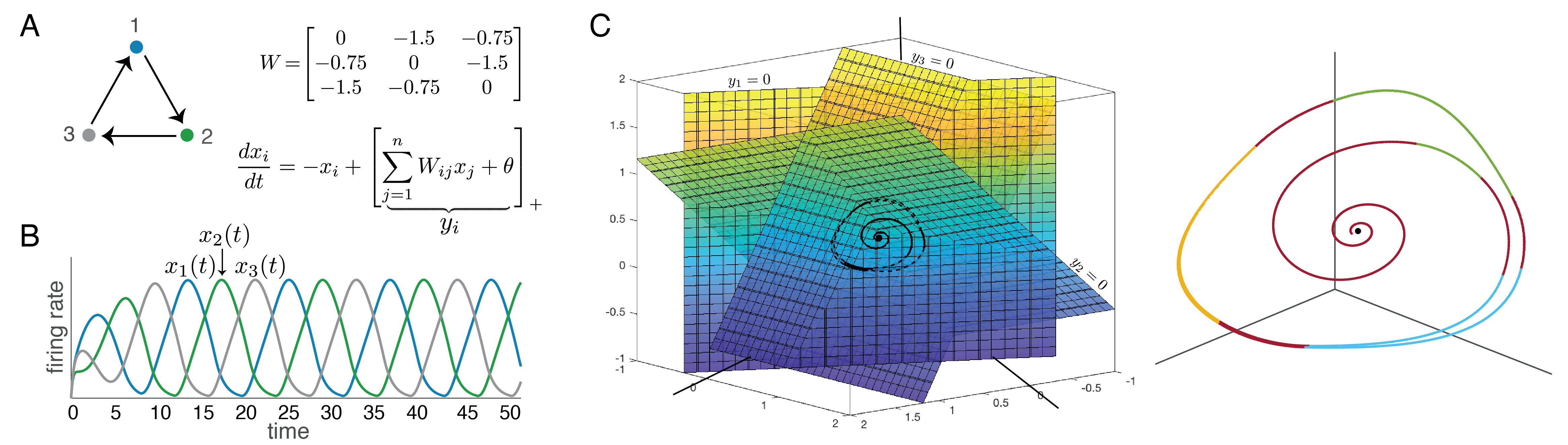}
\end{center}
\vspace{-.1in}
\caption{{\bf A network on $n=3$ neurons, its hyperplane arrangement, and limit cycle.}  (A) A TLN whose connectivity matrix $W$ is dictated by a $3$-cycle graph, together with the TLN equations. (B) The TLN from A produces firing rate activity in a periodic sequence. (C) (Left) The hyperplane arrangement defined by the equations $y_i = 0$, with a trajectory initialized near the fixed point shown in black. (Right) A close-up of the trajectory, spiraling out from the unstable fixed point and falling into a limit cycle. Different colors correspond to different chambers of the hyperplane arrangement through which the trajectory passes.}
\label{fig:hyperplanes}
\vspace{-.1in}
\end{figure*}

Figure~\ref{fig:hyperplanes} shows an example of a TLN on $n=3$ neurons. The $W$ matrix is constructed from a $3$-cycle graph and $b_i = \theta = 1$ for each $i$. The dynamics fall into a limit cycle where the neurons fire in a repeating sequence that follows the arrows of the graph. This time, the TLN equations define a hyperplane arrangement in $\RR^3$, again with each hyperplane $H_i$ defined by $y_i = 0$ (Figure~\ref{fig:hyperplanes}C). An initial condition near the unstable fixed point in the all + chamber (where $y_i > 0$ for each $i$) spirals out and converges to a limit cycle that passes through four distinct chambers. Note that the threshold nonlinearity is critical for the model to produce nonlinear behavior such as limit cycles; without it, the system would be linear. It is, nonetheless, nontrivial to prove that the limit cycle shown in Figure~\ref{fig:hyperplanes} exists. A recent proof was given for a special family of TLNs constructed from any $k$-cycle graph \cite{Horacio-paper}. 

\vspace{-.1in}
\paragraph{The set of all fixed points $\boldsymbol{\FP(W,b)}$.}
A central object that is useful for understanding the dynamics of TLNs is the collection of {\it all} fixed points of the network, both stable and unstable. 
The  {\it support} of a fixed point $x^* \in \RR^n$ is the subset of active neurons, 
\vspace{-.075in}
$$\supp{x^*} \od \{i \mid x^*_i>0\}. \vspace{-.075in}$$ 
Our nondegeneracy condition (that is generically satisfied) guarantees we can have at most one fixed point per chamber of the hyperplane arrangement $\mathcal{H}(W,b)$, and thus at most one fixed point per support.  We can thus label all the fixed points of a given network by their supports:
\begin{small}
\begin{align} 
\FP(W,b) &\od& \{\sigma \subseteq [n] \mid   \sigma = \supp{x^*}, \text{ for some } \label{eq:FP_Wb}\\
& & \text{fixed pt } x^* \text{ of the TLN } (W,b) \}, \nonumber
\end{align}
\end{small}
where 
$$[n] \od \{1,\ldots,n\}.$$
For each support $\sigma \in \FP(W,b)$, the fixed point itself is easily recovered. Outside the support, $x_i^* = 0$ for all $i \not\in \sigma$. Within the support, $x^*$ is given by:
\vspace{-.075in}
$$x_\sigma^* = (I-W_\sigma)^{-1} b_\sigma. \vspace{-.075in}$$
Here $x_\sigma^*$ and $b_\sigma$ are the column vectors obtained by restricting $x^*$ and $b$ to the indices in $\sigma$, and $W_\sigma$ is the induced principal submatrix obtained by restricting rows and columns of $W$ to $\sigma$.

From~\eqref{eq:at-fp}, we see that a fixed point with $\supp{x^*} = \sigma$ must satisfy the ``on-neuron'' conditions, $y_i^*>0$ for all $i \in \sigma$, as well as the ``off-neuron" conditions, $y_k^* \leq 0$ for all $k \notin \sigma$, to ensure that $x_i^*>0$ for each $i \in \sigma$ and $x_k^* = 0$ for each $k \notin \sigma$. Equivalently, these conditions guarantee that the fixed point $x^*$ of $L_\sigma$ lies inside its corresponding chamber, $R_\sigma.$ 
Note that for such a fixed point, the values $x_i^*$ for $i \in \sigma$ depend only on the restricted subnetwork $(W_\sigma,b_\sigma)$. Therefore, the on-neuron conditions for $x^*$ in $(W,b)$ are satisfied if and only if they hold in $(W_\sigma,b_\sigma)$. 
Since the off-neuron conditions are trivially satisfied in $(W_\sigma,b_\sigma)$, it follows that $\sigma \in \FP(W_\sigma,b_\sigma)$ is a necessary condition for $\sigma \in \FP(W,b)$. It is not, however, sufficient, as the off-neuron conditions may fail in the larger network. \carina{Satisfying all the on- and off-neuron conditions, however, is both necessary and sufficient to guarantee $\sigma \in \FP(G)$ \cite{book-chapter, fp-paper}.}

Conveniently, the off-neuron conditions are independent and can be checked one neuron at a time. Thus, 
\vspace{-.05in}
$$\sigma \in \FP(W,b) \Leftrightarrow \sigma \in \FP(W_{\sigma \cup k},b_{\sigma \cup k}) \text{ for all } k \notin \sigma.\vspace{-.05in}$$
When $\sigma \in \FP(W_\sigma,b_\sigma)$ satisfies all the off-neuron conditions, so that $\sigma \in \FP(W,b)$, we say that $\sigma$ {\it survives} to the larger network; otherwise, we say $\sigma$ {\it dies}. 

The fixed point corresponding to $\sigma \in \FP(W,b)$ is {\em stable} if and only if all eigenvalues of $-I+W_\sigma$ have negative real part. For competitive (or inhibition-dominated) TLNs, all fixed points -- whether stable or unstable -- have a stable manifold. This is because competitive TLNs have $W_{ij} \leq 0$ for all $i,j \in [n]$. Applying the Perron-Frobenius theorem to $-I+W_\sigma$, we see that the largest magnitude eigenvalue is guaranteed to be real and negative. The corresponding eigenvector provides an attracting direction into the fixed point. Combining this observation with the nondegeneracy condition reveals that the unstable fixed points are all hyperbolic (i.e., saddle points).

\vspace{-.1in}
\section{Combinatorial threshold-linear\\ networks} \label{sec:ctlns}
\vspace{-.05in}


{\it Combinatorial threshold-linear networks} (CTLNs) are a special case of competitive (or inhibition-dominated) TLNs, with the same threshold nonlinearity, that were first introduced in \cite{CTLN-preprint, book-chapter}. What makes CTLNs special is that we restrict to having only two values for the connection strengths $W_{ij}$, for $i \neq j$. These are obtained as follows from a directed graph $G$, where $j \to i$ indicates that there is an edge from $j$ to $i$ and $j \not\to i$ indicates that there is no such edge:
\begin{equation} \label{eq:binary-synapse}
W_{ij} = \left\{\begin{array}{ll} \phantom{-}0 & \text{ if } i = j, \\ -1 + \varepsilon & \text{ if } j \rightarrow i \text{ in } G,\\ -1 -\delta & \text{ if } j \not\rightarrow i \text{ in } G. \end{array}\right. \quad \quad \quad \quad
\end{equation}
Additionally, CTLNs typically have a constant external input $b_i=\theta$ for all $i$ in order to ensure the dynamics are internally generated rather than inherited from a changing or spatially heterogeneous input.  

A CTLN is thus completely specified by the choice of a graph $G$, together with three real parameters: $\varepsilon, \delta,$ and $\theta$.  We additionally require that $\delta >0$, $\theta>0$, and $0 < \varepsilon < \dfrac{\delta}{\delta+1}$. When these conditions are met, we say the parameters are within the {\em legal range}. Note that the upper bound on $\varepsilon$ implies $\varepsilon < 1$, and so the $W$ matrix is always effectively inhibitory. For fixed parameters, only the graph $G$ varies between networks. The network in Figure~\ref{fig:hyperplanes} is a CTLN
with the \emph{standard parameters} $\varepsilon=0.25$, $\delta=0.5$, and $\theta=1$.

\begin{figure}[!h]
\begin{center}
\vspace{-.05in}
\includegraphics[width=2.7in]{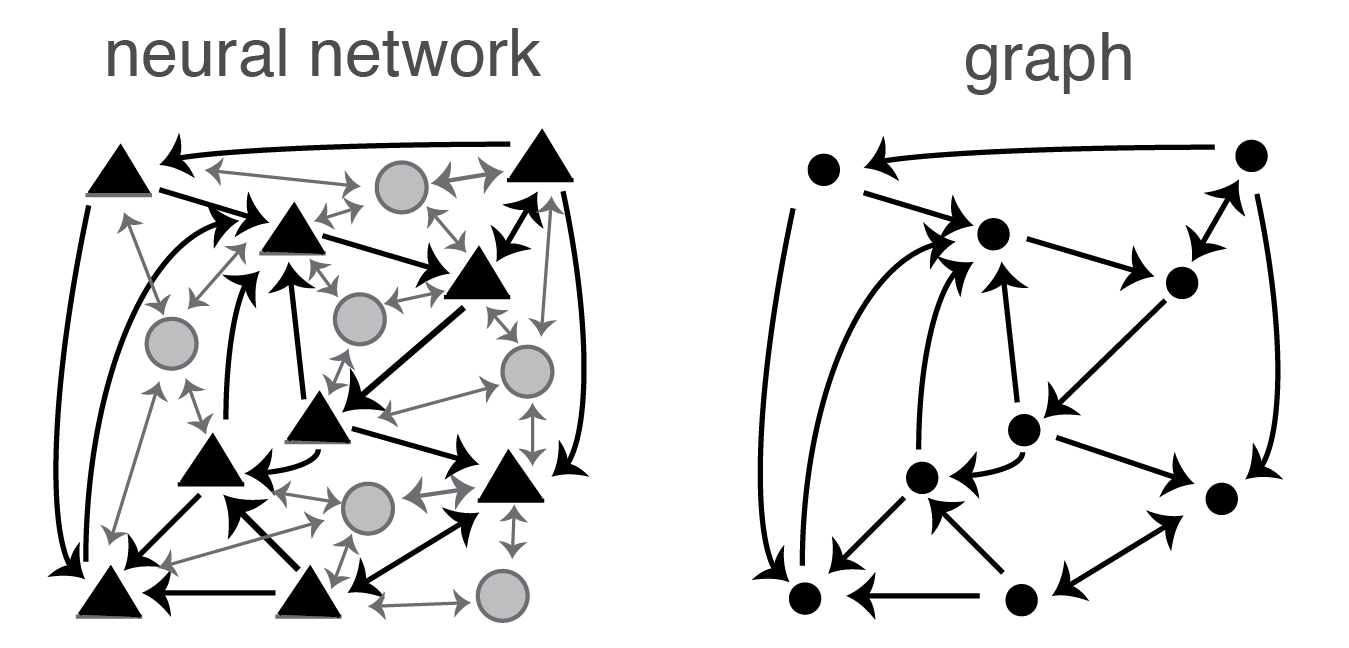}
\end{center}
\caption{{\bf CTLNs.} A neural network with excitatory pyramidal neurons (triangles) and a background network of inhibitory interneurons (gray circles) that produces a global inhibition. The corresponding graph (right) retains only the excitatory neurons and their connections.}
\label{fig:network-setup}
\vspace{-.1in}
\end{figure}

We interpret a CTLN as modeling a network of $n$ excitatory neurons, whose net interactions are effectively inhibitory due to a strong global inhibition (Figure~\ref{fig:network-setup}). When $j \not\to i$, we say $j$ \emph{strongly inhibits} $i$; when $j \to i$, we say $j$ \emph{weakly inhibits} $i$. The weak inhibition is thought of as the sum of an excitatory synaptic connection and the background inhibition.  Note that because $-1-\delta < -1 < -1+\varepsilon$, when $j \not\to i$, neuron $j$ inhibits $i$ \emph{more} than it inhibits itself via its leak term; when $j \to i$, neuron $j$ inhibits $i$ \emph{less} than it inhibits itself.  These differences in inhibition strength cause the activity to follow the arrows of the graph. 

The set of fixed point supports of a CTLN with graph $G$ is denoted as:
\begin{small}
\begin{eqnarray*}
\FP(G,\varepsilon,\delta)  &\od& \{\sigma \subseteq [n] \mid   \sigma = \supp{x^*} \text{ for some }\\
& & \text{fixed pt } x^* \text{ of the associated CTLN} \}.
\end{eqnarray*}
\end{small}
$\FP(G,\varepsilon,\delta)$ is precisely $\FP(W,b)$, where $W$ and $b$ are specified by a CTLN with graph $G$ and parameters $\varepsilon$ and $\delta$. Note that $\FP(G,\varepsilon,\delta)$ is independent of $\theta$, provided $\theta$ is constant across neurons as in a CTLN. It is also frequently independent of $\varepsilon$ and $\delta$. For this reason we often refer to it as $\FP(G)$, especially when a fixed choice of $\varepsilon$ and $\delta$ is understood.

The legal range condition, $\varepsilon < \dfrac{\delta}{\delta+1},$ is motivated by a theorem in \cite{CTLN-preprint}. It ensures that single directed edges $i \to j$ are not allowed to support stable fixed points $\{i,j\} \in \FP(G,\varepsilon,\delta)$. This allows us to prove the following theorem connecting a certain graph structure to the absence of stable fixed points. Note that a graph is {\em oriented} if for any pair of nodes, $i \to j$ implies $j \not\to i$ (i.e., there are no bidirectional edges). A {\it sink} is a node with no outgoing edges.

\begin{theorem}\cite[Theorem 2.4]{CTLN-preprint}\label{thm:oriented-graphs}
Let $G$ be an oriented graph with no sinks. Then for any parameters $\varepsilon, \delta, \theta$ in the legal range, the associated CTLN has no stable fixed points. Moreover, the activity is bounded. 
\end{theorem}

The graph in Figure~\ref{fig:hyperplanes}A is an oriented graph with no sinks. It has a single fixed point, $\FP(G) = \{123\}$, irrespective of the parameters (note that we use ``$123$" as shorthand for the set $\{1, 2, 3\}$). This fixed point is unstable and the dynamics converge to a limit cycle (Figure~\ref{fig:hyperplanes}C).

\begin{figure*}[!ht]
\vspace{-.1in}
\begin{center}
\includegraphics[width=6.5in]{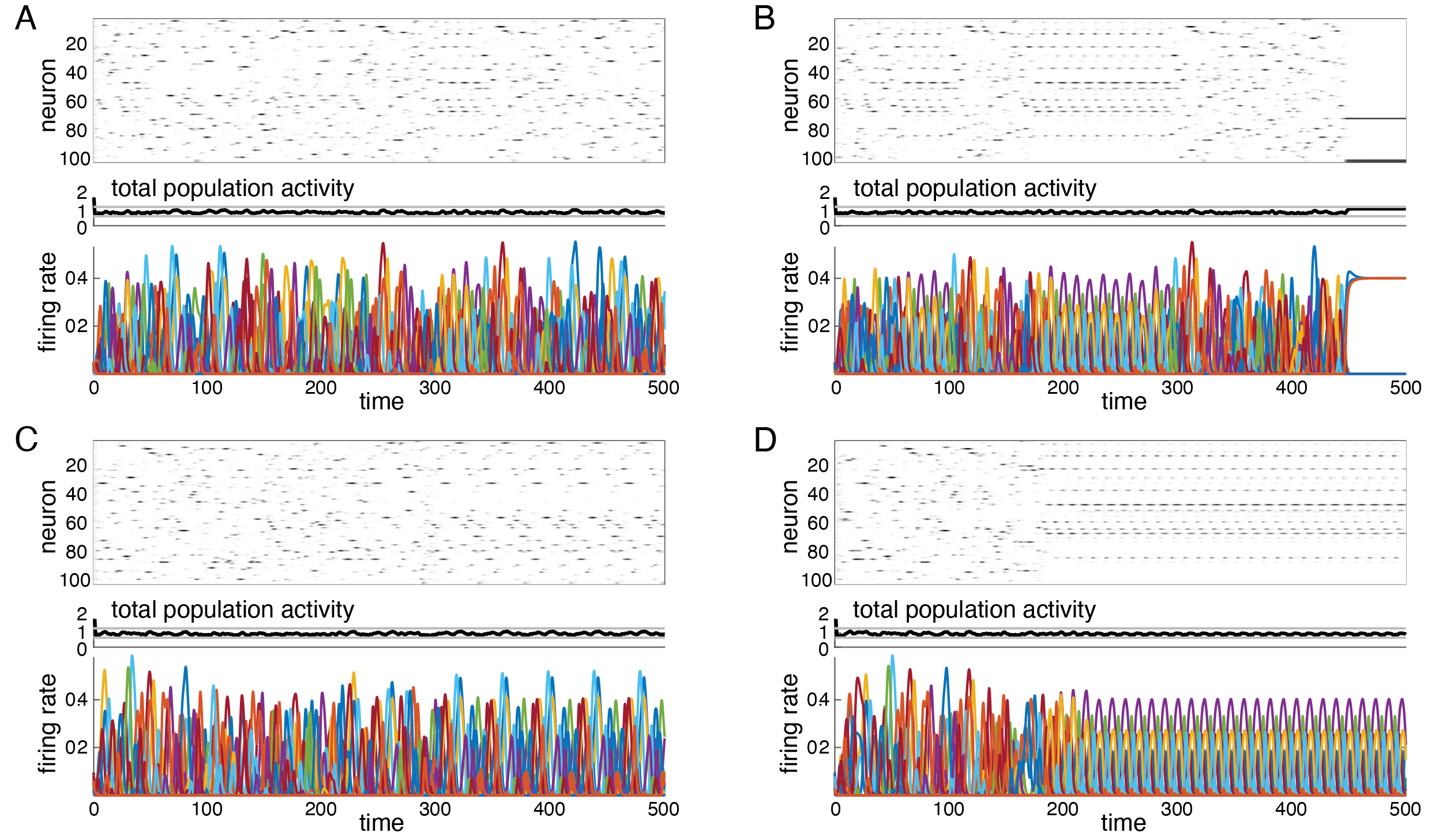}
\caption{{\bf Dynamics of a CTLN network on $n=100$ neurons.} The graph $G$ is a directed Erdos-Renyi random graph with edge probability $p = 0.2$ and no self loops. The CTLN parameters are $\varepsilon = 0.25$, $\delta = 0.5$, and $\theta = 1.$ Initial conditions for each neuron, $x_i(0)$, are randomly and independently chosen from the uniform distribution on $[0,0.1].$ (A-D) Four solutions from the same deterministic network, differing only in the choice of initial conditions. In each panel, the top plot shows the firing rate as a function of time for each neuron in grayscale. The middle plot shows the summed total population activity, $\sum_{i=1}^n x_i$, which quickly becomes trapped between the horizontal gray lines -- the bounds in equation~\eqref{eq:pop-bounds}. The bottom plot shows individual rate curves for all $100$ neurons, in different colors.
(A) The network appears chaotic, with some recurring patterns of activity. 
(B) The solution initially appears to be chaotic, like the one in A, but eventually converges to a stable fixed point supported on a $3$-clique. (C) The solution converges to a limit cycle after $t=300$. (D) The solution converges to a different limit cycle after $t=200$. Note that one can observe brief ``echoes'' of this limit cycle in the transient activity of panel B.}
\label{fig:large-ctln}
\end{center}
\vspace{-.25in}
\end{figure*}  

Even when there are no stable fixed points, the dynamics of a CTLN are always bounded \cite{CTLN-preprint}. In the limit as $t \to \infty$, we can bound the total population activity as a function of the parameters $\varepsilon, \delta,$ and $\theta$:
\vspace{-.05in}
\begin{equation}\label{eq:pop-bounds}
\dfrac{\theta}{1+\delta} \leq \sum_{i=1}^n x_i \leq \dfrac{\theta}{1-\varepsilon}.
\end{equation}
\vspace{-.05in}

In simulations, we observe a rapid convergence to this regime. 
Figure~\ref{fig:large-ctln} depicts four solutions for the same CTLN on $n = 100$ neurons. The graph $G$ was generated as a directed Erdos-Renyi random graph with edge probability $p = 0.2$; note that it is {\it not} an oriented graph. Since the network is deterministic, the only difference between simulations is the initial conditions. While panel A appears to show chaotic activity, the solutions in panels B, C and D all settle into a fixed point or a limit cycle within the allotted time frame. The long transient of panel B is especially striking: around $t = 200$, the activity appears as though it will fall into the same limit cycle from panel D, but then escapes into another period of chaotic-looking dynamics before abruptly converging to a stable fixed point. In all cases, the total population activity rapidly converges to lie within the bounds given in~\eqref{eq:pop-bounds}, depicted in gray.

\vspace{-.1in}
\paragraph{Fun examples.}
Despite their simplicity, CTLNs display a rich variety of nonlinear dynamics. Even very small networks can exhibit interesting attractors with unexpected properties. Theorem~\ref{thm:oriented-graphs} tells us that one way to guarantee that a network will produce dynamic -- as opposed to static -- attractors is to choose 
$G$ to be an oriented graph with no sinks. The following examples are of this type.

\begin{figure*}[!hb]
\begin{center}
\includegraphics[width=6.5in]{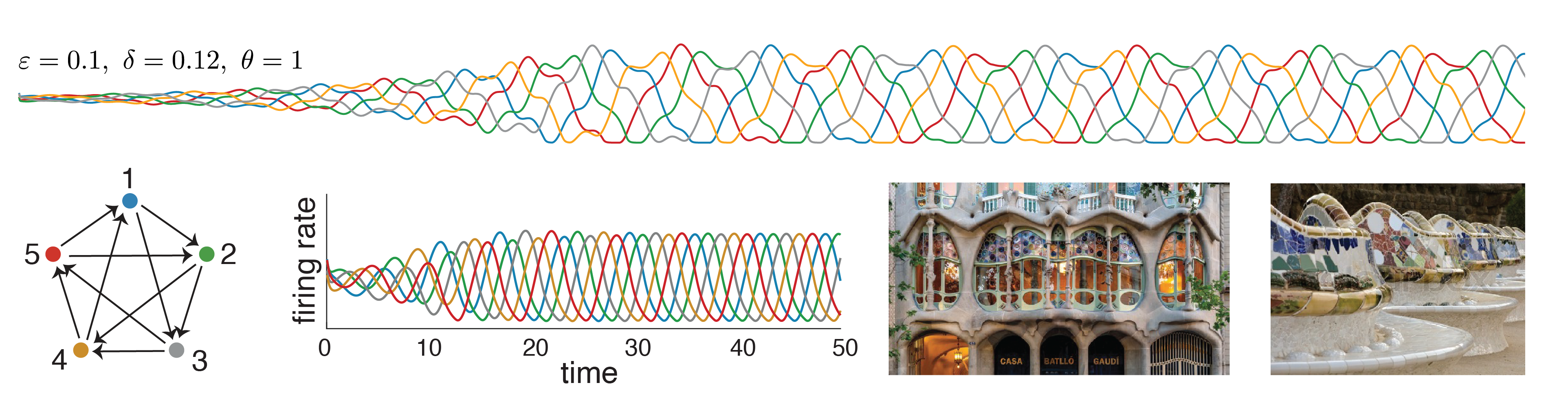}
\end{center}
\vspace{-.1in}
\caption{{\bf Gaudi attractor.} A CTLN for a cyclically symmetric tournament on $n=5$ nodes produces two distinct attractors, depending on initial conditions. We call the top one the Gaudi attractor because the undulating curves are reminiscent of work by the architect from Barcelona.}
\label{fig:gaudi}
\vspace{-.1in}
\end{figure*}

{\it The Gaudi attractor.} 
Figure~\ref{fig:gaudi} shows two solutions to a CTLN for a cyclically symmetric tournament\footnote{A \emph{tournament} is a directed graph in which every pair of nodes has exactly one (directed) edge between them.} graph on $n=5$ nodes. For some initial conditions, the solutions converge to a somewhat boring limit cycle with the firing rates $x_1(t), \ldots, x_5(t)$ all peaking in the expected sequence, $12345$ (bottom middle). For a different set of initial conditions, however, the solution converges to the beautiful and unusual attractor displayed at the top.

{\it Symmetry and synchrony.} Because the pattern of weights in a CTLN is completely determined by the graph $G$, any symmetry of the graph necessarily translates to a symmetry of the differential equations, and hence of the vector field. It follows that the automorphism group of $G$ also acts on the set of all attractors, which must respect the symmetry. For example, in the cyclically symmetric tournament of Figure~\ref{fig:gaudi}, both the Gaudi attractor and the ``boring'' limit cycle below it are invariant under the cyclic permutation $(12345)$: the solution is preserved up to a time translation.

\begin{figure}[!h]
\begin{center}
\includegraphics[width=3.2in]{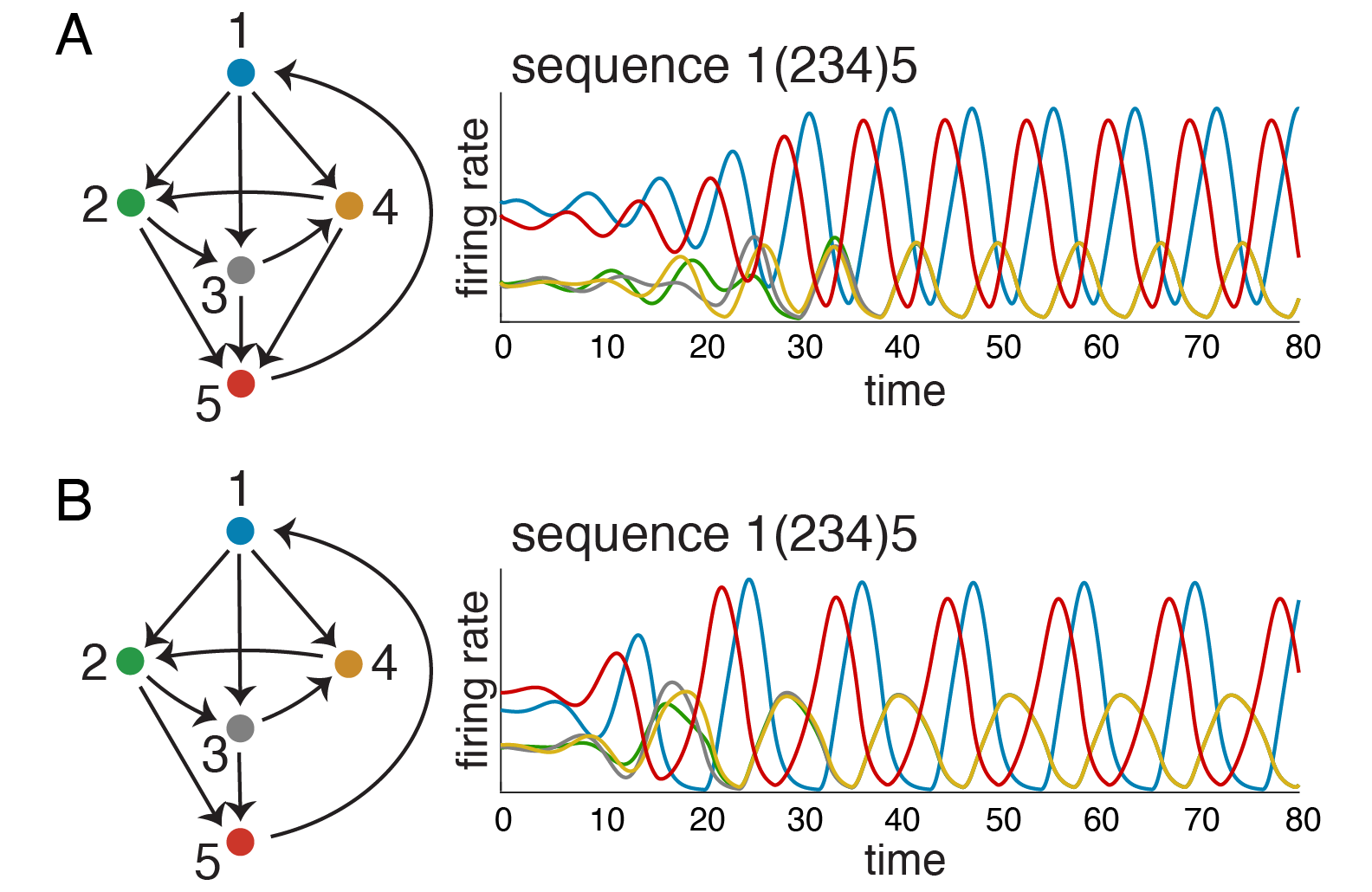}
\end{center}
\caption{{\bf Symmetry and synchrony.} (A) A graph with automorphism group $C_3$ has an attractor where neurons $2, 3,$ and $4$ fire synchronously. The overall sequence of activation is denoted $1(234)5$, indicating that neurons $2,3,4$ fire synchronously after neuron $1$ and before $5$, repeating periodically. (B) The symmetry is broken due to the dropped $4 \to 5$ edge. Nevertheless, the attractor still respects the $(234)$ symmetry with
nodes $2, 3,$ and $4$ firing synchronously. Note that both attractors are very similar limit cycles, but the one in B has longer period. (Simulations used the standard parameters: $\varepsilon=0.25$, $\delta=0.5$, $\theta=1$.)}
\label{fig:outer-neuron}
\vspace{-.1in}
\end{figure}

Another way for symmetry to manifest itself in an attractor is via synchrony. The network in Figure~\ref{fig:outer-neuron}A depicts a CTLN with a graph on $n=5$ nodes that has a nontrivial automorphism group $C_3$, cyclically permuting the nodes $2, 3$ and $4$. In the corresponding attractor, the neurons $2, 3, 4$ perfectly synchronize as the solution settles into the limit cycle. Notice, however, what happens for the network in Figure~\ref{fig:outer-neuron}B. In this case, the limit cycle looks very similar to the one in A, with the same synchrony among neurons $2, 3$ and $4$. However, the graph is missing the $4 \to 5$ edge, and so the graph has no nontrivial automorphisms. We refer to this phenomenon as {\it surprise symmetry}. 

On the flip side, a network with graph symmetry may have multiple attractors that are exchanged by the group action, but do not individually respect the symmetry. This is the more familiar scenario of spontaneous symmetry breaking.

{\it Emergent sequences.} One of the most reliable properties of CTLNs is the tendency of neurons to fire in sequence. Although we have seen examples of synchrony, the global inhibition promotes competitive dynamics wherein only one or a few neurons reach their peak firing rates at the same time. The sequences may be intuitive, as in the networks of Figures~\ref{fig:gaudi} and~\ref{fig:outer-neuron}, following obvious cycles in the graph. However, even for small networks the emergent sequences may be difficult to predict. 

\begin{figure}[!h]
\begin{center}
\includegraphics[width=2.9in]{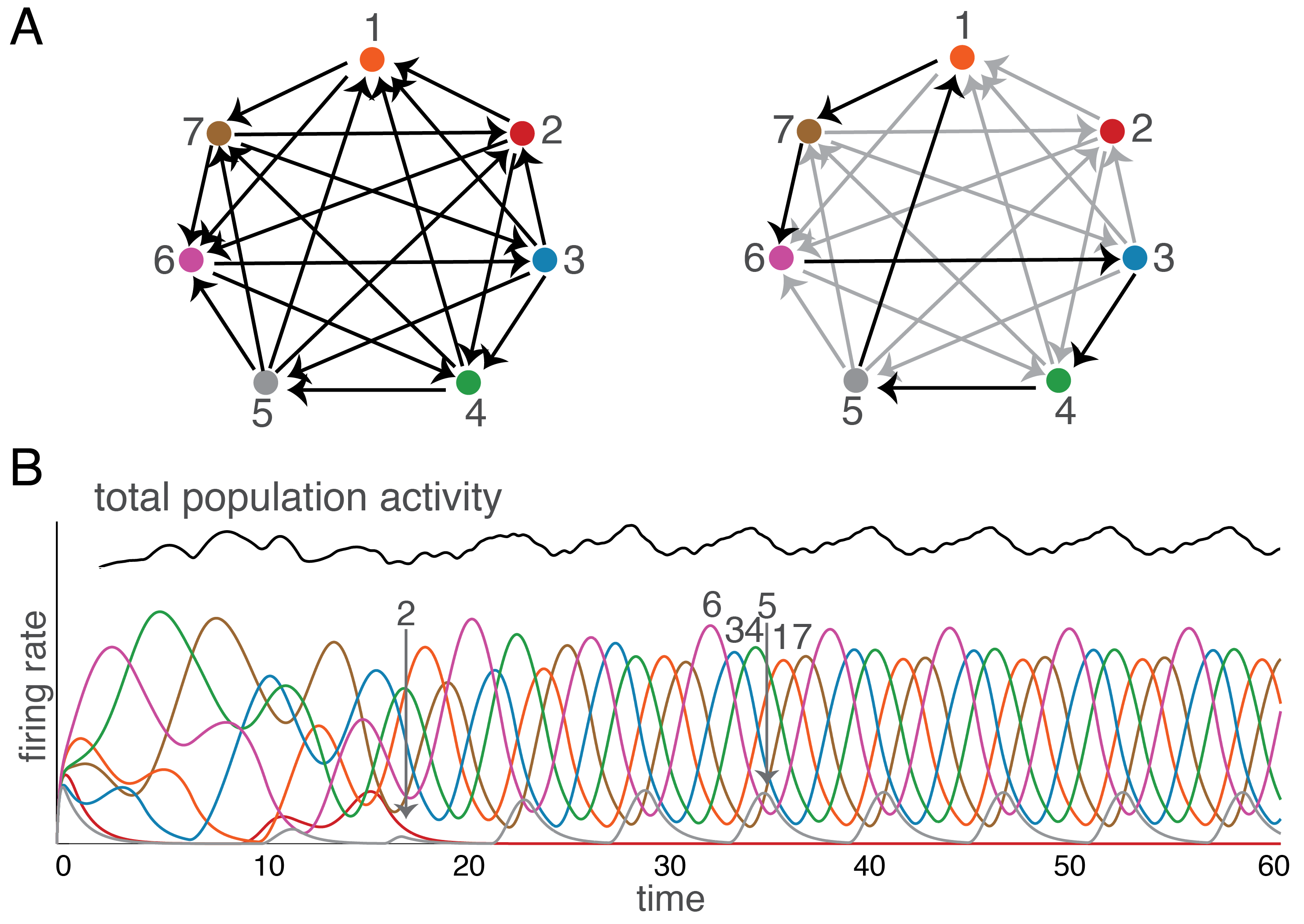}
\end{center}
\vspace{-.15in}
\caption{{\bf Emergent sequences can be difficult to predict.} (A) (Left) The graph of a CTLN that is a tournament on $7$ nodes. (Right) The same graph, but with the cycle corresponding to the sequential activity highlighted in black.
(B) A solution to the CTLN that converges to a limit cycle. This appears to be the only attractor of the network for the standard parameters.
}
\label{fig:n7-sequence}
\vspace{-.1in}
\end{figure}

The network in Figure~\ref{fig:n7-sequence}A has $n=7$ neurons, and the graph is a tournament with no nontrivial automorphisms. The corresponding CTLN appears to have a single, global attractor, shown in Figure~\ref{fig:n7-sequence}B. The neurons in this limit cycle fire in a repeating sequence, 634517, with 5 being the lowest-firing node. This sequence is highlighted in black in the graph, and corresponds to a cycle in the graph. However, it is only one of many cycles in the graph. Why do the dynamics select this sequence and not the others? And why does neuron 2 drop out, while all others persist? This is particularly puzzling given that node 2 has in-degree three, while nodes 3 and 5 have in-degree two. 

Indeed, local properties of a network, such as the in- and out-degrees of individual nodes, are insufficient for predicting the participation and ordering of neurons in emergent sequences. Nevertheless, the sequence is fully determined by the structure of $G$. We just have a limited understanding of how. Recent progress in understanding sequential attractors has relied on special network architectures that are cyclic like the ones in Figures~\ref{fig:gaudi} and~\ref{fig:outer-neuron} \cite{seq-attractors}. Interestingly, although the graph in Figure~\ref{fig:n7-sequence} does not have such an architecture, the induced subgraph generated by the high-firing nodes 1, 3, 4, 6, and 7 is isomorphic to the graph in Figure~\ref{fig:gaudi}. This graph, as well as the two graphs in Figure~\ref{fig:outer-neuron}, have corresponding networks that are in some sense irreducible in their dynamics. These are examples of graphs that we refer to as {\it core motifs} \cite{core-motifs}.

\section{Minimal fixed points, core motifs, and attractors} \label{sec:core-motifs}
Stable fixed points of a network are of obvious interest because they correspond to static attractors \cite{HahnSeungSlotine, net-encoding, pattern-completion, stable-fp-paper}. One of the most striking features of CTLNs, however, is the strong connection between {\it unstable} fixed points and 
dynamic attractors \cite{book-chapter, core-motifs, seq-attractors}. 

\begin{question}
For a given CTLN, can we predict the dynamic attractors of the network from its unstable fixed points?  Can the unstable fixed points be determined from the structure of the underlying graph $G$? 
\end{question}

Throughout this section, $G$ is a directed graph on $n$ nodes.  Subsets $\sigma \subseteq [n]$ are often used to denote both the collection of vertices indexed by $\sigma$ and the induced subgraph $G|_\sigma$. The corresponding network is assumed to be a nondegenerate CTLN with fixed parameters $\varepsilon, \delta,$ and $\theta$.

\begin{figure*}[!ht]
\vspace{-.1in}
\begin{center}
\includegraphics[width=5.75in]{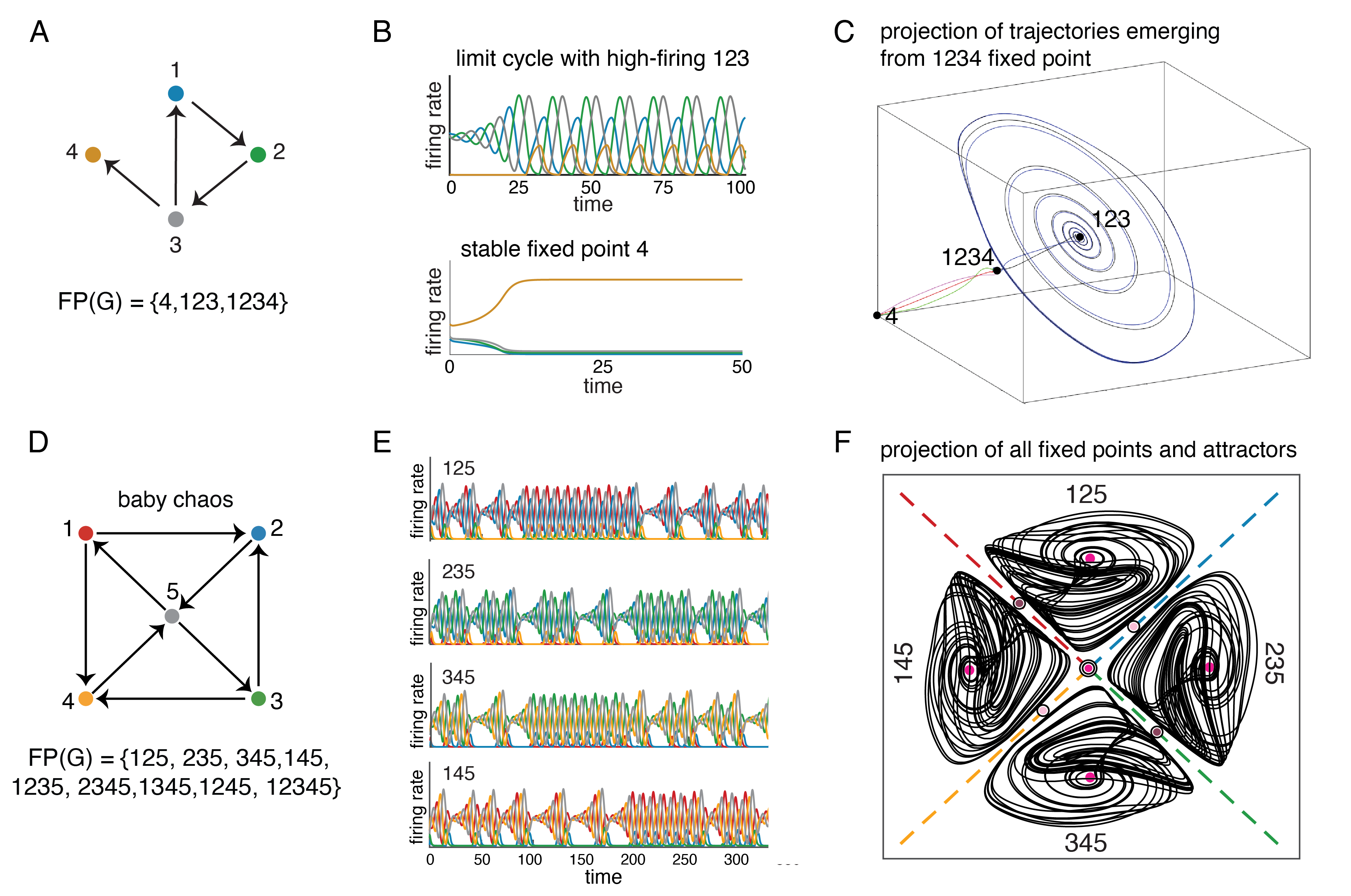}
\caption{\carina{{\bf Core motifs of CTLNs correspond to attractors.}} (A) The graph of a CTLN. The fixed point supports are given by $\FP(G) = \{4, 123, 1234\}$, irrespective of parameters $\varepsilon, \delta, \theta$. (B) Solutions to the CTLN in A using the standard parameters $\theta=1$, 
$\varepsilon=0.25$, and $\delta=0.5$. (Top) The initial condition was chosen as a small perturbation of the fixed point supported on $123$. The activity quickly converges to a limit cycle where the high-firing neurons are the ones in the fixed point support. (Bottom) A different initial condition yields a solution that converges to the static attractor corresponding to the stable fixed point on node $4$. (C) The three fixed points are depicted in a three-dimensional projection of the four-dimensional state space. Perturbations of the fixed point supported on $1234$ produce solutions that either converge to the limit cycle or to the stable fixed point from B. \carina{(D) A network on $n=5$ nodes whose fixed point supports are also independent of the CTLN parameters. (E) The four core motifs, supported on $125, 235, 345$ and $145$, each have a corresponding chaotic attractor. (F) A projection of the four chaotic attractors (black trajectories) together with all nine fixed points of the network (pink dots), which are all unstable.} }
\label{fig:tadpole}
\end{center}
\end{figure*}

Figure~\ref{fig:tadpole} provides \carina{two example networks} to illustrate the relationship between unstable fixed points and dynamic attractors. Any CTLN with the graph in panel A has three fixed points, with supports $\FP(G) = \{4,123,1234\}$. The collection of fixed point supports can be thought of as a partially ordered set, ordered by inclusion. In our example, $4$ and $123$ are thus {\it minimal} fixed point supports, because they are minimal under inclusion. It turns out that the corresponding fixed points each have an associated attractor (Figure~\ref{fig:tadpole}B). The one supported on $4$, a sink in the graph, yields a stable fixed point, while the $123$ (unstable) fixed point, whose induced subgraph $G|_{123}$ is a $3$-cycle, yields a limit cycle attractor with high-firing neurons $1$, $2$, and $3$. Figure~\ref{fig:tadpole}C depicts all three fixed points in the state space. Here we can see that the third one, supported on $1234$, acts as a ``tipping point'' on the boundary of two basins of attraction. Initial conditions near this fixed point can yield solutions that converge either to the stable fixed point or the limit cycle. 

\carina{Figure~\ref{fig:tadpole}D-F provides another example network, called ``baby chaos,'' in which all fixed points are unstable. The minimal fixed point supports, $125, 235, 345$ and $145$, all correspond to core motifs (embedded $3$-cycles in the graph). The corresponding attractors are chaotic, and are depicted as firing rate curves (panel E) and trajectories in the state space (panel F). Note that the graph has an automorphism group that exchanges core motifs and their corresponding attractors.}

Not all minimal fixed points have corresponding attractors. In \cite{core-motifs} we saw that the key property of such a $\sigma \in \FP(G)$ is that it be minimal not only in $\FP(G)$ but also in $\FP(G|_\sigma)$, corresponding to the induced subnetwork restricted to the nodes in $\sigma$. In other words, $\sigma$ is the only fixed point in $\FP(G|_\sigma)$. This motivates the definition of core motifs.

\begin{definition}
Let $G$ be the graph of a CTLN on $n$ nodes. An induced subgraph $G|_\sigma$ is a {\it core motif} of the network if $\FP(G|_\sigma) = \{\sigma\}$.
\end{definition}

\begin{figure*}[!h]
\begin{center}
\includegraphics[width=6.5in]{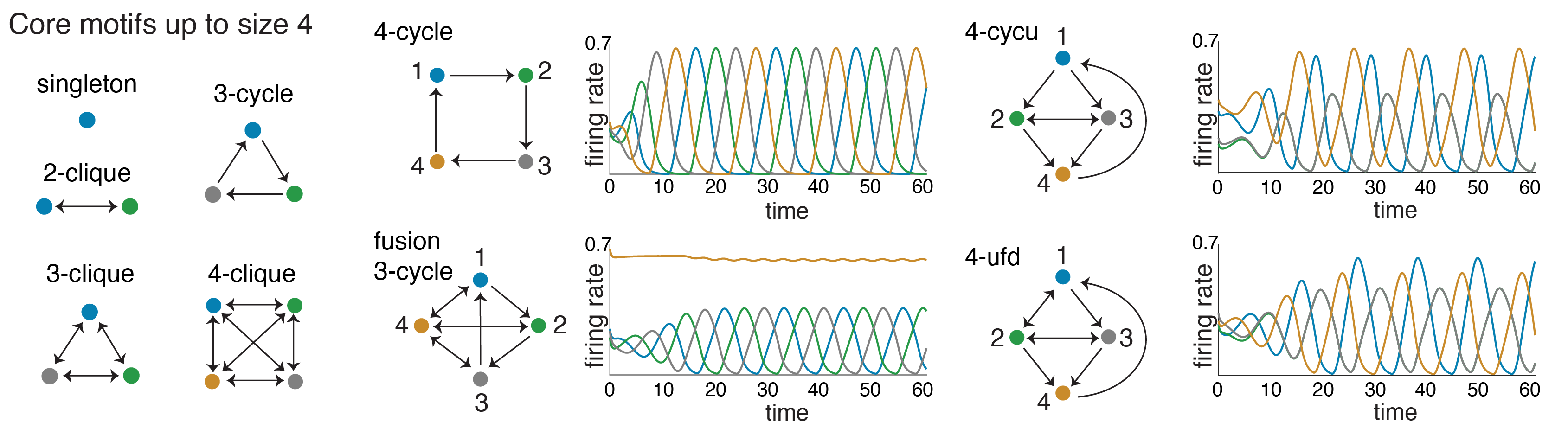}
\end{center}
\vspace{-.15in}
\caption{{\bf Small core motifs.} For each of these graphs, $\FP(G) = \{[n]\}$, where $n$ is the number of nodes. Attractors are shown for CTLNs with the standard parameters $\varepsilon=0.25$, $\delta=0.5$, and $\theta=1$.}
\label{fig:n4-cores}
\end{figure*}

\noindent When the graph $G$ is understood, we sometimes refer to $\sigma$ itself as a core motif if $G|_\sigma$ is one. The associated fixed point is called a {\it core fixed point}. 
Core motifs can be thought of as ``irreducible'' networks because they have a single fixed point which has full support. Since the activity is bounded and must converge to an attractor, the attractor can be said to correspond to this fixed point. A larger network that contains $G|_\sigma$ as an induced subgraph may or may not have $\sigma \in \FP(G)$. When the core fixed point does survive, we say refer to the embedded $G|_\sigma$ as a {\it surviving} core motif, and we expect the associated attractor to survive. In Figure~\ref{fig:tadpole}, the surviving core motifs are $G|_{4}$ and $G|_{123}$, and they precisely predict the attractors of the network.

The simplest core motifs are cliques. When these survive inside a network $G$, the corresponding attractor is always a stable fixed point supported on all nodes of the clique \cite{fp-paper}. In fact, we conjectured that any stable fixed point for a CTLN must correspond to a maximal clique of $G$ -- specifically, a {\it target-free} clique \cite{fp-paper,stable-fp-paper}.

\begin{figure*}[!h]
\begin{center}
\includegraphics[width=6in]{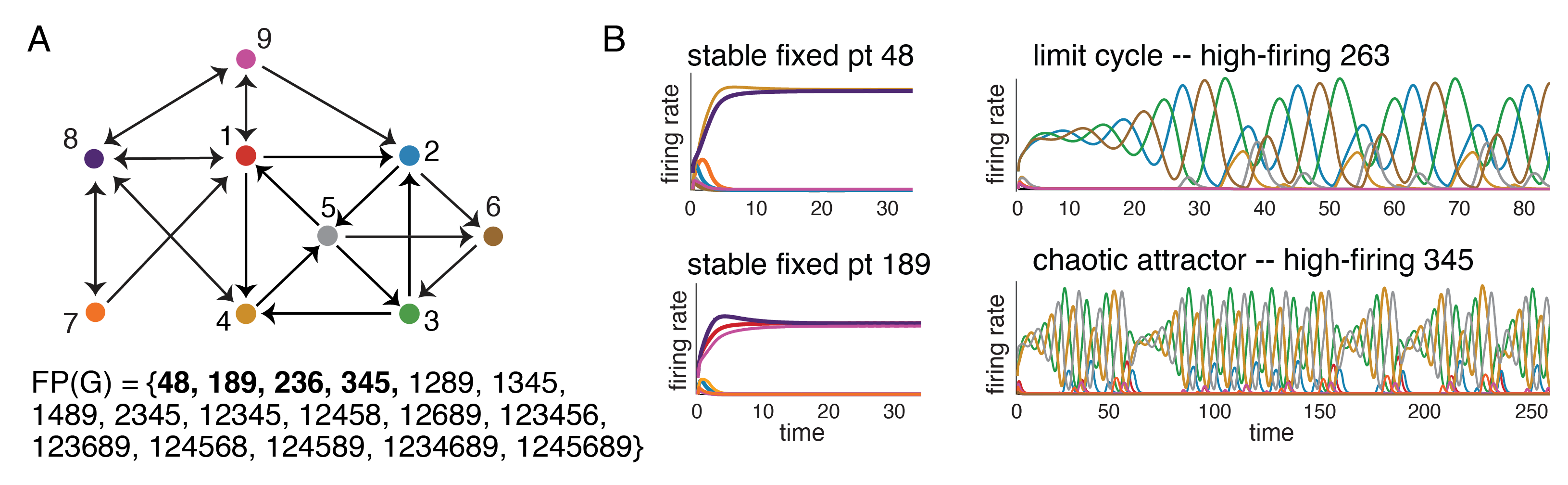}
\caption{\carina{{\bf Coexistence of attractors.} Stable fixed points supported on $48$ and $189$, a limit cycle corresponding to $236$, and a chaotic attractor for $345$. All attractors can be easily accessed via an initial condition near the corresponding fixed point.}}
\label{fig:coexistence}
\end{center}
\vspace{-.25in}
\end{figure*}

Up to size $4$, all core motifs are parameter-independent. For size $5$, $37$ of $45$ core motifs are parameter-independent. 
Figure~\ref{fig:n4-cores} shows the complete list of all core motifs of size $n \leq 4$, together with some associated attractors. The cliques all correspond to stable fixed points, the simplest type of attractor. The $3$-cycle yields the limit cycle attractor in Figure~\ref{fig:hyperplanes}, which may be distorted when embedded in a larger network (see Figure~\ref{fig:tadpole}B). The other core motifs whose fixed points are unstable have dynamic attractors. Note that the $4$-cycu graph has a $(23)$ symmetry, and the rate curves for these two neurons are synchronous in the attractor. This synchrony  is also evident in the $4$-ufd attractor, despite the fact that this graph does not have the $(23)$ symmetry. Perhaps the most interesting attractor, however, is the one for the fusion $3$-cycle graph. Here the $123$ $3$-cycle attractor, which does not survive the embedding to the larger graph, appears to ``fuse'' with the stable fixed point associated to $4$ (which also does not survive). The resulting attractor can be thought of as binding together a pair of smaller attractors.

\carina{Figure~\ref{fig:coexistence}A depicts a larger example of a network whose fixed point structure $\FP(G)$ is predictive of the attractors. Note that only four supports are minimal: $48$, $189$, $236$, and $345$. The first two correspond to surviving cliques, and the last two correspond to $3$-cycles with surviving fixed points. An extensive search of attractors for this network reveals only four attractors, corresponding to the four surviving core motifs. Figure~\ref{fig:coexistence}B shows trajectories converging to each of the four attractors. The cliques yield stable fixed points, as expected, while the $3$-cycles correspond to dynamic attractors: one limit cycle, and one strange or chaotic attractor.}

\begin{figure*}[!h]
\begin{center}
\includegraphics[width=\textwidth]{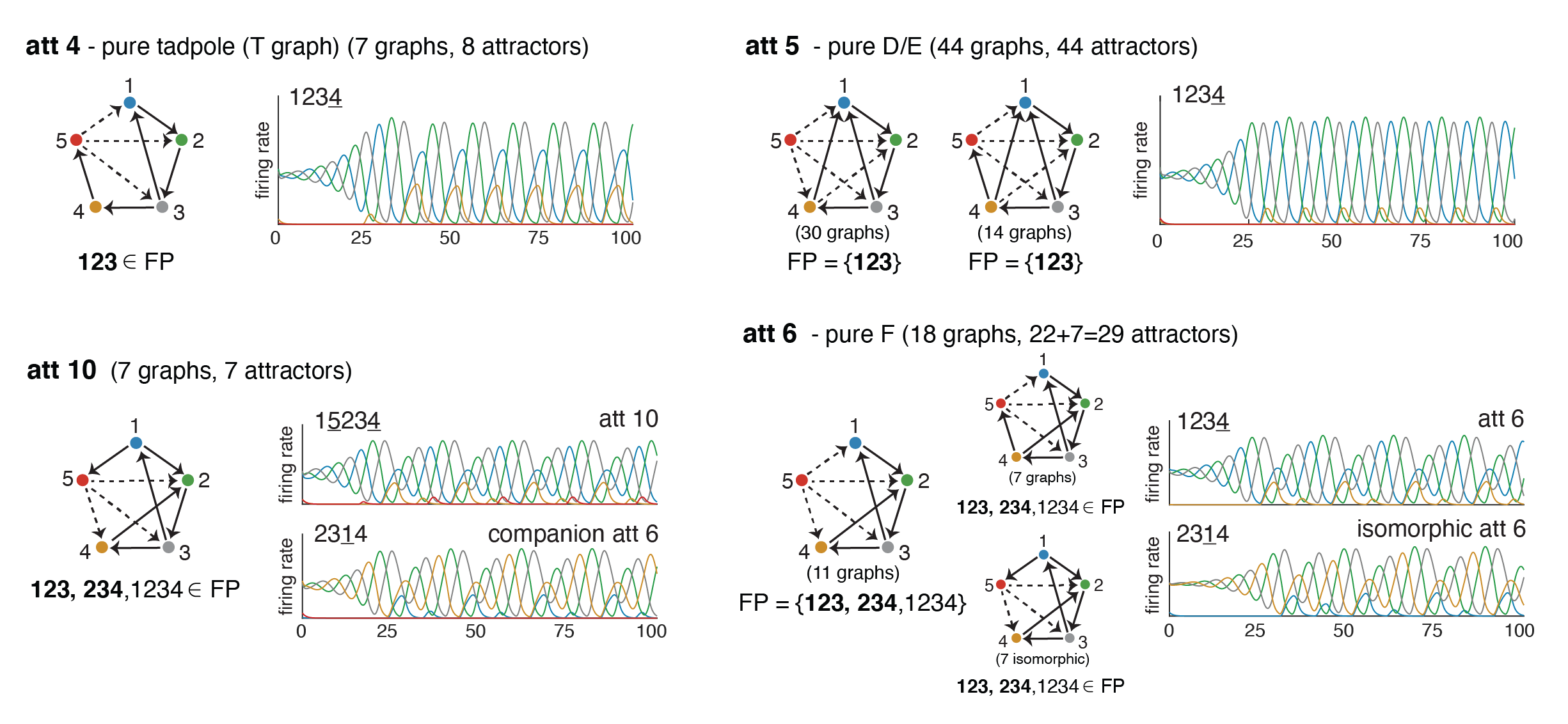}
\end{center}
\vspace{-.2in}
\caption{\carina{{\bf Modularity of attractors.} For each attractor family, one or more ``master graphs'' are shown. The master graphs represent a collection of graphs where the solid edges are shared by all graphs and the dashed edges are optional. For example, the master graph corresponding to att 4 represents 7 distinct graphs, all having the same attractor corresponding to the common core motif $G|_{123}$, embedded so that node $4$ receives an edge from $3$ but does not send any edge back to $G|_{123}$. The other families, att~5, att~6, and att~10, yield attractors supported on the same core motif, $G|_{123}$, but with different embeddings that alter the shape of the attractors. Note that this analysis only considered oriented graphs with no sinks; so, for example, the master graph for att 4 represents only 7 graphs, not 8, as node $5$ is required to have at least one outgoing edge. Adapted from \cite{core-motifs}.}}
\label{fig:n5-dictionary}
\vspace{-.1in}
\end{figure*}

We have performed extensive tests on whether or not core motifs predict attractors in small networks. 
Specifically, we decomposed all 9608 non-isomorphic directed graphs on $n=5$ nodes into core motif components, and used this to predict the attractors \cite{n5-github}.
We found that 1053 of the graphs have surviving core motifs that are not cliques; these graphs were thus expected to support dynamic attractors. The remaining 8555 graphs contain only cliques as surviving core motifs, and were thus expected to have only stable fixed point attractors. Overall, we found that core motifs correctly predicted the set of attractors in 9586 of the 9608 graphs. Of the 22 graphs with mistakes, 19 graphs have a core motif with no corresponding attractor, and 3 graphs have no core motifs for the chosen parameters \cite{n5-github}.

\carina{Across the 1053 graphs with core motifs that are not cliques, we observed a total of 1130 dynamic attractors. Interestingly, these fall into distinct equivalence classes determined by (a) the core motif, and (b) the details of how the core motif is embedded in the larger graph. In the case of {\it oriented graphs} on $n=5$ nodes, we performed a more detailed analysis of the dynamic attractors to determine a set of attractor families \cite{core-motifs}. Here we observed a striking modularity of the embedded attractors, wherein the precise details of an attractor remained nearly identical across large families of non-isomorphic graphs with distinct CTLNs. 
Figure~\ref{fig:n5-dictionary} gives a sampling of these common attractors, together with corresponding graph families. Graph families are depicted via ``master graphs,'' with solid edges being shared across all graphs in the family, and dashed edges being optional. Graph counts correspond to non-isomorphic graphs. See \cite{core-motifs} for more details.}

\section{Graph rules}\label{sec:graph-rules}
We have seen that CTLNs exhibit a rich variety of nonlinear dynamics, and that the attractors are closely related to the fixed points. This opens up a strategy for linking attractors to the underlying network architecture $G$ via the fixed point supports $\FP(G)$. Our main tools for doing this are {\em graph rules}. 

Throughout this section, we will use greek letters $\sigma, \tau, \omega$ to denote subsets of $[n] = \{1,\ldots,n\}$ corresponding to fixed point supports (or potential supports), while latin letters $i,j,k,\ell$ denote individual nodes/neurons. As before, $G|_\sigma$ denotes the induced subgraph obtained from $G$ by restricting to $\sigma$ and keeping only edges between vertices of $\sigma$. The fixed point supports are:
\vspace{-.05in}
\begin{eqnarray*}
\FP(G)  &\od& \{\sigma \subseteq [n] \mid   \sigma = \supp{x^*} \text{ for some }\\
& & \text{fixed pt } x^* \text{ of the associated CTLN} \}.
\vspace{-.05in}
\end{eqnarray*}

\noindent The main question addressed by graph rules is:

\begin{question}\label{ques:FP}
What can we say about $\FP(G)$ from knowledge of $G$ alone?
\end{question}

\begin{figure}[!h]
\begin{center}
\includegraphics[width=2.75in]{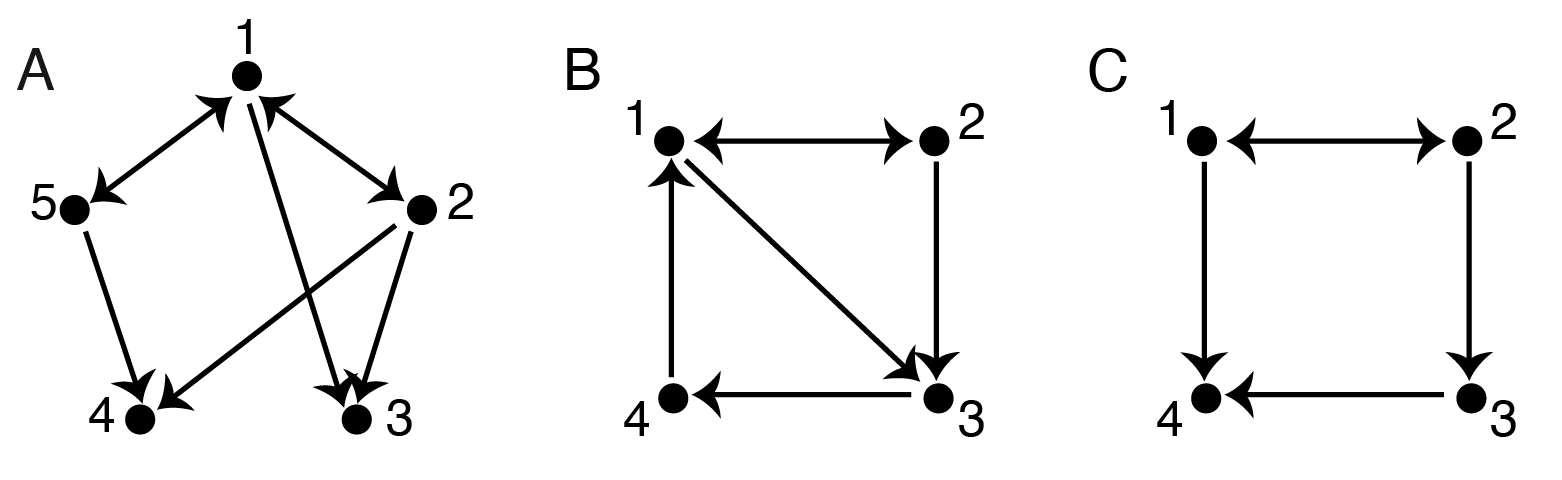}
\end{center}
\vspace{-.1in}
\caption{Graphs for which $\FP(G)$ is completely determined by graph rules.}
\label{fig:fp-examples}
\vspace{-.2in}
\end{figure}

 For example, consider the graphs in Figure~\ref{fig:fp-examples}.  Can we determine from the graph alone which subgraphs will support fixed points?  Moreover, can we determine which of those subgraphs are core motifs that will give rise to attractors of the network?  We saw in Section~\ref{sec:core-motifs} (Figure~\ref{fig:n4-cores}) that cycles and cliques are among the small core motifs; can cycles and cliques produce core motifs of any size?  Can we identify other graph structures that are relevant for either ruling in or ruling out certain subgraphs as fixed point supports?  The rest of Section~\ref{sec:graph-rules} focuses on addressing these questions.

Note that implicit in the above questions is the idea that graph rules are {\em parameter-independent}: that is, they directly relate the structure of $G$ to $\FP(G)$ via results that are valid for all choices of $\varepsilon, \delta,$ and $\theta$ (provided they lie within the legal range). In order to obtain the most powerful results, we also require that our CTLNs be {\it nondegenerate}. As has already been noted, nondegeneracy is generically satisfied for TLNs \cite{fp-paper}. For CTLNs, it is satisfied irrespective of $\theta$ and for almost all legal range choices of $\varepsilon$ and $\delta$ (i.e., up to a set of measure zero in the two-dimensional parameter space for $\varepsilon$ and $\delta$).

\subsection{Examples of graph rules}

We've already seen some graph rules. For example, Theorem~\ref{thm:oriented-graphs} told us that if $G$ is an oriented graph with no sinks, the associated CTLN has no stable fixed points.  Such CTLNs are thus guaranteed to only exhibit dynamic attractors.  Here we present a set of eight simple graph rules, all proven in \cite{fp-paper}, that are easy to understand and give a flavor of the kinds of theorems we have found.

We will use the following graph theoretic terminology. A {\it source} is a node with no incoming edges, while a {\it sink} is a node with no outgoing edges. Note that a node can be a source or sink in an induced subgraph $G|_\sigma$, while not being one in $G$. An {\it independent set} is a collection of nodes with no edges between them, while a {\it clique} is a set of nodes that is all-to-all bidirectionally connected. A {\it cycle} is a graph (or an induced subgraph) where each node has exactly one incoming and one outgoing edge, and they are all connected in a single directed cycle. A {\it directed acyclic graph} (DAG) is a graph with a topological ordering of vertices such that $i \not\to j$ whenever $i > j$; such a graph does not contain any directed cycles. Finally, a {\it target} of a graph $G|_\sigma$ is a node $k$ such that $i \to k$ for all $i \in \sigma \setminus \{k\}$. Note that a target may be inside or outside $G|_\sigma$. 

\begin{figure*}[!h]
\begin{center}
\includegraphics[width=5.9in]{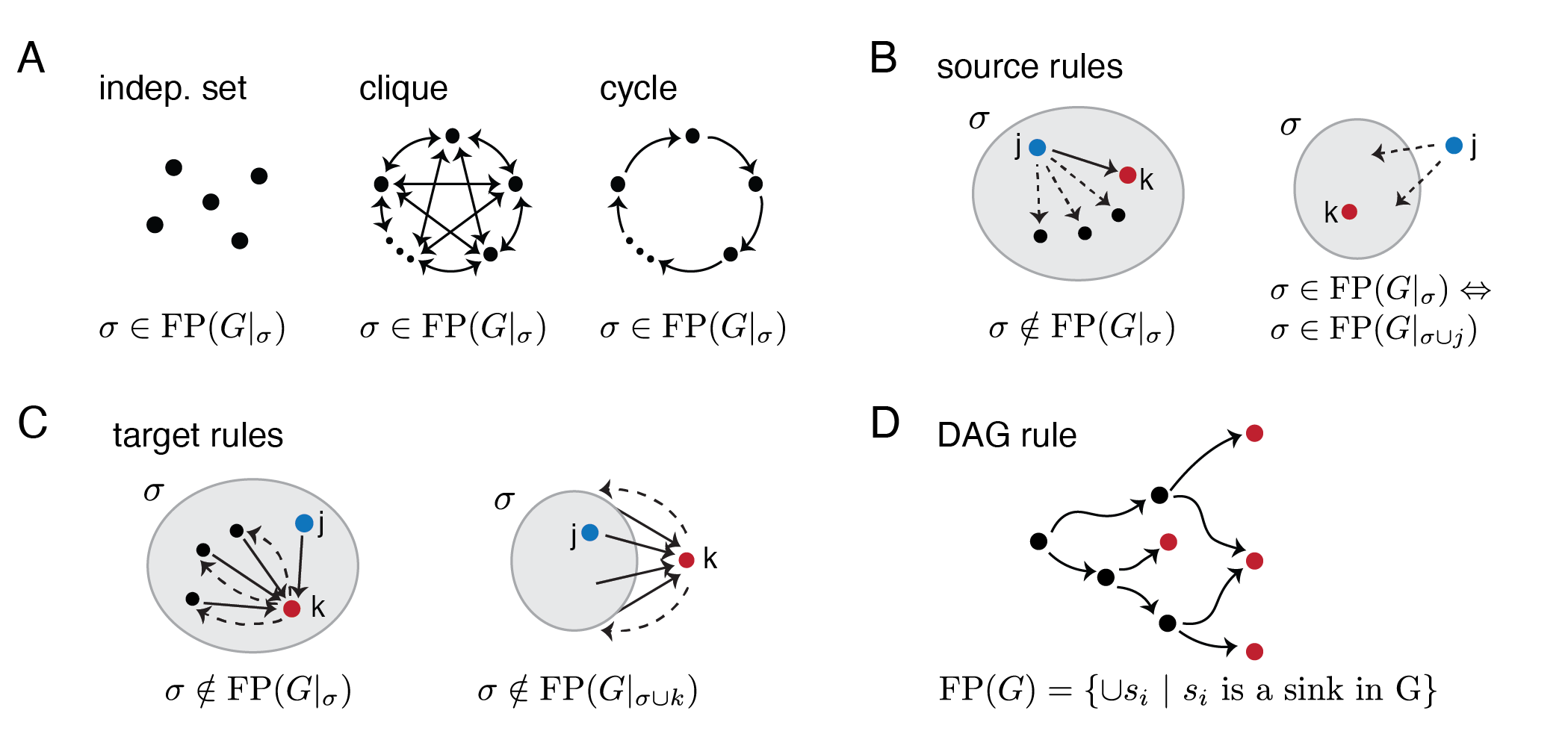}
\end{center}
\caption{\carina{{\bf A sampling of graph rules.} (A) Independent sets, cliques, and cycles all yield full-support fixed points in isolation. When embedded in a larger graph, the survival of these fixed points is dictated by Rules 1-3. (B) Illustration of Rules 4(i) and 4(ii), pertaining to a source node $j$ that lies inside or outside $\sigma$. The solid $j \to k$ edge is mandatory in Rule 4(i); dashed edges are optional. (C) Illustration of Rules 5(i) and 5(ii), pertaining to a target node $k$ that lies inside or outside of $\sigma$. (D) The only fixed point supports in a DAG are sinks and unions of sinks.}}
\label{fig:basic-graph-rules}
\vspace{-.2in}
\end{figure*}

\carina{The graph rules presented here can be found, with detailed proofs, in \cite{fp-paper}. We also summarize them in Table~\ref{table:graph-rules} and Figure~\ref{fig:basic-graph-rules}.}
\medskip

\noindent Examples of graph rules: 
\medskip

\noindent {\bf Rule 1} (independent sets): If $G|_\sigma$ is an independent set, then $\sigma \in \FP(G)$ if and only if
each $i \in \sigma$ is a sink in $G$.
\medskip

\noindent {\bf Rule 2} (cliques): If $G|_\sigma$ is a clique, then $\sigma \in \FP(G)$ if and only if there is no node $k$ of $G$, $k \notin \sigma$, such that $i \to k$ for all $i \in \sigma.$ In other words, $\sigma \in \FP(G)$ if and only if $G|_\sigma$ is a target-free clique. If $\sigma \in \FP(G)$, the corresponding fixed point is stable. 
\medskip

\noindent {\bf Rule 3} (cycles): If $G|_\sigma$ is a cycle, then $\sigma \in \FP(G)$ if and only if there is no node $k$ of $G$, $k \notin \sigma$, such that $k$ receives two or more edges from $\sigma$. If $\sigma \in \FP(G)$, the corresponding fixed point is unstable. 
\medskip

\noindent {\bf Rule 4} (sources): (i) If $G|_\sigma$ contains a source $j \in \sigma$, with $j \to k$ for some $k \in [n]$, then 
$\sigma \notin \FP(G)$. (ii) Suppose $j \notin \sigma$, but $j$ is a source in $G$. Then $\sigma \in  \FP(G|_{\sigma \cup j})$ if and only if $\sigma \in \FP(G|_\sigma)$. 
\medskip

\noindent {\bf Rule 5} (targets): (i) If $\sigma$ has target $k$, with $k \in \sigma$ and $k \not\to j$ for some $j \in \sigma$ ($j \neq k$), then 
$\sigma \notin \FP(G|_\sigma)$ and thus $\sigma \notin \FP(G)$. (ii) If $\sigma$ has target $k \not\in \sigma$, then 
$\sigma \notin \FP(G|_{\sigma \cup k})$ and thus $\sigma \notin \FP(G)$. 
\medskip

\noindent {\bf Rule 6} (sinks): If $G$ has a sink $s \notin \sigma$, then 
$\sigma \cup \{s\} \in \FP(G)$ if and only if $\sigma \in \FP(G)$.
\medskip

\noindent {\bf Rule 7} (DAGs): If $G$ is a directed acyclic graph with sinks $s_1,\ldots,s_\ell$, then 
$\FP(G) = \{\cup s_i \mid s_i \text{ is a sink in } G\}$, the set of all $2^\ell-1$ unions of sinks. 
\medskip

\noindent {\bf Rule 8} (parity): For any $G$, $|\FP(G)|$ is odd. 
\medskip

\begin{table*}[!h]
\vspace{.15in}
\centering
\setstretch{1.5}
\begin{tabular}{l|l|l}
 Rule name & $G|_\sigma$ structure & graph rule\\
 \hline
 \hline
Rule 1 & independent set  &  $\sigma \in \FP(G|_\sigma),$ and $\sigma \in \FP(G) \Leftrightarrow \sigma$ is a union of sinks \\
 \hline
Rule 2 & clique  & $\sigma \in \FP(G|_\sigma),$ and $\sigma \in \FP(G) \Leftrightarrow \sigma$ is target-free \\
 \hline
Rule 3 & cycle & $\sigma \in \FP(G|_\sigma),$ and $\sigma \in \FP(G) \Leftrightarrow $ each $k \notin \sigma$ \\
& & receives at most one edge $i \to k$ with $i \in \sigma$ \\
\hline
Rule 4(i) & $\exists$ a source $j \in \sigma$ & $\sigma \notin \FP(G)$ if $j \to k$ for some $k \in [n]$\\
\hline
Rule 4(ii) & $\exists$ a source $j \not\in \sigma$ & $\sigma \in  \FP(G|_{\sigma \cup j}) \Leftrightarrow \sigma \in \FP(G|_\sigma)$\\
\hline
Rule 5(i) & $\exists$ a target $k \in \sigma$ & $\sigma \notin \FP(G|_\sigma)$ and $\sigma \notin \FP(G)$ if $k \not\to j$ for some $j \in \sigma$\\
\hline
Rule 5(ii) & $\exists$ a target $k \not\in \sigma$ & $\sigma \not\in \FP(G|_{\sigma \cup k})$ and $\sigma \notin \FP(G)$ \\
\hline
Rule 6 & $\exists$ a sink $s \notin \sigma$ & $\sigma \cup \{s\} \in \FP(G) \Leftrightarrow \sigma \in \FP(G)$ \\
\hline
Rule 7 & DAG & $\FP(G) = \{\cup s_i \mid s_i \text{ is a sink in } G\}$ \\
\hline
Rule 8 & arbitrary & $|\FP(G)|$ is odd \\
\hline
\end{tabular}
\vspace{.15in}
\setstretch{1}
\caption{\carina{Graph rules connect properties of a graph $G$ to the fixed point supports, $\FP(G),$ of the associated CTLN. Each rule refers to the structure of the induced subgraph $G|_\sigma$ in order to determine whether
$\sigma \in \FP(G|_\sigma)$ and/or $\sigma \in \FP(G)$.}}
\label{table:graph-rules}
\vspace{-.1in}
\end{table*}

In many cases, particularly for small graphs, our graph rules are complete enough that they can be used to fully work out $\FP(G)$. In such cases, $\FP(G)$ is guaranteed to be parameter-independent (since the graph rules do not depend on $\varepsilon$ and $\delta$).  As an example, consider the graph on $n = 5$ nodes in Figure~\ref{fig:fp-examples}A; we will show that $\FP(G)$ is completely determined by graph rules. Going through the possible subsets $\sigma$ of different sizes, we find that for $|\sigma| = 1$ only $3,4 \in \FP(G)$ (as those are the sinks). Using Rules 1, 2, and 4, we see that the only $|\sigma| = 2$ elements in $\FP(G)$ are the clique $15$ and the independent set $34$. A crucial ingredient for determining the fixed point supports of sizes $3$ and $4$ is the sinks rule, which guarantees that $135$, $145$, and $1345$ are the only supports of these sizes. Finally, notice that the total number of fixed points up through size $|\sigma| = 4$ is odd. Using Rule 8 (parity), we can thus conclude that there is no fixed point of full support -- that is, with $|\sigma| = 5$. It follows that $\FP(G) = \{3,4,15,34,135,145,1345\}$; moreover, this result is parameter-independent because it was determined purely from graph rules. 
Although the precise values of the fixed points will change for different choices of the parameters $\varepsilon, \delta$ and $\theta$, the set of supports $\FP(G)$ is invariant.

We leave it as an exercise to use graph rules to show that $\FP(G) = \{134\}$ for the graph in Figure~\ref{fig:fp-examples}B, and $\FP(G) = \{4, 12, 124 \}$ for the graph in Figure~\ref{fig:fp-examples}C.  For the graph in C, it is necessary to appeal to a more general rule for \emph{uniform in-degree} subgraphs, which we review next.

Rules 1-7, and many more, all emerge as corollaries of more general rules. In the next few subsections, we will introduce the uniform in-degree rule, graphical domination, and simply-embedded subgraphs. 
Then, in Section~\ref{sec:elem-rules}, we will pool together the more general rules into a complete set of {\it elementary graph rules} from which all others follow.


\subsection{Uniform in-degree rule}

It turns out that Rules 1, 2, and 3 (for independent sets, cliques, and cycles) are all corollaries of a single rule for graphs of {\it uniform in-degree}.

\begin{definition}
We say that $G|_\sigma$ has {\em uniform in-degree} $d$ if every node $i \in \sigma$ has $d$ incoming edges from within
$G|_\sigma$. 
\end{definition}

Note that an independent set has uniform in-degree $d = 0$, a cycle has uniform in-degree $d = 1$, and an $n$-clique is uniform in-degree with $d = n-1$. 
But, in general, uniform in-degree graphs need not be symmetric. For example, the induced subgraph $G|_{145}$ in Figure~\ref{fig:fp-examples}A is uniform in-degree, with $d=1$.

\begin{figure}[!h]
\begin{center}
\includegraphics[width=2.75in]{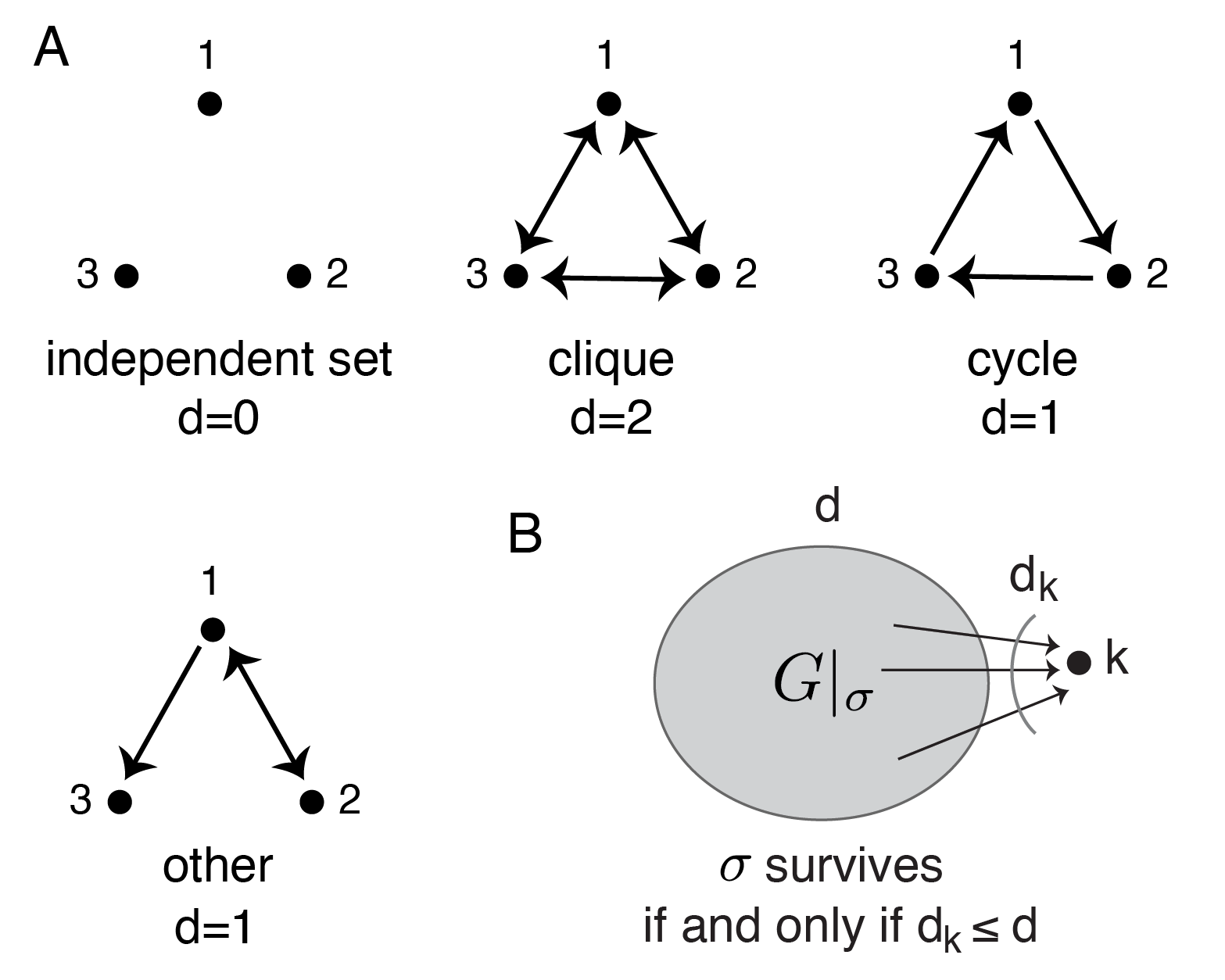}
\end{center}
\caption{\carina{(A) All uniform in-degree graphs of size $n=3$. (B) The fixed point survival rule in Theorem~\ref{thm:ufd}.}}
\label{fig:uniform-in-deg}
\end{figure}

\carina{For CTLNs, a fixed point $x^*$ with support $\sigma$ satisfies:
$$(I-W_\sigma)x_\sigma^* = \theta 1_\sigma,$$
where $1_\sigma$ is a vector of all $1$'s restricted to the index set $\sigma$.
If $G|_\sigma$ has uniform in-degree $d$, then the row sums of $I-W_\sigma$ are identical, and so $1_\sigma$ is an eigenvector. In particular, 
$$x_\sigma^* = \dfrac{\theta}{R}1_\sigma,$$
where $R$ is the (uniform) row sum for the matrix $I-W_\sigma$. For in-degree $d$, we
compute 
$$R = 1+d(1-\varepsilon) + (|\sigma|-d-1)(1+\delta).$$
Uniform in-degree fixed points with support $\sigma$ thus have the same value for all $i \in \sigma$:
\begin{eqnarray}\label{eq:ufd}
x_i^* = \dfrac{\theta}{|\sigma| + \delta(|\sigma|-d-1)-\varepsilon d}.
\end{eqnarray}
(See also \cite[Lemma 18]{fp-paper}.) From the derivation, it is clear that this formula holds for all uniform in-degree graphs, even those that are not symmetric.}

\carina{We can use the formula~\eqref{eq:ufd} to verify that the on-neuron conditions, $x_i^*>0$ for each $i \in \sigma$, are satisfied for $\varepsilon, \delta, \theta$ within the legal range. Using it to check the off-neuron conditions, we find that for $k \notin \sigma$,
\begin{eqnarray*}
y_k^* &=& \sum_{i \in \sigma} W_{ki} x_i^* + \theta,\\
&=& \sum_{i \to k} (-1+\varepsilon) x_i^* + \sum_{i \not\to k} (-1-\delta) x_i^* + \theta,\\
&=& \theta\left(\dfrac{d_k(-1+\varepsilon) + (|\sigma|-d_k)(-1-\delta)}{|\sigma| + \delta(|\sigma|-d-1)-\varepsilon d}+1\right),
\end{eqnarray*}
where $d_k = |\{i \in \sigma \mid i \to k\}|$. From here, it is not difficult to see that the off-neuron condition, $y_k^* \leq 0$, will be satisfied if and only if $d_k \leq d$. This gives us the following theorem.
}

\begin{theorem}[\cite{fp-paper}]\label{thm:ufd}
Let $G|_\sigma$ be an induced subgraph of $G$ with uniform in-degree $d$. For $k \notin \sigma$, let $d_k$ denote the number of edges $i \to k$ for $i \in \sigma$. Then $\sigma \in \FP(G|_\sigma)$, and
\vspace{-.075in}
$$\sigma \in \FP(G|_{\sigma \cup k}) \; \Leftrightarrow \; d_k \leq d. \vspace{-.075in}$$
In particular, $\sigma \in \FP(G)$ if and only if there does not exist $k \notin \sigma$ such that $d_k > d$.
\end{theorem}

Figure~\ref{fig:uniform-in-deg} gives examples of uniform in-degree graphs and illustrates the survival condition in Theorem~\ref{thm:ufd}.

\subsection{Graphical domination}

We have seen that uniform in-degree graphs support fixed points that have uniform firing rates (equation~\eqref{eq:ufd}). More generally, fixed points can have very different values across neurons. However, there is some level of ``graphical balance'' that is required of $G|_\sigma$ for any fixed point support $\sigma$. For example, it can be shown that if $\sigma$ contains a pair of neurons $j,k$ that have the property that all neurons sending edges to $j$ also send edges to $k$, and $j \to k$ but $k \not\to j$, then $\sigma$ cannot be a fixed point support. Intuitively, this is because $k$ is receiving a strict superset of the inputs to $j$, and this imbalance rules out their ability to coexist in the same fixed point support.  This property motivates the following definition.

\begin{definition} We say that $k$ {\it graphically dominates} $j$ {\it with respect to} $\sigma$ in $G$ if the following three conditions all hold:
\vspace{-.1in}
\begin{enumerate}
\item For each $i \in \sigma \setminus \{j,k\}$, if $i \to j$ then $i \to k$.
\vspace{-.1in}
\item If $j \in \sigma$, then $j \to k$.
\vspace{-.1in}
\item If $k \in \sigma$, then $k \not\to j$.
\end{enumerate}
\end{definition}

\begin{figure}[!h]
\vspace{-.1in}
\begin{center}
\includegraphics[width=3in]{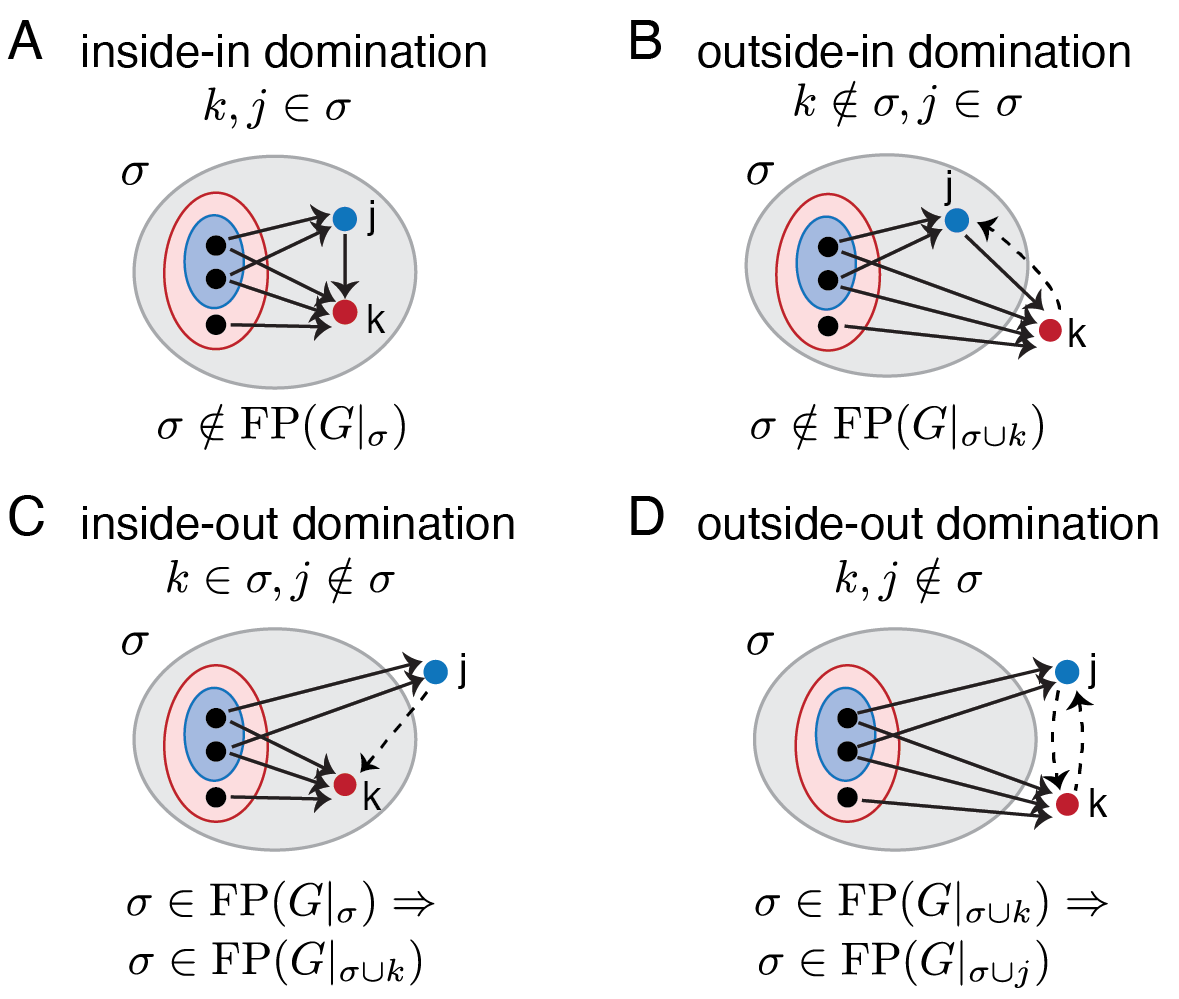}
\end{center}
\vspace{-.1in}
\caption{{\bf Graphical domination: four cases.} In all cases, $k$ graphically dominates $j$ with respect to $\sigma$. In particular, the set of vertices of $\sigma \setminus \{j,k\}$ sending edges to $k$ (red ovals) always contains the set of vertices sending edges to $j$ (blue ovals).}
\label{fig:domination}
\end{figure}

\noindent We refer to this as ``inside-in'' domination if $j,k \in \sigma$ (see Figure~\ref{fig:domination}A). In this case, we must have $j \to k$ and $k \not\to j$. If $j \in \sigma$, $k \notin \sigma$, we call it ``outside-in'' domination (Figure~\ref{fig:domination}B). On the other hand, ``inside-out'' domination is the case where $k \in \sigma$, $j \notin \sigma$, and ``outside-out'' domination refers to $j,k \notin \sigma$ (see Figure~\ref{fig:domination}C-D).

What graph rules does domination give us? Intuitively, when inside-in domination is present, the ``graphical balance'' necessary to support a fixed point is violated, and so $\sigma \notin \FP(G)$.  When $k$ outside-in dominates $j$, with $j \in \sigma$ and $k \notin \sigma$, again there is an imbalance, and this time it guarantees that neuron $k$ turns on, since it receives all the inputs that were sufficient to turn on neuron $j$.  Thus, there cannot be a fixed point with support $\sigma$ since node $k$ will violate the off-neuron conditions.  
We can draw interesting conclusions in the other cases of graphical domination as well, as Theorem~\ref{thm:domination} shows. 

\begin{theorem}[\cite{fp-paper}]\label{thm:domination}
Suppose $k$ graphically dominates $j$ with respect to $\sigma$ in $G$. Then the following all hold:
\begin{enumerate}
\item (inside-in) If $j,k \in \sigma$, then $\sigma \notin \FP(G|_\sigma)$ and thus $\sigma \notin \FP(G)$.
\item (outside-in) If $j \in \sigma$, $k \notin \sigma$, then $\sigma \notin \FP(G|_{\sigma\cup k})$ and thus $\sigma \notin \FP(G)$.
\item (inside-out) If $k \in \sigma$, $j \notin \sigma$, then 
$\sigma \in \FP(G|_{\sigma}) \; \Rightarrow \; \sigma \in \FP(G|_{\sigma \cup j})$.
\item (outside-out) If $j,k \not\in \sigma$, then $\sigma \in \FP(G|_{\sigma \cup k}) \; \Rightarrow \; \sigma \in \FP(G|_{\sigma \cup j})$.
\end{enumerate}
\end{theorem}

\carina{The four cases of Theorem~\ref{thm:domination} are illustrated in Figure~\ref{fig:domination}.  This theorem was originally proven in \cite{fp-paper}. Here we provide a more elementary proof, using only the definition of CTLNs and ideas from Section~\ref{sec:hyperplanes}.} 

\begin{proof}
\carina{Suppose that $k$ graphically dominates $j$ with respect to $\sigma$ in $G$. To prove statements 1 and 2 in the theorem, we will also assume that there exists a fixed point $x^*$ of the associated CTLN with support $\supp(x^*) = \sigma$. This will allow us to arrive at a contradiction.}

\carina{If $x^*$ is a fixed point, we must
have $x_i^* = [y_i^*]_+$ for all $i \in [n]$ (see equation~\eqref{eq:at-fp} from Section~\ref{sec:hyperplanes}).
Recalling that $W_{jj} = W_{kk} = 0$, and that $x_i^* = 0$ for $i \notin \sigma$, it follows that for any $j,k \in [n]$,
we have:
\begin{eqnarray*}
 y_j^* & = & \sum_{i \in \sigma \setminus \{j,k\}} W_{ji} x^*_i + W_{jk} x^*_k + \theta,\\
 y_k^* & = & \sum_{i \in \sigma \setminus \{j,k\}} W_{ki} x^*_i + W_{kj} x^*_j + \theta.
\end{eqnarray*}}

%
\carina{Since $k$ graphically dominates $j$ with respect to $\sigma$, we know that $W_{ji} \leq W_{ki}$ for all $i \in \sigma~\setminus~\{j,k\}$. This is because the off-diagonal values $W_{\ell i}$ are either $-1+\varepsilon$, for $i \to \ell$, or $-1-\delta$, for $i \not\to \ell$; and $-1+\varepsilon > -1-\delta$.
It now follows from the above equations that $y_j^* - W_{jk}x^*_k \leq y_k^* - W_{kj}x^*_j$. Equivalently,
\begin{equation}\label{eq:dom}
y_j^*+W_{kj}x^*_j \leq y_k^*+W_{jk}x^*_k.
\end{equation}
We will refer frequently to~\eqref{eq:dom} in what follows. There are four cases of domination to consider. 
We begin with the first two:
\begin{enumerate}
\item[1.] (inside-in) If $j, k \in \sigma$, then $x^*_j = y_j^* > 0$ and $x^*_k = y_k^* > 0$, and so at the fixed point we must have $(1+W_{kj})x^*_j \leq (1+W_{jk})x^*_k.$ But domination in this case implies $j \to k$ and $k \not\to j$, so that $W_{kj} = -1+\varepsilon$ and $W_{jk} = -1-\delta$. Plugging this in, we obtain $\varepsilon x^*_j \leq -\delta x^*_k$. This results in a contradiction, since $x^*_j, x^*_k > 0$ and $\varepsilon, \delta > 0$. We conclude that $\sigma \notin \FP(G)$. More specifically, since the contradiction involved only the on-neuron conditions, it follows that $\sigma \notin \FP(G|_\sigma)$.
\item[2.] (outside-in) If $j \in \sigma$ and $k \notin \sigma$, then $x^*_j = y_j^* > 0$ and $x^*_k = 0,$ with $y_k^* \leq 0$. It follows from~\eqref{eq:dom} that
$(1+W_{kj})x^*_j \leq~0$. Since this case of domination also has $j\to k$, we obtain $(1+W_{kj})x^*_j = \varepsilon x_j^* \leq 0$, a contradiction. Again, we can conclude that $\sigma \notin \FP(G)$, and more specifically that 
$\sigma \notin \FP(G|_{\sigma \cup k})$. 
\end{enumerate}}
\noindent\carina{This completes the proof of statements 1 and 2.}

\carina{To prove statements 3 and 4, we assume only that $\sigma \in \FP(G|_\sigma)$, so that a fixed point $x^*$
with support $\sigma$ exists in the restricted network $G|_\sigma$, but does not necessarily extend to larger networks.
Whether or not it extends depends on whether $y_i^* \leq 0$ for all $i \notin \sigma$. 
\begin{enumerate}
\item[3.] (inside-out) If $j \not\in \sigma$ and $k \in \sigma$, then $x^*_j = 0$ and $x^*_k = y_k^* > 0$, and so~\eqref{eq:dom} becomes $y_j^* \leq (1+W_{jk})x^*_k.$ Domination in this case implies $k \not\to j$, so we obtain $y_j^* \leq -\delta x^*_k < 0$. This shows that $j$ is guaranteed to satisfy the required off-neuron condition. We can thus conclude that $\sigma \in \FP(G|_{\sigma \cup j})$. 
\item[4.] (outside-out) If $j,k \notin \sigma$, then $x^*_j = x^*_k = 0$, and so~\eqref{eq:dom} tells us that $y_j^* \leq y_k^*$. This is true irrespective of whether or not $j \to k$ or $k \to j$ (and both are optional in this case). 
Clearly, if $y_k^* \leq 0$ then $y_j^* \leq 0$. We can thus conclude that if
$\sigma \in \FP(G|_{\sigma \cup k})$, then $\sigma \in \FP(G|_{\sigma \cup j})$.
\end{enumerate}}
\end{proof}

\carina{Rules 4, 5, and 7 are all consequences of Theorem~\ref{thm:domination}.}
To see how, consider a graph with a source $j \in \sigma$ that has an edge $j \to k$ for some $k \in [n]$. Since $j$ is a source, it has no incoming edges from within $\sigma$. If $k \in \sigma$, then $k$ inside-in dominates $j$ and so $\sigma \notin \FP(G)$. If $k \notin \sigma$, then $k$ outside-in dominates $j$ and again $\sigma \notin \FP(G)$. Rule 4(i) immediately follows.  We leave it as an exercise to prove Rules 4(ii), 5(i), 5(ii), and 7.

\subsection{Simply-embedded subgraphs and covers}
Finally, we introduce the concept of simply-embedded subgraphs. 
\carina{This is the last piece we need before presenting the complete set of elementary graph rules.}

\begin{definition}[simply-embedded]
We say that a subgraph $G|_\tau$ is {\em simply-embedded in} $G$ if for each $k \notin \tau$, either 
\begin{itemize}
\item[(i)] $k \to i$ for all $i \in \tau$, or 
\vspace{-.05in}
\item[(ii)] $k \not\to i$ for all $i \in \tau$. 
\end{itemize}
\end{definition}

\begin{figure}[!h]
\begin{center}
\includegraphics[width=3in]{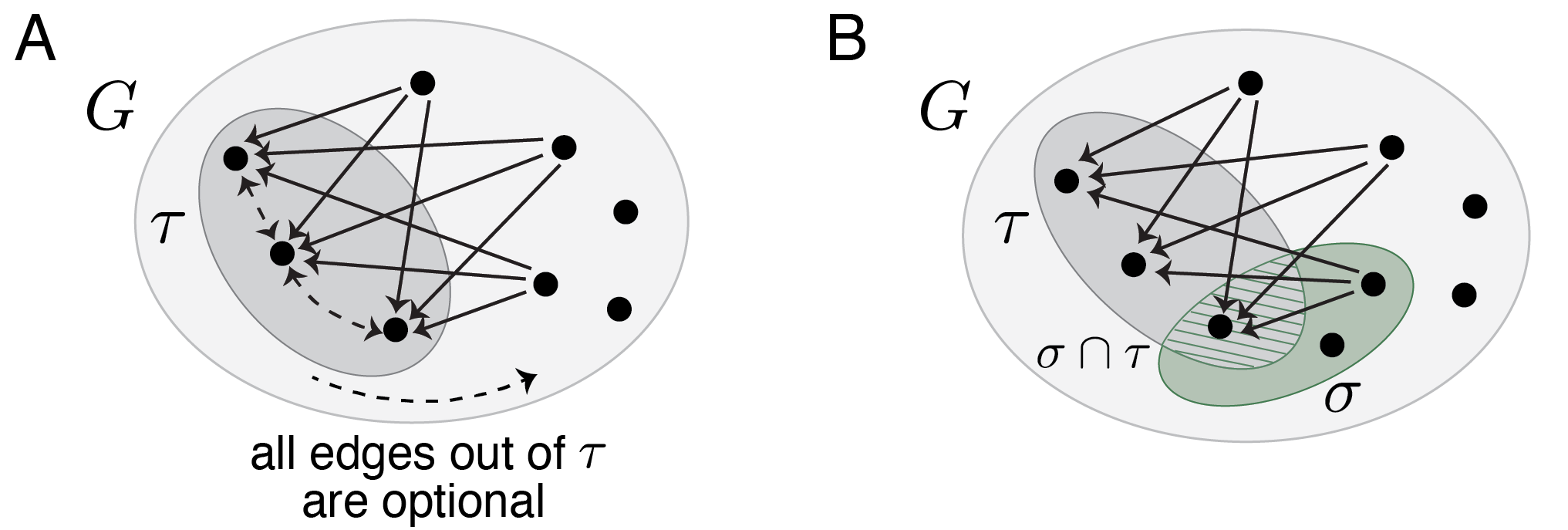}
\end{center}
\caption{\bf Simply-embedded subgraphs.}
\label{fig:simply-embedded}
\end{figure}

\noindent In other words, while $G|_\tau$ can have any internal structure, the rest of the network treats all nodes in $\tau$ equally (see Figure~\ref{fig:simply-embedded}A). By abuse of notation, we sometimes say that the corresponding subset of vertices $\tau \subseteq [n]$ is simply-embedded in $G$.

\carina{We allow $\tau = [n]$ as a trivial case, meaning that $G$ is simply-embedded in itself. At the other extreme, all singletons $\tau = \{i\}$ and the empty set $\tau = \emptyset$ are simply-embedded in $G$, also for trivial reasons. Note that a subset of a simply-embedded set, $\omega \subset \tau$, need not be simply-embedded. This is because nodes in $\tau \setminus \omega$ may not treat those in $\omega$ equally.}

\carina{Now let's consider the CTLN equations for neurons in a simply-embedded subgraph $G|_\tau$, for $\tau \subset [n]$. For each $i \in \tau$, the equations for the dynamics can be rewritten as:
$$\dfrac{dx_i}{dt} = -x_i + \left[\sum_{j \in \tau} W_{ij} x_j + \sum_{k \not\in \tau} W_{ik} x_k + \theta\right]_+,$$
where the term $\sum_{k \not\in \tau} W_{ik} x_k$ is identical for all $i \in \tau$. This is because $W_{ik} = -1+\varepsilon$, if $k \to i$, and $W_{ik} = -1-\delta$ if $k \not\to i$; so the fact that $k$ treats all $i \in \tau$ equally means that the matrix entries $\{W_{ik}\}_{i \in \tau}$ are identical for fixed $k$.
We can thus define a single time-varying input function,
$$\mu_\tau(t) = \sum_{k \not\in \tau} W_{ik} x_k(t) + \theta, \;\; \text{for } \; i \in \tau,$$
that is the same independent of the choice of $i \in \tau$. This gives us:
\begin{small}
$$\dfrac{dx_i}{dt} = -x_i + \left[\sum_{j \in \tau} W_{ij} x_j + \mu_\tau(t) \right]_+, \text{ for each } i \in \tau.$$
\end{small}
In particular, the neurons in $\tau$ evolve according to the dynamics of the local network $G|_\tau$ in the presence of a time-varying input $\mu_\tau(t)$, in lieu of the constant $\theta$.}

\carina{Suppose we have a fixed point $x^*$ of the full network $G$, with support $\sigma \in \FP(G)$. At the fixed point, 
$$\mu_\tau^* = \sum_{k \not\in \tau} W_{ik} x_k^* + \theta = \sum_{k \in \sigma \setminus \tau} W_{ik} x_k^* + \theta,$$
which is a constant. We can think of this as a new choice of the CTLN input parameter, $\widetilde{\theta} = \mu_\tau^*$, with the caveat that we may have $\widetilde{\theta} \leq 0$. It follows that the restriction of the fixed point to $\tau$, $x_\tau^*$, must be a fixed point of subnetwork $G|_\tau$. If $\widetilde{\theta} \leq 0$, this will be the zero fixed point corresponding to $\emptyset$ support. If $\widetilde{\theta} >0$, this fixed point will have nonempty support $\sigma \cap \tau \in \FP(G|_\tau)$. 
From these observations,} we have the following key lemma (see Figure~\ref{fig:simply-embedded}B):


\begin{lemma}\label{lemma:se-restriction}
Let $G|_\tau$ be simply-embedded in $G$. Then for any $\sigma \subseteq [n]$, 
\vspace{-.05in}
$$\sigma \in \FP(G) \; \Rightarrow \; \sigma \cap \tau \in \FP(G|_\tau) \cup \{\emptyset\}.\vspace{-.05in}$$
\end{lemma}

What happens if we consider more than one simply-embedded subgraph? \carina{Lemma~\ref{lemma:simply-embedded-intersections-unions} shows that} intersections of simply-embedded subgraphs are also simply-embedded. However, the union of two simply-embedded subgraphs is only guaranteed to be simply-embedded if the intersection is nonempty.  \carina{(It is easy to find a counterexample if the intersection is empty.)}

\carina{
\begin{lemma}\label{lemma:simply-embedded-intersections-unions}
Let $\tau_1, \tau_2 \subseteq [n]$ be simply-embedded in $G$. Then $\tau_1 \cap \tau_2$ is simply-embedded in $G$. If $\tau_1 \cap \tau_2 \neq \emptyset,$ then $\tau_1 \cup \tau_2$ is also simply-embedded in $G$.
\end{lemma}
}

\begin{proof}
\carina{If $\tau_1 \cap \tau_2 = \emptyset$, then the intersection is trivially simply-embedded. Assume $\tau_1 \cap \tau_2 \neq \emptyset$, and consider $k \notin \tau_1 \cap \tau_2$. If $k \notin \tau_1$, then $k$ treats all vertices in $\tau_1$ equally and must therefore treat all  vertices in $\tau_1 \cap \tau_2$ equally. By the same logic, if $k \notin \tau_2$ then it must treat all vertices in $\tau_1 \cap \tau_2$ equally. It follows that $\tau_1 \cap \tau_2$ is simply-embedded in $G$. }

\carina{Next, consider $\tau_1 \cup \tau_2$ for a pair of subsets $\tau_1, \tau_2$ such that $\tau_1\cap\tau_2 \neq \emptyset.$ Let
$j \in \tau_1 \cap \tau_2$ and $k \notin \tau_1 \cup \tau_2$. If $k \to j$, then $k \to i$ for all $i \in \tau_1$ since $k \notin \tau_1$; moreover, $k \to \ell$ for all $\ell \in \tau_2$ since $k \notin \tau_2$. If, on the other hand, $k \not\to j$, then by the same logic $k \not\to i$ for any $i \in \tau_1$ and $k \not\to \ell$ for any $\ell \in \tau_2$. It follows that $\tau_1 \cup \tau_2$ is simply-embedded in $G$. }
\end{proof}

%
%
%
%
%
If we have two simply-embedded subgraphs, $G|_{\tau_i}$ and $G|_{\tau_j}$, we know that for any $\sigma \in \FP(G)$, $\sigma$ must restrict to a fixed point $\sigma_i = \sigma \cap \tau_i$ and $\sigma_j = \sigma \cap \tau_j$ in each of those subgraphs.  But when can we \emph{glue} together such a $\sigma_i \in \FP(G|_{\tau_i})$ and $\sigma_j \in \FP(G|_{\tau_j})$ to produce a larger fixed point support $\sigma_i \cup \sigma_j$ in $\FP(G|_{\tau_i \cup \tau_j})$?  

Lemma~\ref{lemma:rule6b} precisely answers this question.  It uses the following notation: 
$$\tilFP(G) \od \FP(G) \cup \{\emptyset \}.$$

\begin{lemma}[pairwise gluing]\label{lemma:rule6b}
Suppose $G|_{\tau_i}, G|_{\tau_j}$ are simply-embedded in $G$, and consider $\sigma_i \in \tilFP(G|_{\tau_i})$ and $\sigma_j \in \tilFP(G|_{\tau_j})$ that satisfy 
$\sigma_i \cap \tau_j = \sigma_j \cap \tau_i$ (so that $\sigma_i, \sigma_j$ agree on the overlap $\tau_i \cap \tau_j$).  Then
$$\sigma_i \cup \sigma_j \in \tilFP(G|_{\tau_i \cup \tau_j})$$ 
if and only if one of the following holds:
\begin{enumerate}
\item[(i)] $\tau_i \cap \tau_j = \emptyset$ and $\sigma_i, \sigma_j \in  \tilFP(G|_{\tau_i \cup \tau_j}),$ or
\vspace{-.075in}
\item[(ii)] $\tau_i \cap \tau_j = \emptyset$ and $\sigma_i, \sigma_j \notin  \tilFP(G|_{\tau_i \cup \tau_j}),$ or
\vspace{-.075in}
\item[(iii)] $\tau_i \cap \tau_j \neq \emptyset.$
\end{enumerate}
\end{lemma}

\carina{Parts (i-ii) of Lemma~\ref{lemma:rule6b} are essentially the content of  \cite[Theorem 14]{fp-paper}. Part (iii) can also be proven with similar arguments.}

\subsection{\carina{Elementary graph rules}}\label{sec:elem-rules}

In this section we collect a set of elementary graph rules from which all other graph rules can be derived. The first two elementary rules arise from general arguments about TLN fixed points stemming from the hyperplane arrangement picture. They hold for all competitive/inhibition-dominated nondegenerate TLNs, as does Elem Rule 3 (aka Rule 8). The last three elementary graph rules are specific to CTLNs, and recap results from the previous three subsections.
As usual, $G$ is a graph on $n$ nodes and $\FP(G)$ is the set of fixed points supports.

There are six elementary graph rules:

\begin{enumerate}
\item[Elem Rule 1] (unique supports): For a given $G$, there is at most one fixed point per support $\sigma \subseteq [n]$. The fixed points can therefore be labeled by the elements of $\FP(G)$.

\item[Elem Rule 2]  (restriction/lifting): Let $\sigma \subseteq [n]$. Then
\begin{eqnarray*}
\sigma \in \FP(G) &\Leftrightarrow& \sigma \in \FP(G|_{\sigma}) \text{ and } \\
&& \sigma \in \FP(G|_{\sigma \cup k}) \text{ for all } k \notin \sigma.
\end{eqnarray*}
Moreover, whether $\sigma \in \FP(G|_\sigma)$ survives to $\sigma \in \FP(G|_{\sigma \cup k})$ depends only on the outgoing edges $i \to k$ for $i \in \sigma$, not on the backward edges $k \to i$.

\item[Elem Rule 3]  (parity): The total number of fixed points, $|\FP(G)|,$ is always odd.

\item[Elem Rule 4] (uniform in-degree): If $G|_\sigma$ has uniform in-degree $d$, then
\begin{enumerate}
\item $\sigma \in \FP(G|_\sigma)$, and
\item $\sigma \in \FP(G|_{\sigma \cup k}) \; \Leftrightarrow \; d_k \leq d$ in $G|_{\sigma \cup k}.$
\end{enumerate}
In particular, $\sigma \in \FP(G) \; \Leftrightarrow \; $ there does not exist $k \notin \sigma$ that receives more than $d$ edges from $\sigma$.

\item[Elem Rule 5]  (domination): Suppose $k$ graphically dominates $j$ with respect to $\sigma$. 
\begin{enumerate}
\item (inside-in) If $j,k \in \sigma$, then $\sigma \notin \FP(G|_\sigma)$ and thus $\sigma \notin \FP(G)$.
\item (outside-in) If $j \in \sigma$, $k \notin \sigma$, then \\ 
$\sigma \notin \FP(G|_{\sigma\cup k})$ and thus $\sigma \notin \FP(G)$.
\item (inside-out) If $k \in \sigma$, $j \notin \sigma$, then \\
$\sigma \in \FP(G|_{\sigma}) \; \Rightarrow \; \sigma \in \FP(G|_{\sigma \cup j})$.
\item (outside-out) If $j,k \not\in \sigma$, then \\
$\sigma \in \FP(G|_{\sigma \cup k}) \; \Rightarrow \; \sigma \in \FP(G|_{\sigma \cup j})$.
\end{enumerate}

\item[Elem Rule 6]  (simply-embedded): Suppose that $G|_{\tau_i}, G|_{\tau_j}$ are simply-embedded in $G$, and recall the notation 
$\tilFP(G) = \FP(G) \cup \{\emptyset \}.$ We have the following restriction and gluing rules:
\begin{enumerate}
\item (restriction) $\sigma \in \FP(G)   \Rightarrow \sigma \cap \tau_i \in \tilFP(G|_{\tau_i}).$
\item (pairwise gluing) If $\sigma_i \in \tilFP(G|_{\tau_i})$, $\sigma_j \in \tilFP(G|_{\tau_j}),$ and
$\sigma_i \cap \tau_j = \sigma_j \cap \tau_i$ (so that $\sigma_i, \sigma_j$ agree on the overlap $\tau_i \cap \tau_j$), then
$\sigma_i \cup \sigma_j \in \tilFP(G|_{\tau_i \cup \tau_j})$ if and only if one of the following holds:
\begin{enumerate}
\item $\tau_i \cap \tau_j = \emptyset$ and $\sigma_i, \sigma_j \in  \tilFP(G|_{\tau_i \cup \tau_j}),$
\item $\tau_i \cap \tau_j = \emptyset$ and $\sigma_i, \sigma_j \notin  \tilFP(G|_{\tau_i \cup \tau_j}),$
\item $\tau_i \cap \tau_j \neq \emptyset.$
\end{enumerate}
Moreover, if $\tau_i \cap \tau_j \neq \emptyset,$ we are also guaranteed that $G|_{\tau_i \cup \tau_j}$ and $G|_{\tau_i \cap \tau_j}$ are simply-embedded in $G$. Thus, $\sigma_i \cap \sigma_j \in \tilFP(G|_{\tau_i \cap \tau_j})$. If, additionally, $\sigma_i \cap \sigma_j \neq \tau_i \cap \tau_j$, then $\sigma_i, \sigma_j \in  \tilFP(G|_{\tau_i \cup \tau_j}).$
\item (lifting) If $\{\tau_1, \ldots, \tau_N\}$ is a simply-embedded cover of $G$ and $\sigma \cap \tau_i  \in \FP(G|_{\tau_i})$ for each $i \in [N]$, then
$$ \sigma \in \FP(G) \; \Leftrightarrow \; \sigma \in \FP(G|_\sigma).$$
\end{enumerate}
\end{enumerate}

\begin{figure*}[!ht]
\begin{center}
\includegraphics[width=5.75in]{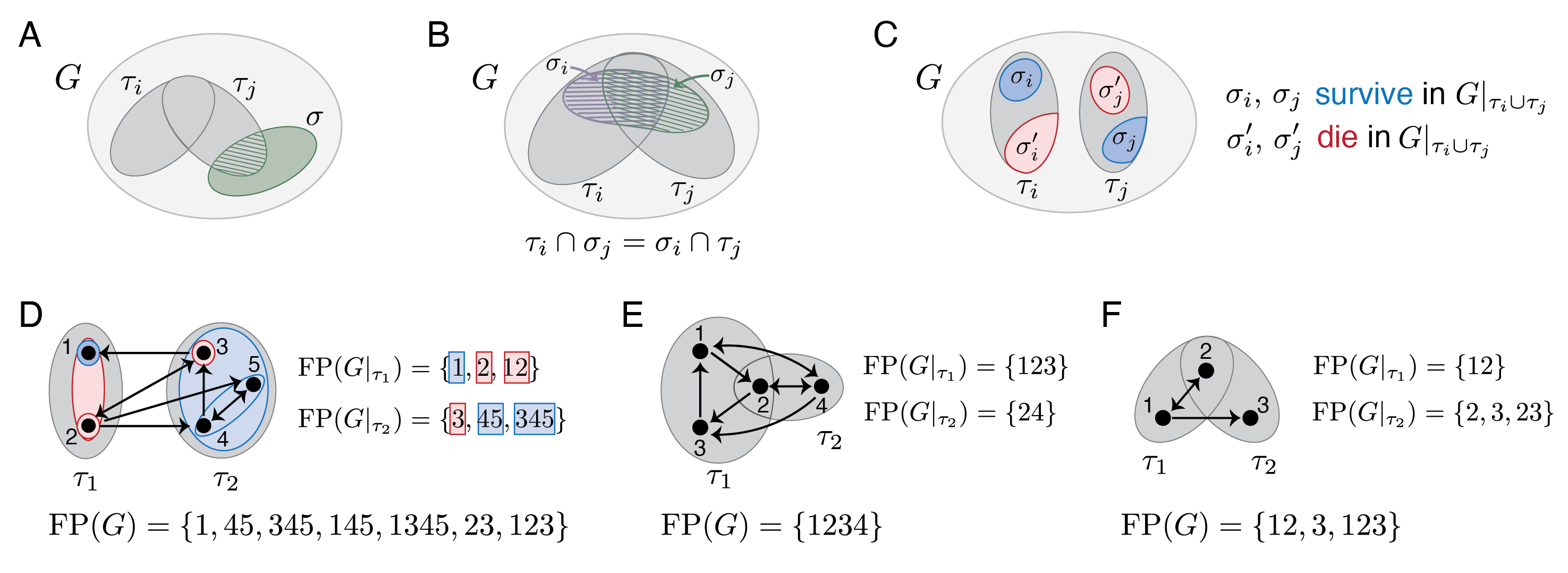}
\end{center}
\caption{\carina{{\bf Elementary Rule 6.} (A) Sets $\tau_i, \tau_j$ are from a simply-embedded cover of $G$. If $\sigma \in \FP(G)$, then $\sigma \cap \tau_j \in \tilFP(G|_{\tau_j})$, per Elem Rule 6(a). Note that we also have $\sigma \cap \tau_i \in \tilFP(G|_{\tau_i})$, since $\sigma \cap \tau_i = \emptyset$ is included in $\tilFP(G|_{\tau_i})$. (B) Two sets, $\sigma_i \in \tilFP(G|_{\tau_i})$ and $\sigma_j \in \tilFP(G|_{\tau_j})$, that agree on the nonempty overlap $\tau_i \cap \tau_j$. We thus have the pairwise gluing, $\sigma_i \cup \sigma_j \in \tilFP(G|_{\tau_i \cup \tau_j})$, per Elem Rule 6(b)iii. (C) When $\tau_i \cap \tau_j = \emptyset$, we obtain pairwise gluing $\sigma_i \cup \sigma_j \in \tilFP(G|_{\tau_i \cup \tau_j})$ if either $\sigma_i, \sigma_j$ both survive to be elements of $\tilFP(G|_{\tau_i \cup \tau_j})$, per Elem Rule 6(b)i, or if they both die so that $\sigma_i, \sigma_j \notin  \tilFP(G|_{\tau_i \cup \tau_j}),$ 
per Elem Rule 6(b)ii. (D) A concrete example of Elem Rule 6(b)i-ii at work. We can fully determine $\FP(G)$ in this case, via pairwise gluing. (E) A concrete example of Elem Rule 6(b)iii at work. Note that this graph is the same as the core motif 4-ufd in Figure~\ref{fig:n4-cores}. (F) Another example where $\FP(G)$ can be fully determined using Elem Rule 6(b)iii.}}
\label{fig:elem-graph-rules}
\end{figure*}

Elem Rule 6 is illustrated in Figure~\ref{fig:elem-graph-rules}. It collects several results related to simply-embedded graphs.
Elem Rule 6(a) is the same as Lemma~\ref{lemma:se-restriction}, while Elem Rule 6(b) is given by 
Lemmas~\ref{lemma:simply-embedded-intersections-unions} and~\ref{lemma:rule6b}. Note that this rule is valid even if $\sigma_i$ or $\sigma_j$ is empty. 
\carina{Elem Rule 6(c) applies to {\it simply-embedded covers} of $G$, a notion we will define in the next section (see Definition~\ref{def:s-e-covers}, below).
The forward direction, $\sigma \in \FP(G) \Rightarrow \sigma \in \FP(G|_\sigma)$, follows from Elem Rule 2. The backwards direction is the content of \cite[Lemma 8]{fp-paper}.}

\section{Gluing rules}\label{sec:gluing-rules}

So far we have seen a variety of graph rules and the elementary graph rules from which they are derived. These rules allow us to rule in and rule out potential fixed points in $\FP(G)$ from purely graph-theoretic considerations. In this section, we consider networks whose graph $G$ is composed of smaller induced subgraphs, $G|_{\tau_i}$, for $i \in [N] = \{1,\ldots,N\}$. What is the relationship between $\FP(G)$ and the fixed points of the components, $\FP(G|_{\tau_i})$? 

It turns out we can obtain nice results if the induced subgraphs $G|_{\tau_i}$ are all simply-embedded in $G$. In this case, we say that $G$ has a simply-embedded cover.

\begin{definition}[simply-embedded covers]\label{def:s-e-covers}
We say that $\U = \{\tau_1,\ldots,\tau_N\}$ is a {\em simply-embedded cover} of $G$ if each $\tau_i$ is simply-embedded in $G$, and for every vertex $j \in [n],$ there exists an $i \in [N]$ such that $j \in \tau_i$. In other words, the $\tau_i$'s are a vertex cover of $G$. If the $\tau_i$'s are all disjoint, we say that 
$\U$ is a {\em simply-embedded partition} of $G$.
\end{definition}

\carina{Every graph $G$ has a trivial simply-embedded cover, with $N = n$, obtained by taking $\tau_i = \{i\}$ for each $i \in [n]$. This is also a simply-embedded partition. At the other extreme, since the full set of vertices $[n]$ is a simply-embedded set, we also have the trivial cover with $N = 1$ and $\tau_1 = [n]$. These covers, however, do not yield useful information about $\FP(G)$.  In contrast, nontrivial simply-embedded covers can provide strong constraints on, and in some cases fully determine, the set of fixed points $\FP(G)$. Some of these constraints can be described via {\it gluing rules}, which we explain below.}

In the case that $G$ has a simply-embedded cover, Lemma~\ref{lemma:se-restriction} tells us that all ``global'' fixed point supports in $\FP(G)$ must be unions of ``local'' fixed point supports in the $\FP(G|_{\tau_i})$, since every $\sigma \in \FP(G)$ restricts to $\sigma \cap \tau_i \in \FP(G|_{\tau_i}) \cup \{\emptyset\}$.  But what about the other direction? 

\begin{question}\label{ques:gluing}
When does a collection of local fixed point supports $\{\sigma_i\}$, with each nonempty $\sigma_i \in  \FP(G|_{\tau_i})$, glue together to form a global fixed point support $\sigma = \cup \sigma_i \in \FP(G)$? 
\end{question}

To answer this question, we develop some notions inspired by sheaf theory. 
For a graph $G$ on $n$ nodes, with a simply-embedded cover $\U = \{\tau_1, \ldots, \tau_N\}$, we define the 
{\it gluing complex} as:
\begin{small}
\begin{eqnarray*}
\F_G(\U) &\od& \{\sigma = \cup_i \sigma_i \mid \sigma \neq \emptyset, 
\sigma_i \in \FP(G|_{\tau_i}) \cup \{\emptyset\},\\
&& \text{ and } \sigma_i \cap \tau_j = \sigma_j \cap \tau_i \text{ for all } i,j \in [N]\}.
\end{eqnarray*}
\end{small}

\noindent In other words, $\F_G(\U)$ consists of all $\sigma \subseteq [n]$ that can be obtained by gluing together local fixed point supports $\sigma_i \in \FP(G|_{\tau_i})$. Note that in order to guarantee that $\sigma_i = \sigma \cap \tau_i$ for each $i$, it is necessary that the $\sigma_i$'s agree on overlaps $\tau_i \cap \tau_j$ (hence the last requirement). This means that $\F_G(\U)$ is equivalent to:
$$\F_G(\U) = \{\sigma \neq \emptyset \mid  
\sigma \cap \tau_i \in \tilFP(G|_{\tau_i}) \; \forall \: \tau_i \in \U \},$$
using the notation $\tilFP(G|_{\tau_i}) = \FP(G|_{\tau_i}) \cup \{\emptyset \}.$

It will also be useful to consider the case where $\sigma \cap \tau_i$ is not allowed to be empty for any $i$. In this case, we define
$$\F^*_G(\U) \od \{\sigma \subseteq [n] \mid \sigma \cap \tau_i \in \FP(G|_{\tau_i}) \; \forall \: \tau_i \in \U\}.$$

Translating Lemma~\ref{lemma:se-restriction} into the new notation yields the following:

\begin{lemma}\label{lemma:rule6a}
A CTLN with graph $G$ and simply-embedded cover $\U$ satisfies
$$\FP(G) \subseteq \F_G(\U).$$
\end{lemma}

The central question addressed by gluing rules (Question~\ref{ques:gluing}) thus translates to:
What elements of $\F_G(\U)$ are actually in $\FP(G)$? 
\medskip

\noindent \carina{\bf Some examples.}
\carina{Before delving into this question, we make a few observations. First, note that although $\F_G(\U)$ is never empty (it must contain $\FP(G)$), the set $\F^*_G(\U)$ may be empty.}

\begin{figure}[!h]
\begin{center}
\includegraphics[width=2.75in]{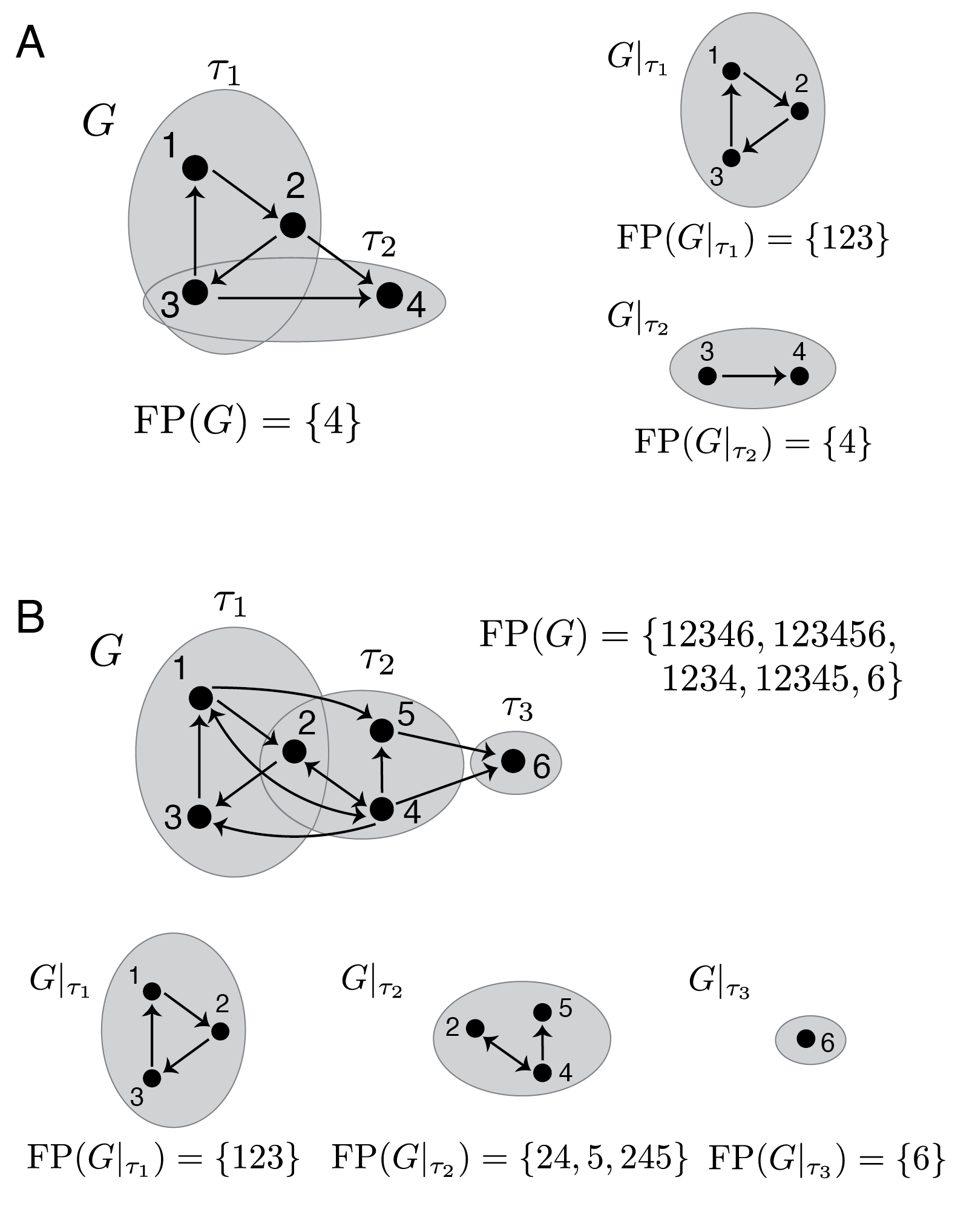}
\end{center}
\vspace{-.2in}
\caption{\carina{Two networks with simply-embedded covers.}}
\label{fig:curlyF-examples}
\end{figure}

\carina{For example, in Figure~\ref{fig:curlyF-examples}A, $\F^*_G(\U) = \emptyset$, because the only option for $\sigma \cap \tau_1$ is $\{123\}$, and this would imply $3 \in \sigma \cap \tau_2$; but there is no such option in $\FP(G|_{\tau_2}).$ On the other hand, if we are allowed $\sigma \cap \tau_i = \emptyset$, we can choose $\sigma = \{4\}$ and satisfy both $\sigma \cap \tau_1 \in \tilFP(G|_{\tau_1})$ and $\sigma \cap \tau_2 \in \tilFP(G|_{\tau_2})$. In fact, this is the only such choice and therefore $\F_G(\U) = \{4\}$. Since $|\FP(G)| \geq 1$, it follows from Lemma~\ref{lemma:rule6a} that $\FP(G) = \{4\}$. In this case, $\FP(G) = \F_G(\U)$.}

\carina{Figure~\ref{fig:curlyF-examples}B displays another graph, $G$, that has a simply-embedded cover $\U$ with three components, $\tau_1, \tau_2,$ and $\tau_3$. Each set of local fixed point supports, $\FP(G|_{\tau_i})$ (shown at the bottom of Figure~\ref{fig:curlyF-examples}B), can easily be computed using graph rules. Applying the definitions, we obtain:
\begin{eqnarray*}
\F^*_G(\U) &=& \{12346, 123456\},\\
\F_G(\U) &=& \{12346, 123456, 1234, 12345, 56, 5, 6\}.
\end{eqnarray*}
Since $\FP(G) \subseteq \F_G(\U)$, this narrows down the list of candidate fixed point supports in $\FP(G)$. Using Elem Rule 5 (domination), we can eliminate supports $56$ and $5$, since $6$ dominates $5$ with respect to every $\sigma \subseteq [n]$. On the other hand, Elem Rule 4 (uniform in-degree) allows us to verify that $1234, 12345,$ and $123456$ are all fixed point supports of $G$, while Rule 1 and Rule 6 (sinks) tell us that $6, 12346 \in \FP(G)$. We can thus conclude that  $\FP(G) = \{12346, 123456, 1234, 12345, 6\} \subsetneq \F_G(\U).$}

\carina{Note that for both graphs in Figure~\ref{fig:curlyF-examples}, we have $\F^*_G(\U) \subseteq \FP(G) \subseteq \F_G(\U)$. While the second containment is guaranteed by Lemma~\ref{lemma:rule6a}, the first one need not hold in general.}
\medskip

\carina{As mentioned above, the central gluing question is to identify what elements of $\F_G(\U)$ are in $\FP(G)$.}
Our strategy to address this question will be to identify architectures where we can iterate the pairwise gluing rule, Lemma~\ref{lemma:rule6b} (a.k.a. Elem Rule 6(b)). Iteration is possible in a simply-embedded cover $\U = \{\tau_i\}$ provided the unions at each step, $\tau_1 \cup \tau_2 \cup \cdots \cup \tau_\ell,$ are themselves simply-embedded (this may depend on the order). Fortunately, this is the case for several types of natural constructions, including {\it connected unions}, {\it disjoint unions}, {\it clique unions}, and {\it linear chains}, which we consider next. 
Finally, we will examine the case of {\it cyclic unions}, where pairwise gluing rules cannot be iterated, but for which we find an equally clean characterization of $\FP(G)$. 
\carina{All five architectures result in theorems, which we call {\em gluing rules}, that are summarized in Table~\ref{table:gluing-rules}.}

\subsection{\carina{Connected unions}}

Recall that the {\it nerve} of a cover $\U = \{\tau_i\}_{i=1}^N$ is the simplicial complex:
$$\N(\U) \od \{\alpha \subseteq [N] \mid \bigcap_{i \in \alpha} \tau_i \neq \emptyset\}.$$
The nerve keeps track of the intersection data of the sets in the cover. 
We say that a vertex cover $\U = \{\tau_i\}_{i=1}^N$ of $G$ is {\it connected} if its nerve is a connected simplicial complex. This means one can ``walk'' from any $\tau_i$ to any other $\tau_j$ through a sequence of steps between $\tau_i$'s that overlap. (Note that a connected nerve does not imply a connected $G$, or vice versa.) 

\begin{figure*}[!h]
\begin{center}
\includegraphics[width=6.25in]{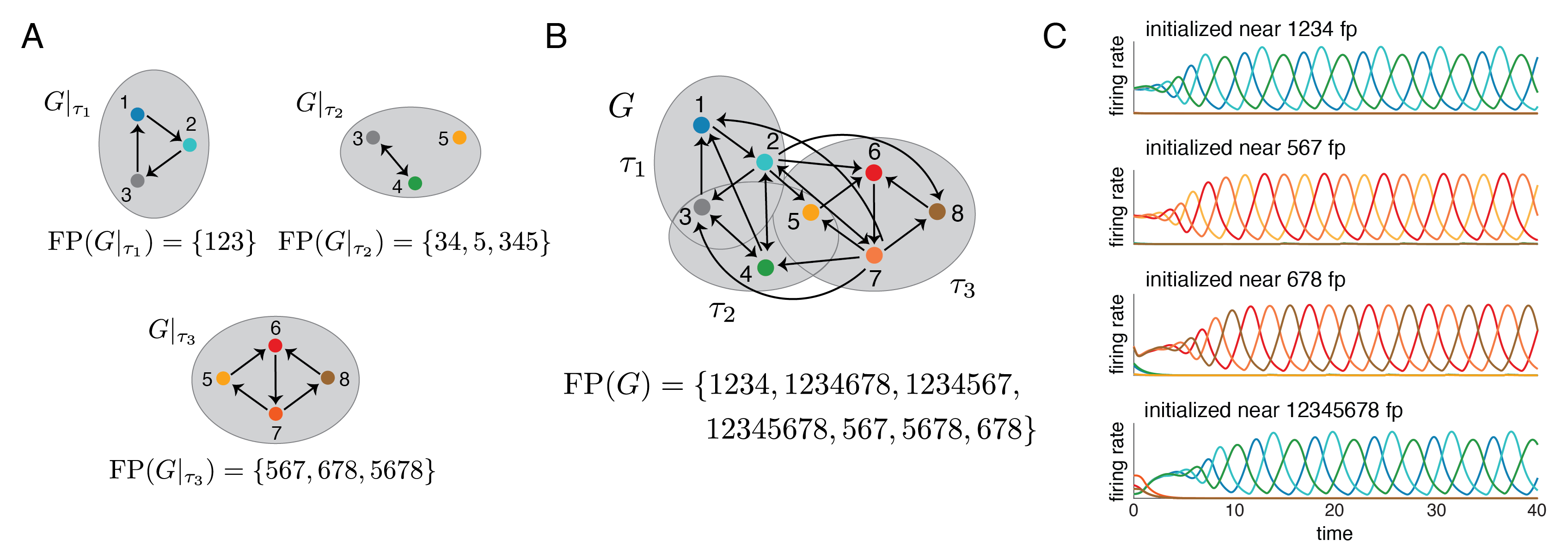}
\end{center}
\vspace{-.1in}
\caption{{\bf Connected union example.} (A) Component subgraphs and their fixed point supports. (B) The full network $G$, with $\FP(G)$ computed using Theorem~\ref{thm:connected-union}. The minimal fixed point supports, 1234, 567, and 678, all correspond to core motifs. Vertices are colored to match the rate curves in C. (C) Several solutions to a CTLN with graph $G$ and parameters $\varepsilon = 0.51, \delta = 1.76,$ and $\theta = 1$. The top three panels show that initial conditions near each of the minimal (core) fixed points produce solutions $x(t)$ that fall into corresponding attractors. The bottom panel shows the solution for an initial condition near the full-support fixed point. Interestingly, even though the initial conditions for $x_1, x_2, x_3$ and $x_4$ are lower than those of the other nodes, the solution quickly converges to the attractor corresponding to the core motif $G|_{1234}$ (same as in the top panel).}
\vspace{-.1in}
\label{fig:rule6-connected}
\end{figure*}

Any graph $G$ admits vertex covers that are connected. Having a connected cover that is also simply-embedded, however, is quite restrictive. We call such architectures {\it connected unions}:

\begin{definition}
A graph $G$ is a {\em connected union} of induced subgraphs $\{G|_{\tau_i}\}$ if $\{\tau_1,\ldots,\tau_N\}$ is a simply-embedded cover of $G$ that is also connected.
\end{definition}

If $G$ has a connected simply-embedded cover, then without loss of generality we can enumerate the sets $\tau_1,\ldots,\tau_N$ in such a way that each partial union $\tau_1 \cup \tau_2 \cup \cdots \cup \tau_\ell$ is also simply-embedded in $G$, by ensuring that $\tau_\ell \cap (\tau_1 \cup \cdots \cup \tau_{\ell-1}) \neq \emptyset$ for each $\ell$ (see Lemma~\ref{lemma:simply-embedded-intersections-unions}). This allows us to iterate the pairwise gluing rule, Elem Rule 6(b)iii. In fact, by analyzing the different cases with the $\sigma_i$ empty or nonempty, we can determine that all gluings of compatible fixed points supports $\{\sigma_i\}$ are realized in $\FP(G)$. This yields our first gluing rule theorem:

\begin{theorem}\label{thm:connected-union}
If $G$ is a connected union of subgraphs $\{G|_{\tau_i}\}_{i=1}^N$, with $\U = \{\tau_i\}_{i=1}^N$, then
$$\FP(G) = \F_G(\U).$$
\end{theorem}

\carina{It is easy to check that this theorem exactly predicts $\FP(G)$ for the graphs in Figure~\ref{fig:elem-graph-rules}E,F and Figure~\ref{fig:curlyF-examples}A.}

\paragraph{Example.}
To see the power of Theorem~\ref{thm:connected-union}, consider the graph $G$ on $n=8$ nodes in Figure~\ref{fig:rule6-connected}. $G$ is a rather complicated graph, but it has a connected, simply-embedded cover $\{\tau_1 = 123, \tau_2 = 345, \tau_3 = 5678\}$ with subgraphs $G|_{\tau_i}$ given in Figure~\ref{fig:rule6-connected}A. 
\carina{Note that for this graph, the simply-embedded requirement automatically determines all additional edges in $G$. For example, since $2 \to 3$ in $G|_{\tau_1}$, and $3 \in \tau_2$, we must also have $2 \to 4,5$. In contrast, $1 \not\to 3$ in $G|_{\tau_1}$, and hence we must have $1 \not\to 4$ and $1 \not\to 5$.}

\carina{Using simple graph rules, it is easy to compute 
$\FP(G|_{\tau_1}) = \{123\},$  $\FP(G|_{\tau_2}) = \{34, 5, 345\},$ and $\FP(G|_{\tau_3}) = \{567, 678, 5678\},$ as these are small graphs.}
It would be much more difficult to compute the full network's $\FP(G)$ in this way. However, because $G$ is a connected union, Theorem~\ref{thm:connected-union} tells us that $\FP(G) = \F_G(\U).$ By simply checking compatibility on overlaps of the possible $\sigma_i = \sigma \cap \tau_i \in \FP(G|_{\tau_i})$, we can easily compute: 
\begin{eqnarray*}
\FP(G) \; = \; \F_G(\U) &=& \{1234, 1234678, 1234567, \\
&& 12345678, 567, 5678, 678\}.
\end{eqnarray*}

Note that the minimal fixed point supports, $1234, 567,$ and $678,$ are all core motifs: $G|_{1234}$ is a $4$-ufd graph, while the others are $3$-cycles. Moreover, they each have corresponding attractors, as predicted from our previous observations about core motifs \cite{core-motifs}. The attractors are shown in Figure~\ref{fig:rule6-connected}C.

\subsection{\carina{Disjoint unions, clique unions, cyclic unions, and linear chains}}

\carina{Theorem~\ref{thm:connected-union} gave us a nice gluing rule in the case where $G$ has a connected simply-embedded cover. At the other extreme are simply-embedded {\it partitions}. If $\U = \{\tau_i\}_{i=1}^N$ is a simply-embedded partition, then all $\tau_i$'s are disjoint and the nerve $\N(\U)$ is completely disconnected, consisting of the isolated vertices $1,\ldots,N$.}

The following graph constructions all arise from simply-embedded {partitions}. 

\begin{figure*}[!h]
\begin{center}
\includegraphics[width=6.5in]{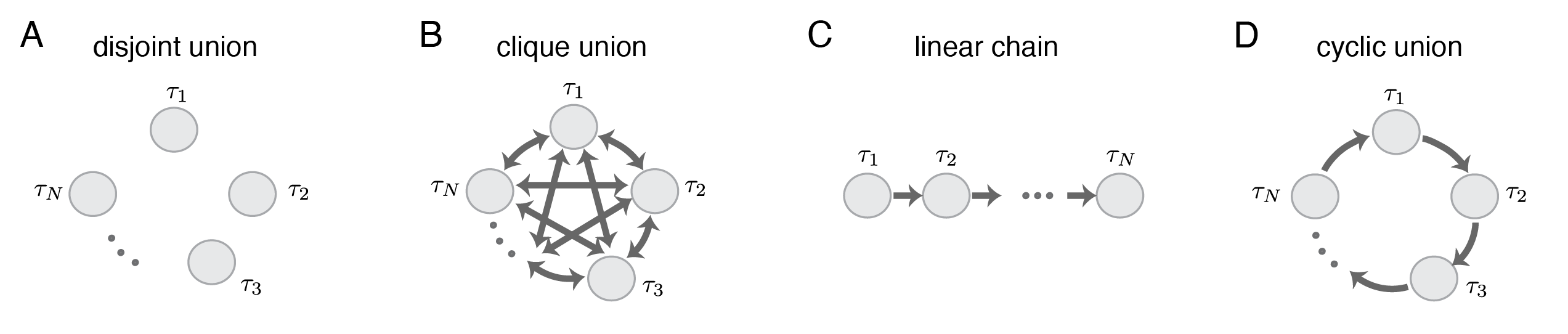}
\end{center}
\caption{{\bf Disjoint unions, clique unions, cyclic unions, \carina{and linear chains}.} In each architecture, the $\{\tau_i\}$ form a simply-embedded partition of $G$.  Thick edges between components indicate directed edges between every pair of nodes in the components.}
\label{fig:composite-graphs}
\end{figure*}

\begin{definition}
Consider a graph $G$ with induced subgraphs $\{G|_{\tau_i}\}$ corresponding to a vertex partition $\U = \{\tau_1,\ldots,\tau_N\}$. Then
\begin{itemize}[--]
\item $G$ is a \emph{disjoint union} if there are no edges between $\tau_i$ and $\tau_j$ for $i \neq j$. 
(See Figure~\ref{fig:composite-graphs}A.)
\item $G$ is a \emph{clique union} if it contains all possible edges between $\tau_i$ and $\tau_j$ for $i \neq j$.
(See Figure~\ref{fig:composite-graphs}B.)
\item \carina{$G$ is a {\em linear chain} if it contains all possible edges from $\tau_i$ to $\tau_{i+1}$, for $i = 1,\ldots,N-1$, and no other edges between distinct $\tau_i$ and $\tau_j$. (See Figure~\ref{fig:composite-graphs}C.)}
\item $G$ is a {\em cyclic union} if it contains all possible edges from $\tau_i$ to $\tau_{i+1}$, for $i = 1,\ldots,N-1$, as well as all possible edges from $\tau_N$ to $\tau_1$, but no other edges between distinct components $\tau_i$, $\tau_j$. (See Figure~\ref{fig:composite-graphs}D.)
\end{itemize}
\vspace{-.05in}
Note that in each of these cases, $\U$ is a simply-embedded partition of $G$.
\end{definition}

\begin{table*}[!h]
\centering
\setstretch{1.5}
\begin{tabular}{l|l|l|l}
 {s-e architecture} & fixed point supports &$|\FP(G)|$ & theorem\\
 \hline
 \hline
connected union  & $\FP(G) = \F_G(\U)$ 
& depends on overlaps
& Thm~\ref{thm:connected-union} \\
 \hline
disjoint union  & $\FP(G) = \F_G(\U)$ 
& $\prod_{i=1}^N (|\FP(G|_{\tau_i})| + 1) -1$
& Thm~\ref{thm:disjoint-union} \\
& = $\{\cup_i\,\sigma_i \mid \sigma_i \in \tilFP(G|_{\tau_i}) \; \forall i \} \setminus \{\emptyset\}$ & \\
 \hline
clique union & $\FP(G) = \F^*_G(\U)$ 
& $\prod_{i=1}^N |\FP(G|_{\tau_i})|$
& Thm~\ref{thm:clique-union}\\
 & = $\{\cup_i\,\sigma_i \mid \sigma_i \in \FP(G|_{\tau_i}) \; \forall i \in [N] \}$ & \\
\hline
linear chain & $\FP(G) = \FP(G|_{\tau_N})$ 
& $|\FP(G|_{\tau_N})|$
& Thm~\ref{thm:linear-chain}\\
\hline
cyclic union & $\FP(G) = \F^*_G(\U)$
& $\prod_{i=1}^N |\FP(G|_{\tau_i})|$
& Thm~\ref{thm:clique-union}\\
& = $\{\cup_i\,\sigma_i \mid \sigma_i \in \FP(G|_{\tau_i}) \; \forall i \in [N] \}$ & \\
\hline
\end{tabular}
\vspace{.15in}
\setstretch{1.0}
\caption{\carina{{\bf Summary of gluing rules.} For each simply-embedded architecture, $\FP(G)$ is given in terms
of the $\FP(G|_{\tau_i})$'s for component subgraphs.}}
\label{table:gluing-rules}
\end{table*} 

Since the simply-embedded subgraphs in a partition are all disjoint, Lemma~\ref{lemma:rule6b}(i-ii) applies. Consequently, fixed point supports $\sigma_i \in \FP(G|_{\tau_i})$ and $\sigma_j \in \FP(G|_{\tau_j})$ will glue together if and only if either $\sigma_i$ and $\sigma_j$ both survive to yield fixed points in $\FP(G)$, or neither survives. 
For both disjoint unions and clique unions,  
it is easy to see that all larger unions of the form $\tau_1 \cup \tau_2 \cup \cdots \cup \tau_\ell$ are themselves simply-embedded. We can thus iteratively use the pairwise gluing Lemma~\ref{lemma:rule6b}.
For disjoint unions, Lemma~\ref{lemma:rule6b}(i) applies, since every $\sigma_i \in \FP(G|_{\tau_i})$ survives in $G$. This yields our first gluing theorem.  Recall that $\tilFP(G) = \FP(G) \cup \{\emptyset\}$.

\begin{theorem}\cite[Theorem 11]{fp-paper}\label{thm:disjoint-union}
If $G$ is a disjoint union of subgraphs $\{G|_{\tau_i}\}_{i=1}^N$, with $\U = \{\tau_i\}_{i=1}^N$, then  
\begin{small}
\begin{eqnarray*}
\FP(G) &=& \F_G(\U)\\
&=& \{\cup_{i=1}^N \sigma_i \mid \sigma_i \in \tilFP(G|_{\tau_i}) \: \forall \: i \in [N] \} \setminus \{\emptyset\}.
\end{eqnarray*}
\end{small}
\end{theorem}

\carina{\noindent Note that this looks identical to the result for connected unions, Theorem~\ref{thm:connected-union}.
One difference is that compatibility of $\sigma_i$'s need not be checked, since the $\tau_i$'s are disjoint, so $\F_G(\U)$ is 
particularly easy to compute. In this case the size of $\FP(G)$ is also the maximum possible for a graph with a simply-embedded cover $\U$: 
$$|\FP(G)| = \prod_{i=1}^N (|\FP(G|_{\tau_i})| + 1) -1.$$}

On the other hand, for clique unions, we must apply Lemma~\ref{lemma:rule6b}(ii), which shows that only gluings involving a \emph{nonempty} $\sigma_i$ from each component are allowed.  Hence $\FP(G) = \F^*_G(\U)$.  Interestingly, the same result holds for cyclic unions, but the proof is different because the simply-embedded structure does {\it not} get preserved under unions, and hence Lemma~\ref{lemma:rule6b} cannot be iterated.  These results are combined in the next theorem.

\begin{theorem}\cite[Theorems 12 and 13]{fp-paper}\label{thm:clique-union}
If $G$ is a clique union or a cyclic union of subgraphs $\{G|_{\tau_i}\}_{i=1}^N$, with $\U = \{\tau_i\}_{i=1}^N$, then
\begin{eqnarray*}
\FP(G) &=& \F^*_G(\U)\\
&=& \{\cup_{i=1}^N \sigma_i \mid \sigma_i \in \FP(G|_{\tau_i}) \: \forall \: i \in [N] \}.
\end{eqnarray*}
\end{theorem}

\noindent \carina{In this case, 
$|\FP(G)| = \prod_{i=1}^N |\FP(G|_{\tau_i})|.$}

\carina{Finally, we consider linear chain architectures. In the case of a linear chain (Figure~\ref{fig:composite-graphs}C), the gluing sequence must respect the ordering $\tau_1,\ldots,\tau_N$ in order to guarantee that the unions $\tau_1 \cup \tau_2 \cup \cdots \cup \tau_\ell$ are all simply-embedded. (In the case of disjoint and clique unions, the order didn't matter.) 
Now consider the first pairwise gluing, with $\tau_1$ and $\tau_2$. Each $\sigma_1 \in \FP(G|_{\tau_1})$ has a target in $\tau_2$, and hence does not survive to $\FP(G|_{\tau_1 \cup \tau_2})$ (by Rule 5(ii)). On the other hand, any $\sigma_2 \in \FP(G|_{\tau_2})$ has no outgoing edges to $\tau_1$, and is thus guaranteed to survive. Elem Rule 6(b) thus tells us that $\sigma_1 \cup \sigma_2 \notin \FP(G|_{\tau_1 \cup \tau_2})$ unless $\sigma_1 = \emptyset$. Therefore, $\FP(G|_{\tau_1 \cup \tau_2}) = \FP(G|_{\tau_2})$. Iterating this procedure, adding the next $\tau_i$ at each step, we see that $\FP(G|_{\tau_1 \cup \cdots \cup \tau_\ell}) = \FP(G|_{\tau_\ell})$. 
In the end, we obtain our fourth gluing theorem:
\begin{theorem}\cite{seq-attractors} \label{thm:linear-chain}
If $G$ is a linear chain of subgraphs $\{G|_{\tau_i}\}_{i=1}^N$, then
$$\FP(G) = \FP(G|_{\tau_N}).$$
\end{theorem}}
\noindent \carina{Clearly, $|\FP(G)|  = |\FP(G|_{\tau_N})|$ in this case.}

\carina{Table~\ref{table:gluing-rules} summarizes the gluing rules for connected unions, disjoint unions, clique unions, cyclic unions, and linear chains.}

\subsection{Applications of gluing rules to core motifs}\label{sec:gluing-cores}

\carina{Using the above results, it is interesting to revisit the subject of core motifs.}
Recall that core motifs of CTLNs are subgraphs $G|_\sigma$ that support a unique fixed point, which has full-support: $\FP(G|_\sigma) = \{\sigma\}$. 
We denote the set of surviving core motifs by
\begin{small}
$$\FPcore(G) \od \{\sigma \in \FP(G)~|~ G|_\sigma \text{ is a core motif of } G\}.$$
\end{small}

\noindent For small CTLNs, we have seen that core motifs are predictive of a network's attractors \cite{core-motifs}. We also saw this in Figure~\ref{fig:rule6-connected}, with attractors corresponding to the core motifs in a CTLN for a connected union.

\carina{What can gluing rules tell us about core motifs? Consider the architectures in Table~\ref{table:gluing-rules}. In the case of disjoint unions, we know that we can never obtain a core motif, since $|\FP(G)| = |\F_G(\U)| \geq 3$ whenever there is more than one component subgraph. In the case of connected unions, however, we have a nice result in the situation where all components $\tau_i$ are core motifs. In this case, 
the additional compatibility requirement on overlaps forces $\FP(G) = \F_G(\U) = \{[n]\}$.}

\begin{corollary}
\carina{If $G$ is a connected union of core motifs, then $G$ is a core motif.}
\end{corollary}

\begin{proof}
\carina{Let $G|_{\tau_1}, \ldots, G|_{\tau_N}$ be the component core motifs for the connected union $G$, a graph on $n$ nodes. Since $\U = \{\tau_i\}$ is a connected cover, and each component has $\FP(G|_{\tau_i}) = \{\tau_i\}$, the only possible $\sigma \in \F_G(\U)$ arises from taking $\sigma_i = \tau_i$ in each component, so that $\sigma = [n]$. (By compatibility, taking an empty set in any component forces choosing an empty set in all components, yielding $\sigma = \cup \sigma_i = \emptyset$, which is not allowed in $\F_G(\U)$.) Applying Theorem~\ref{thm:connected-union},  we see that $\FP(G) = \F_G(\U) = \{[n]\}$. Hence, $G$ is a core motif.}
\end{proof}

\carina{As of this writing, we have no good reason to believe the converse is true. However, we have yet to find a counterexample.}

\carina{In the case of clique unions and cyclic unions, however, $\FP(G) = \F_G^*(\U)$, and gluing in empty sets is again not allowed on components. 
In these cases, we obtain a similar result, and the converse is also true. }

\begin{corollary}\label{cor:cyclic-union-cores}
Let $G$ be a clique union or a cyclic union of components $\tau_1, \ldots, \tau_N$. Then 
$$\FPcore(G) = \{ \cup_{i=1}^N \sigma_i ~|~ \sigma_i \in \FPcore(G|_{\tau_i}) \}. \vspace{-.05in}$$
In particular, $G$ is a core motif if and only if every $G|_{\tau_i}$ is a core motif.
\end{corollary}

\begin{proof}
\carina{We will prove the second statement. The expression for $\FPcore(G)$ easily follows from this together with Elem Rule 6(c).
Let $G$ be a clique union or a cyclic union for a simply-embedded partition $\U = \{\tau_i\}$. Theorem~\ref{thm:clique-union} tells us that 
$\FP(G) = \F_G^*(\U)$. Observe that any $\sigma \in \F_G^*(\U)$ must have nonempty $\sigma_i = \sigma \cap \tau_i \in \FP(G|_{\tau_i})$ for each $i$. 
($\Leftarrow$) If each $G|_{\tau_i}$ is a core motif, it follows that $\sigma_i = \tau_i$ for each $i$, and hence $\FP(G) = \{[n]\}$. ($\Rightarrow$) 
If the component graphs are not all core, then $\FP(G)$ will necessarily have more than one fixed point and $G$ cannot be core.}\end{proof}

\carina{Going back to Figure~\ref{fig:n4-cores}, we can now see that all core motifs up to size $n=4$ are either clique unions, cyclic unions, or connected unions of smaller core motifs. For example, the $4$-cycu graph is the cyclic union of a singleton (node $1$), a $2$-clique (nodes $2$ and $3$), and another singleton (node $4$). The fusion $3$-cycle is a clique union of a $3$-cycle and a singleton. Finally, the $4$-ufd is the connected union of a $3$-cycle and a $2$-clique. Infinite families of core motifs can be generated in this way, each having their own particular attractors.}

\subsection{Modeling with cyclic unions}

The power of graph rules is that they enable us to reason mathematically about the graph of a CTLN and make surprisingly accurate predictions about the dynamics.  
This is particularly true for cyclic unions, where the dynamics consistently appear to traverse the components in cyclic order.
Consequently, these architectures are useful for modeling a variety of phenomena that involve sequential attractors.  This includes the storage and retrieval of sequential memories, as well as CPGs responsible for rhythmic activity, such as locomotion \cite{Marder-CPG,Yuste-CPG}.

\begin{figure}[!h]
\begin{center}
\includegraphics[width=2.5in]{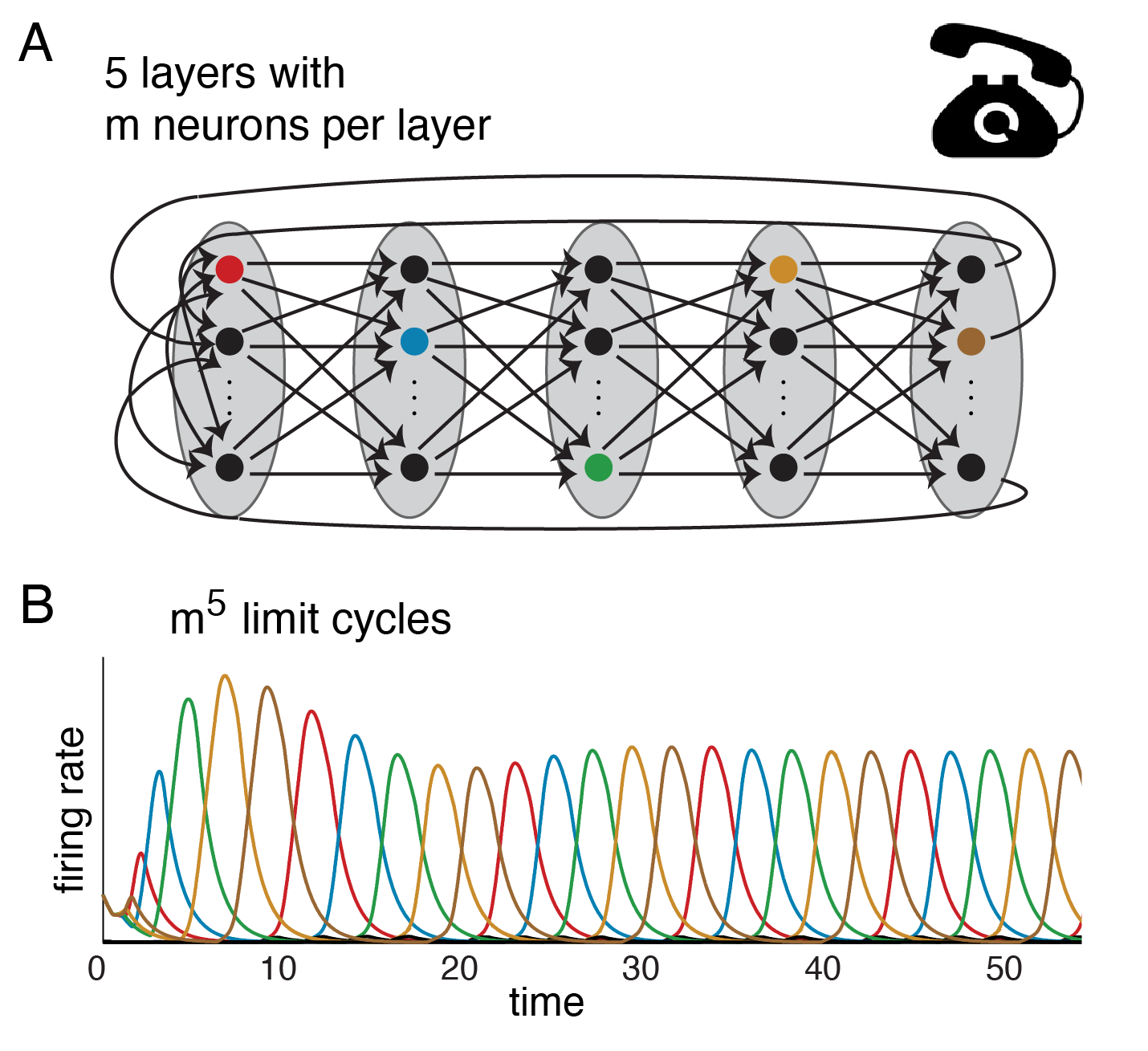}
\caption{{\bf The phone number network.} (A) A cyclic union with $m$ neurons per layer (component), and all $m^2$ feedforward connections from one layer to the next. (B) A limit cycle for the corresponding CTLN (with parameters $\varepsilon=0.75$, $\delta=4$). 
}
\label{fig:phone-number}
\end{center}
\vspace{-.3in}
\end{figure}

Recall that the attractors of a network tend to correspond to core motifs in $\FPcore(G)$.  Using Corollary~\ref{cor:cyclic-union-cores}, we can easily engineer cyclic unions that have multiple sequential attractors.  For example, consider the cyclic union in Figure~\ref{fig:phone-number}A, with $\FPcore(G)$ comprised of all cycles of length 5 that contain exactly one node per component.  
For parameters $\varepsilon=0.75$, $\delta=4$, the CTLN yields a limit cycle (Figure~\ref{fig:phone-number}B), corresponding to one such core motif, with sequential firing of a node from each component.  By symmetry, there must be an equivalent limit cycle for every choice of 5 nodes, one from each layer, and thus the network is guaranteed to have $m^5$ limit cycles.  Note that this network architecture, increased to 7 layers, could serve as a mechanism for storing phone numbers in working memory ($m = 10$ for digits $0-9$). 

\begin{figure}[!h]
\begin{center}
\includegraphics[width=3.2in]{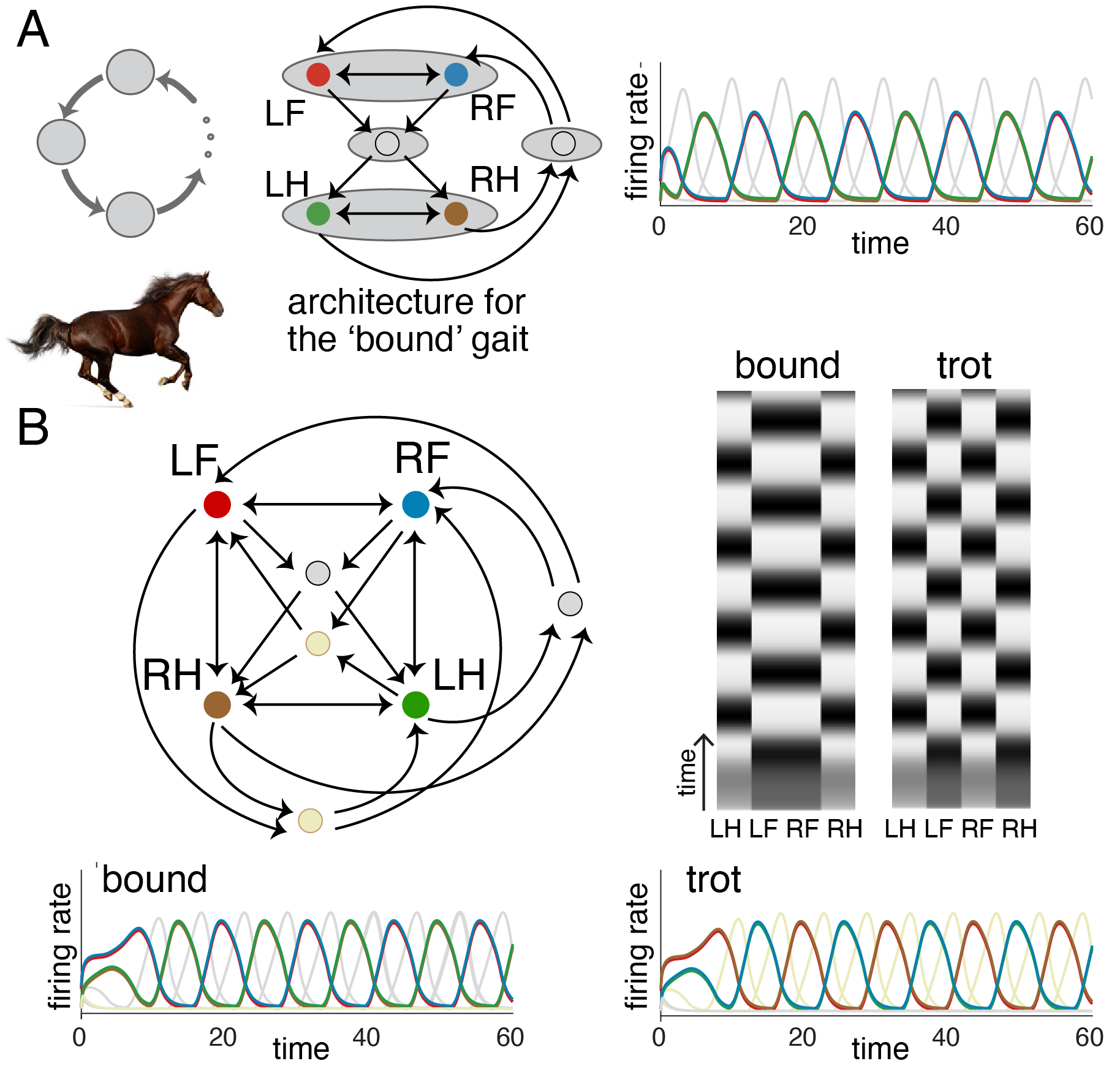}
\caption{{\bf A Central Pattern Generator circuit for quadruped motion.} (A) (Left) A cyclic union architecture on 6 nodes that produces the `bound' gait.  (Right) The limit cycle corresponding to the bound gait.
(B) The graph on 8 nodes is formed from merging together architectures for the individual gaits, `bound' and `trot'.  Note that the positions of the two hind legs (LH, RH) are flipped for ease of drawing the graph.}
\label{fig:gallop-trot}
\end{center}
\vspace{-.2in}
\end{figure}

As another application of cyclic unions, consider the graph in Figure~\ref{fig:gallop-trot}A which produces the quadruped gait `bound' (similar to gallop), where we have associated each of the four colored nodes with a leg of the animal.  Notice that the clique between pairs of legs ensures that those nodes co-fire, and the cyclic union structure guarantees that the activity flows forward cyclically.    
A similar network was created for the `trot' gait, with appropriate pairs of legs joined by cliques.  

Figure~\ref{fig:gallop-trot}B shows a network in which both the `bound' and `trot' gaits can coexist, with the network selecting one pattern (limit cycle) over the other based solely on initial conditions.  This network was produced by essentially overlaying the two architectures that would produce the desired gaits, identifying the two graphs along the nodes corresponding to each leg.  Notice that within this larger network, the induced subgraphs for each gait are no longer perfect cyclic unions (since they include additional edges between pairs of legs), and are no longer core motifs.  And yet the combined network still produces limit cycles that are qualitatively similar to those of the isolated cyclic unions for each gait.  It is an open question when this type of merging procedure for cyclic unions (or other types of subnetworks) will preserve the original limit cycles within the larger network.

\section{Conclusions}

Recurrent network models such as TLNs have historically played an important role in theoretical neuroscience; they give mathematical grounding to key ideas about neural dynamics and connectivity, and provide concrete examples of networks that encode multiple attractors. These attractors represent the possible responses, e.g. stored memory patterns, of the network.

In the case of CTLNs, we have been able to prove a variety of results, such as graph rules, about the fixed point supports $\FP(G)$ -- yielding valuable insights into the attractor dynamics. Many of these results can be extended beyond CTLNs to more general families of TLNs, and potentially to other threshold nonlinearities. The reason lies in the combinatorial geometry of the hyperplane arrangements. In addition to the arrangements discussed in Section~\ref{sec:hyperplanes}, there are closely related hyperplane arrangements given by the {\it nullclines} of TLNs, defined by $dx_i/dt = 0$ for each $i$. It is easy to see that fixed points correspond to intersections of nullclines, and thus the elements of $\FP(W,b)$ are completely determined by the combinatorial geometry of the nullcline arrangement. Intuitively, the combinatorial geometry of such an arrangement is preserved under small perturbations of $W$ and $b$. This allows us to extend CTLN results and study how $\FP(W,b)$ changes as we vary the TLN parameters $W_{ij}$ and~$b_i$. \carina{These ideas, including  connections to oriented matroids, were further developed in \cite{oriented-matroids-paper}.}

\carina{In addition to gluing rules, we have also studied graphs with simply-embedded covers and related structures in order to predict the sequential attractors of a network \cite{seq-attractors}. This has led us to introduce the notions of 
{\it directional graphs} and {\it directional covers}, allowing us to generalize cyclic unions and DAGs. In particular, we were able to prove various {\it nerve theorems} for CTLNs, wherein the dynamics of a network with a directional cover can be described via the dynamics of a reduced network defined on the nerve \cite{nerve-thms-ctlns}.}

\carina{Finally, although the theory of TLNs and CTLNs has progressed significantly in recent years, many open questions remain. We end with a partial list.}

\subsection{\carina{Open Questions}}

We group our open questions into four categories.\medskip

\noindent The first category concerns the bifurcation theory of TLNs, focusing on changes in $\FP(W,b)$ as one varies $W$ or $b$:

\begin{enumerate}
\item[1.] Recall the definition, in equation~\eqref{eq:FP_Wb}, of $\FP(W,b)$ for an arbitrary TLN $(W,b)$. How does the set of fixed point supports change as we vary $W$ or $b$? What are the possible bifurcations? For example, what pairs of supports, $\{\sigma, \tau\}$, can disappear or co-appear at the same time? 

This first question is very general. The next two questions focus on special cases where partial progress has already been made.

\item[2.] If we look beyond CTLNs, but constrain the $W$ matrix to respect a given architecture $G$, how does this constrain the possibilities for $\FP(W,b)$? 

In the case of constant $b_i = \theta$ across neurons, we have identified {\it robust motifs}, graphs for which $\FP(W,b)$ is invariant across all compatible choices of $W$ \cite{robust-motifs}. What graphs allow only a few possibilities for $\FP(W,b)$? What are the most flexible graphs for which $\FP(W,b)$ can vary the most? 

\item[3.] What happens if we fix $W$ and vary $b \in \RR^n$? What features of the connectivity matrix $W$ control the repertoire of possible fixed point regimes, $\FP(W,b)$? What $W$ matrices allow a {\it core motif region}, for which $\FP(W,b) = \{[n]\}$? And how do the dynamic attractors of a network change as we transition between different regions in $b$-space?
\end{enumerate}

\noindent The second category concerns the relationship between TLNs and the geometry of the associated hyperplane arrangements:

\begin{enumerate}
\item[4.] To what extent does the hyperplane arrangement of a TLN, as described in Section~\ref{sec:hyperplanes}, determine its dynamics? What are all the $(W,b)$ choices that have the same hyperplane arrangement? Same nullcline arrangement? 

\item[5.] What happens if we change the nonlinearity in equation~\eqref{network-setup} from $\varphi(y) = [y]_+$ to a sigmoid function, a threshold power-law nonlinearity \cite{ken-miller-thresh-power-law}, or something else? 
Can we adapt the proofs and obtain similar results for $\FP(W,b)$ and $\FP(G)$ in these cases? 

Note that the combinatorial geometry approach in \cite{oriented-matroids-paper} suggests that the results should not depend too heavily on the details of the nonlinearity. Instead, it is the resulting arrangement of nullclines that is essential for determining the fixed points. 
\end{enumerate}

\noindent The third category concerns graph rules, core motifs, and the corresponding attractors:

\begin{enumerate}
\item[6.] What other graph rules or gluing rules follow from the elementary graph rules? We believe our current list is far from exhaustive.

\item[7.] Classify all core motifs for CTLNs. We already have a classification for graphs up to size $n=5$ \cite{n5-github}, but beyond this little is known. Note that gluing rules allow us to construct infinite families of core motifs from gluing together smaller component cores (see Section~\ref{sec:gluing-cores}). Are there other families of core motifs that cannot be obtained via gluing rules? What can we say about the corresponding attractors?

\item[8.] Computational evidence suggests a strong correspondence between core motifs and the attractors of a network, at least in the case of small CTLNs \cite{core-motifs, n5-github}. Can we make this correspondence precise? Under what conditions does the correspondence between surviving core fixed points and attractors hold?

\item[9.] How does symmetry affect the attractors of a network? The automorphism group of a graph $G$ naturally acts on an associated CTLN by permuting the variables, $\{x_i\}$. This translates to symmetries of the defining vector field~\eqref{eq:TLN-dynamics}, and a group action on the set of attractors. The automorphism group can either fix attractors or permute them. Moreover, a network may also have ``surprise symmetry,'' as in Figure~\ref{fig:outer-neuron}, where the attractors display additional symmetry that was not present in the original graph $G$. How do we make sense of these various phenomena?
\end{enumerate}

\noindent Finally, the fourth category collects various conjectures about dynamic behaviors that we have observed in simulations.

\begin{enumerate}

\item[10.] In \cite{CTLN-preprint, stable-fp-paper} we conjectured that all stable fixed points of a CTLN correspond to {\it target-free} cliques. While \cite{stable-fp-paper} provides proofs of this conjecture in special cases, the general question remains open.

\item[11.] The Gaudi attractor from Figure~\ref{fig:gaudi} appears to have constant total population activity. In other words, $\sum_{i=1}^5 x_i(t)$ appears to be constant in numerical experiments, once the trajectory has converged to the attractor. Can we prove this? For what other (non-static) TLN/CTLN attractors is the total population activity conserved?

\item[12.] Prove that the ``baby chaos'' network in Figure~\ref{fig:tadpole}D-F is chaotic. I.e., prove that the individual attractors are chaotic (or strange), in the same sense as the Lorenz or Rossler attractors.

\item[13.] A {\it proper source} of a graph $G$ is a source node $j$ that has at least one outgoing edge, $j \to k$ for $k \neq j$. In numerical experiments, we have observed that proper sources of CTLNs always seem to ``die'' -- that is, their activity $x_j(t)$ tends to zero as $t \to \infty$, regardless of initial conditions. Can we prove this?

Some progress on this question was made in \cite{caitlin-thesis}, but the general conjecture remains open. Note that although the sources Rule 4(i) guarantees that proper sources do not appear in any fixed point support of $\FP(G)$, this alone does not imply that the activity at such nodes converges to zero. 

\item[14.] In our classification of attractors for small CTLNs, we observed that if two CTLNs with distinct graphs have the ``same'' attractor, as in Figure~\ref{fig:n5-dictionary}, then this attractor is preserved for the entire family of TLNs whose $W$ matrices linearly interpolate between the two CTLNs (and have the same constant $b_i = \theta$ for all $i$). In other words, the attractor persists for all TLNs $(W_t, \theta)$ with $W_t = (1-t)W_0 + tW_1$ and $t \in [0,1]$, where $W_0$ and $W_1$ are the two CTLN connectivity matrices. (Note that the interpolating networks $W_t$ for $t \in (0,1)$ are {\it not} CTLNs.) Can we prove this? 

More generally, we conjecture that if the same attractor is present for a set of TLNs $(W_1,b), \ldots, (W_m,b)$, then it is present for all TLNs $(W,b)$ with $W$ in the convex hull of the $W_i$ matrices.
\end{enumerate}

\vspace{-.2in}
\section*{Acknowledgments} We would like to thank Zelong Li, Nicole Sanderson, and Juliana Londono Alvarez for a careful reading of the manuscript. We also thank Caitlyn Parmelee, Caitlin Lienkaemper, Safaan Sadiq, Anda Degeratu, Vladimir Itskov, Christopher Langdon, Jesse Geneson, Daniela Egas Santander, Stefania Ebli, Alice Patania, Joshua Paik, Samantha Moore, Devon Olds, and Joaquin Casta\~{n}eda for many useful discussions. The first author was supported by NIH R01 EB022862, NIH R01 NS120581, NSF DMS-1951165, and a Simons Fellowship. The second author was supported by NIH R01 EB022862 and NSF DMS-1951599.
\bigskip

\bibliographystyle{unsrt}
\bibliography{CTLN-refs-jan2023}

\end{document}